\documentclass{article}
\usepackage[margin=1in]{geometry}

\usepackage[backend=biber]{biblatex}
\usepackage{graphicx}
\usepackage{mathtools, amsthm, amssymb} 
\usepackage[font=small,labelfont=bf]{subcaption}
\usepackage{bm}
\usepackage{array}
\usepackage{adjustbox}
\usepackage{relsize}
\usepackage{nicefrac,xfrac}
\newcommand{\rrvert}{\vert}
\newcommand{\llvert}{\vert}
\newcommand{\rem}[1]{}

\usepackage{tikz}
\usepackage{tikz-network}
\usetikzlibrary{positioning}
\usetikzlibrary{circuits.logic.US,circuits.logic.IEC,fit}
\newcommand\addvmargin[1]{
  \node[fit=(current bounding box),inner ysep=#1,inner xsep=0]{};
}
\usepackage{xcolor}
\tikzset{global scale/.style={
    scale=#1,
    every node/.style={scale=#1}
  }
}

\usepackage{tikz-cd}
\newtheorem{definition}{Definition}

\newtheorem{prop}{Proposition}
\usepackage[mathscr]{euscript}
\usepackage[pdftex,bookmarksnumbered,hidelinks,breaklinks]{hyperref}
\makeatletter
\g@addto@macro\UrlBreaks{\do\-}
\makeatother
\usepackage{booktabs}
\usepackage{enumitem}
\usepackage{kbordermatrix}
\usepackage[font=footnotesize,labelfont=bf]{caption}

\renewcommand{\kbldelim}{(}
\renewcommand{\kbrdelim}{)}

\makeatletter
\newcommand{\subalign}[1]{%
  \vcenter{%
    \Let@ \restore@math@cr \default@tag
    \baselineskip\fontdimen10 \scriptfont\tw@
    \advance\baselineskip\fontdimen12 \scriptfont\tw@
    \lineskip\thr@@\fontdimen8 \scriptfont\thr@@
    \lineskiplimit\lineskip
    \ialign{\hfil$\m@th\scriptstyle##$&$\m@th\scriptstyle{}##$\crcr
      #1\crcr
    }%
  }
}
\makeatother

\definecolor{mycolor}{rgb}{1,0.2,0.3}
\definecolor{myBlue}{rgb}{0.12156863, 0.46666667, 0.70588235}
\definecolor{myOrange}{rgb}{1, 0.49803922, 0.05490196}
\definecolor{myGreen}{rgb}{0.17254902, 0.62745098, 0.17254902}
\definecolor{myRed}{rgb}{0.83921569, 0.15294118, 0.15686275}
\definecolor{myPurple}{rgb}{0.58039216, 0.40392157, 0.74117647}

\definecolor{salmon}{RGB}{250,128,114}

\title{Hypernetwork Science via High-Order Hypergraph Walks}

\author{Sinan G. Aksoy\thanks{Pacific Northwest National Laboratory, Richland, WA 99354, {\tt sinan.aksoy@pnnl.gov}, {\tt carlos.ortizmarrero@pnnl.gov} }, Cliff Joslyn\thanks{Pacific Northwest National Laboratory, Seattle, WA 98109, {\tt cliff.joslyn@pnnl.gov},  {\tt Brenda.Praggastis@pnnl.gov}, {\tt Emilie.Purvine@pnnl.gov}}, Carlos Ortiz Marrero\footnotemark[1], Brenda Praggastis\footnotemark[2], Emilie Purvine\footnotemark[2]}

\date{\today}

\begin{document}

\newcolumntype{C}[1]{%
 >{\vbox to 5ex\bgroup\vfill\centering}%
 p{#1}%
 <{\egroup}}

\maketitle

\abstract{We propose high-order hypergraph walks as a framework to generalize graph-based
network science techniques to hypergraphs. Edge incidence in hypergraphs
is quantitative, yielding hypergraph walks with both length and width.
Graph methods which then generalize to hypergraphs include connected component
analyses, graph distance-based metrics such as closeness centrality, and
motif-based measures such as clustering coefficients. We apply high-order
analogs of these methods to real world hypernetworks, and show they reveal
nuanced and interpretable structure that cannot be detected by graph-based
methods. Lastly, we apply three generative models to the data and find
that basic hypergraph properties, such as density and degree distributions,
do not necessarily control these new structural measurements. Our work
demonstrates how analyses of hypergraph-structured data are richer when
utilizing tools tailored to capture hypergraph-native phenomena, and suggests
one possible avenue towards that end.
}

\section{Introduction}\label{sec1}

In the study of complex systems, graph theory is often perceived as the
mathematical scaffold underlying network science \cite{Barabasi2016}. Systems
studied in biology, sociology, telecommunications, and physical infrastructure
often afford a representation as a set of entities (``vertices'') with
binary relationships (``edges''), and hence may be analyzed utilizing graph
theoretic methods. Graph models benefit from simplicity and a degree of
universality. But as abstract mathematical objects, graphs are limited
to representing \textit{pairwise} relationships between entities. However,
real-world phenomena in these systems can be rich in \emph{multi-way} relationships
involving interactions among more than two entities, dependencies between
more than two variables, or properties of collections of more than two
objects.

Hypergraphs are generalizations of graphs in which edges may connect any
number of vertices, thereby representing $k$-way relationships. As such,
hypergraphs are the natural representation of a broad range of systems,
including those with the kinds of multi-way relationships mentioned above.
Indeed, hypergraph-structured data (i.e. hypernetworks) are ubiquitous,
occurring whenever information presents naturally as set-valued, tabular,
or bipartite data. Additionally, as finite set systems, hypergraphs have
identities related to a number of other mathematical structures important
in data science, including finite topologies, simplicial complexes, and
Sperner systems. This enables use of a wider range of mathematical methods,
such as those from computational topology, to identify features specific
to the high-dimensional complexity in hypernetworks, but not available
using graphs. Although an expanding body of research attests to the increased
utility of hypergraph-based analyses, many network science methods have
been historically developed explicitly (and often, exclusively) for graph-based
analyses. Moreover, it is common that data arising from hypernetworks are
reduced to graphs.

\begin{figure}[t!]
\centering
\includegraphics[scale=0.3]{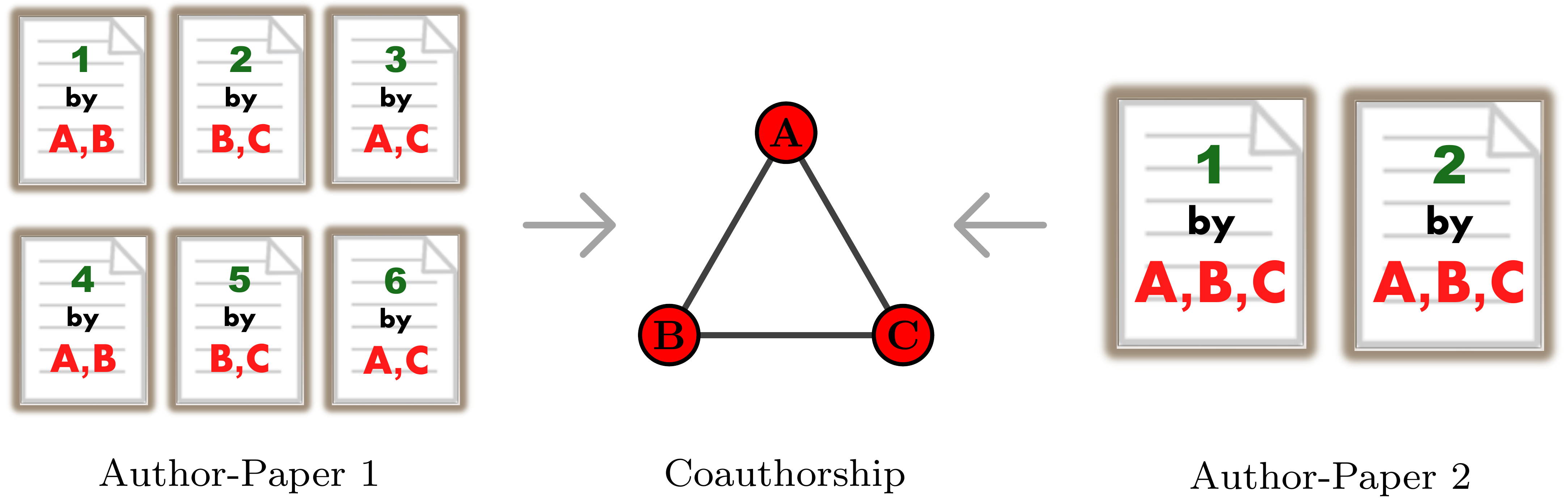}
\caption{Two author-paper networks and their coauthorship
graph. Letters denote authors
 and numbers denote paper titles. These networks may be structured as hypergraphs
on vertices
 $A$, $B$, $C$ with hyperedges $e_{1}=e_{2}=\{A,B,C\}$ for the rightmost network,
and hyperedges
 $e_{1}=e_{4}=\{A,B\}$, $e_{2}=e_{5}=\{B,C\}$, $e_{3}=e_{6}=\{A,C\}$ for
the leftmost. }\label{fig:authPap}
\end{figure}

Before proceeding, let us consider an example. Figure~\ref{fig:authPap} illustrates two author-paper datasets, which may be naturally
structured as a hypergraph by representing authors as vertices, and the
set of authors appearing on each paper as hyperedges.\footnote{One could also have formed a hypergraph by taking papers as vertices
and hyperedges as the set of papers each author has written. In this case,
by virtue of having authors with 4 papers, the hypergraph derived from
the leftmost network exhibits higher-order relationships. The hypergraph
obtained by swapping the roles of vertices is called the \textit{dual} hypergraph.
Duality is an essential consideration in hypernetwork science, which we
discuss further in the Preliminaries section.} The hypergraph derived from the
rightmost network exhibits higher-order relationships by
virtue of having papers with 3 authors. Comparing these
examples highlights structural information retained and lost between graph
and hypergraph representations. For instance, both networks are similar
in that each pair of authors $A$, $B$, $C$ has co-authored a paper (in fact,
exactly two papers) together. This is captured by the coauthorship graph
(center), which is therefore identical for these two networks. However,
there are also clear differences not captured by the graph representation.
For instance, each author appears on 4 versus 2 papers, and each paper
features 2 versus 3 authors. Beyond these basic counts, these networks
also exhibit more subtle differences: for any pair of authors in the leftmost
network, the set of papers they've coauthored is different from the set
of joint papers between any other pair of authors, whereas in the rightmost,
every pair of authors has coauthored exactly the same set of papers. As
this toy example suggests, while graphs do capture some properties of hypernetworks,
they are insufficient as hypergraph substitutes.

In spite of this incongruity between graph and hypergraph
analyses, effectively extending graph theoretical tools to hypergraphs
has sometimes lagged or proven elusive. A~critical aspect of this is axiomatization: as a generalization there are many, sometimes mutually
inconsistent, sets of possible definitions of hypergraph concepts which
can yield the same results consistent with graph theory when instantiated
to the graph case. In some cases, developing \textit{any} coherent hypergraph
analog poses significant theoretical obstacles. For example, extending
the spectral theory of graph adjacency matrices to hypergraphs poses an
immediate challenge in that hyperedges may contain more than two vertices,
thereby rendering the usual (two-dimensional) adjacency matrix insufficient
for encoding adjacency relations. In other cases, graph theoretical concepts
may be trivially extended to hypergraphs, but in doing so ignore structural
nuance native to hypergraphs which are unobservable in graphs. For instance,
while edge incidence and vertex adjacency can occur in at most one vertex
or edge for graphs, these notions are set-valued and hence \textit{quantitative}
for hypergraphs. Consequently, while subsequent graph walk based notions,
such as connectedness, are immediately applicable to hypergraphs, they
ignore high-order structure in failing to account for the varying ``widths''
associated with hypergraph walks.

Due to these challenges, scientists seeking tools to study hypergraph-structured
data are frequently left to contend with disparate approaches towards hypergraph
research. One approach for grappling with hypergraph complexity is to limit
attention to hypergraphs with only uniformly sized edges containing the
same number of vertices. Much of the hypergraph research in the mathematics
literature, such as in hypergraph coloring
\cite{dinur2005hardness,krivelevich2003approximate}, the aforementioned
spectral theory of hypergraphs
\cite{chung1993laplacian,cooper2012spectra}, hypergraph transversals
\cite{alon1990transversal}, and extremal problems
\cite{rodl2004regularity}, focus on this $k$-uniform case only. While imposing
this assumption facilitates more mathematically sophisticated and structurally
faithful analysis of the hypergraphs in question, real-world hypergraph
data is unfortunately very rarely $k$-uniform. Consequently, such tools
are problematic in lacking applicability to real hypernetwork data. Another
approach towards hypergraph research is to limit attention to transformations
of (potentially non-uniform) hypergraphs to graphs. Sometimes called the
hypergraph line graph, 2-section, clique expansion, or one-mode projection,
such transformations clearly enable the application of graph-theoretic
tools to the data. Yet, unsurprisingly, such hypergraph-to-graph reductions
are inevitably lossy \cite{Dewar2018,Kirkland2017}. Hence, although affording
simplicity, such approaches are of limited utility in uncovering hypergraph
structure.

To enable analyses of hypernetwork data that better reflect their complexity
but remain tractable and applicable, we believe striking a balance between
this faithfulness-simplicity tradeoff is essential. With this goal at heart,
we extend a number of graph analytic tools popular in network science to
hypergraphs under the framework of \textit{high-order hypergraph walks}. We
characterize a hypergraph walk as an ``$s$-walk'', where the order
$s$ controls the minimum walk ``width'' in terms of edge overlap size. High-order
$s$-walks ($s>1$) are possible on hypergraphs whereas for graphs, all walks
are 1-walks. The hypergraph walk-based methods we develop include connected
component analyses, graph-distance based metrics such as closeness-centrality,
and motif-based measures such as clustering coefficients. As each of these
methods is based fundamentally on the graph-theoretic notion of a walk,
we extend them to hypergraphs by using hypergraph walks. Ultimately, our
goal is not only to formulate these generalizations in a cogent manner,
but to probe whether these tools reveal \emph{prevalent} and \emph{meaningful}
structure in real hypernetwork data. To the latter end, we compute these
measures based on hypergraph walks on three real datasets from different
domains and discuss the results.

Our work is organized as follows: in Sect.~\ref{sec:Prelim}, we provide
background definitions and review preliminary topics relevant to hypernetwork
theory. In Sect.~\ref{sec:swalk}, we define the $s$-walk notion underpinning
our subsequent work,
discuss related prior research, and reiterate our contributions.
In Sect.~\ref{sec:sMetrics}, we introduce $s$-walk based analytical measures,
apply them to the aforementioned datasets, and analyze the results. In
Sect.~\ref{sec:genModels}, we consider three generative hypergraph models,
and experimentally test the extent to which the structural properties observed
in Sect.~\ref{sec:sMetrics} can be replicated by synthetic models. Finally,
in Sect.~\ref{sec:conc} we conclude and outline several directions for
future research.

\section{Preliminaries} \label{sec:Prelim}

Hypergraphs are generalizations of graphs in which edges may link any number
of vertices together. Just as ``network'' is often used to refer to processes
or systems yielding data streams which are graph-structured, we will
use the term ``hypernetwork'' to refer to those yielding hypergraph-structured
data. More formally, we define a hypergraph as follows:

\begin{definition}%
\label{def:hyp}
A~\emph{hypergraph} $H=(V,E)$ is a set $V=\{v_{1},\ldots,v_{n}\}$ of elements
called \textit{vertices}, and an indexed family of sets
$E=(e_{1},\ldots,e_{m})$ called \textit{hyperedges} in which
$e_{i} \subseteq V$ for $i=1,\ldots, m$.
\end{definition}

When the hypergraph is clear from context,
we call its hyperedges simply ``edges''. The degree
of a vertex is the number of hyperedges to which it belongs,
$d(v)= \llvert  \{e: v \in e\} \rrvert  $, and the size of a hyperedge is its cardinality,
$ \llvert  e \rrvert  $. A~hypergraph in which all hyperedges have size $k$ is called
$k$-uniform, and a $2$-uniform hypergraph is simply a graph.\footnote{More precisely, if a 2-uniform hypergraph contains duplicated hyperedges, it is a multigraph.} Definitions of hypergraphs given in the literature may differ slightly
from author to author. For instance, Bretto's hypergraph definition
\cite{Bretto2013} is identical to ours, apart from prohibiting
empty edges ($e_{i}$ such that $e_{i}=\varnothing $). Berge
\cite{Berge1984} similarly prohibits empty edges, as well as isolated vertices
($v_{i}$ such that $v_{i} \notin \bigcup_{i=1}^{m} e_{i}$). In contrast,
Katona \cite{Katona1975} allows empty edges and isolated vertices, but
defines $E=\{e_{1},\ldots,e_{m}\}$ as a \emph{set} and explicitly prohibits
pairs of duplicated edges $e_{i}=e_{j}$ for $i\neq j$. In defining
$E$ as an (indexed) family of sets, we allow for duplicated edges but require
edges be distinguishable by index.
Returning to the leftmost author-paper network in Fig.~\ref{fig:authPap}, in the corresponding hypergraph with authors as vertices,
the hyperedges corresponding to papers 1 and 4 are examples of duplicate
edges: they are equivalent as sets yet distinguishable by paper title.
The generality of Definition~\ref{def:hyp} in permitting isolated
vertices, as well as empty, duplicated, and singleton edges is intended
to facilitate the application of hypergraphs to real data, which commonly
possess such features.

\begin{definition}
The \emph{incidence matrix} $S$ of a hypergraph $H=(V,E)$, is a
$ \llvert  V \rrvert   \times  \llvert  E \rrvert  $ matrix defined by
\begin{equation*}
S(i,j)= %
\begin{cases}
1 & \text{if }v_{i} \in e_{j},
\\
0 & \text{otherwise}.
\end{cases} %
\end{equation*}
\end{definition}

\label{def:inc}

Under Definition~\ref{def:hyp}, any rectangular
Boolean matrix uniquely defines a labeled hyper\-graph;\footnote{By ``labeled hypergraph'' we mean a hypergraph in which each vertex and edge
are distinguishable via an assignment of distinct labels---this is not
meant to be confused with so-called attributed hypergraphs in which the
vertices and edges have associated metadata.} conversely any labeled hypergraph
uniquely defines an incidence matrix. Consequently,
there is a bijection between hypergraphs and \textit{bicolored graphs}. Recall
a bicolored graph is a triple $(V,E,f)$ where $V$ is a set of vertices,
$E$ is a set of pairs of vertices, and $f: V \to \{0,1\}$ satisfies
$f(v_{i}) \neq f(v_{j})$ for all $v_{i},v_{j} \in V$ where
$\{v_{i},v_{j}\} \in E$. Indexing rows and columns by vertices such that
$f(v_{i})=0$ and $f(v_{j})=1$, respectively, incidence matrices may
be uniquely associated with bicolored graphs by defining $S(i,j)=1$ for
$\{v_{i},v_{j}\}\in E$. 
Note bicolored graphs specify a fixed bicoloring $f$ and differ from bipartite graphs, which are graphs admitting \textit{some} bicoloring. 
Accordingly, a bipartite graph with $k$ connected components has $2^k$ possible bicolorings, each of which may correspond to a distinct hypergraph. 
In applications, however, the data often comes with a bicoloring (e.g. in an author-paper
network) and hence the terms ``bipartite'' and ``bicolored'' graphs have
been used synonymously.
For the purposes of this work, gearing our exposition towards hypergraphs rather than bicolored graphs is more natural because our approach is set-theoretic.\footnote{That is, our focus in this work is on hyperedge
incidences and hyperwalks that arise from sequences of incident hyperedges.
Hyperedges themselves are defined explicitly for hypergraphs, but only
implicitly for bicolored graphs (as the neighborhood of a vertex in the
color class designated for hyperedges). For this reason, framing our exposition
using the language of arbitrary set systems is natural, whereas adopting
the constrained language of bicolored graphs would be cumbersome and confusing.}
Beyond these bijective correspondences, mathematical research
on hypergraph categories and their isomorphisms requires careful consideration
\cite{Dorfler1980,Fong2018,Schmidt2019}.

As an upshot of the hypergraph-bicolored graph correspondence, a number
of complex network analytics for bipartite data extend naturally
to hypergraphs, and \textit{vice versa}. However, interpreting this
in light of the fact that bicolored graphs are \textit{graphs} does not mean graph theoretic methods suffice for studying hypergraphs.
Whether interpreted as bicolored graphs or hypergraphs,
data with this structure often require entirely different network science
methods than (general) graphs. An obvious example is triadic measures
like the graph clustering coefficient: these cannot be applied to bicolored
graphs since (by definition) bicolored graphs have no triangles. Detailed
work developing bipartite analogs of modularity \cite{Barber2007}, community
structure inference techniques \cite{Larremore2014}, and other graph-based
network science topics \cite{Latapy2008} further attests that bipartite
graphs (and hypergraphs) require a different network science toolset than
for graphs.

Another important topic highlighted by the bicolored graphs-hypergraph
correspondence is the duality of hypergraphs. That is, just as
it may be arbitrary to label one partition in a bicolored graph ``left''
and the other partition ``right'', which class of objects one designates
as ``vertices'' versus ``hyperedges'' in a hypernetwork may also be arbitrarily
chosen. However, hypergraph properties and methods may be vertex-based
or edge-based, and hence differ depending on which choice is made. To avoid
limiting one's analysis towards either a vertex-centric or edge-centric
approach, it may be prudent to consider both the hypergraph and its \textit{dual
hypergraph}. Loosely speaking, the dual of a hypergraph is the hypergraph
constructed by swapping the roles of vertices and edges. More precisely:

\begin{definition}
Let $H=(V,E)$ be a hypergraph with vertex set
$V=\{v_{1},\ldots,v_{n}\}$ and family of edges
$E=(e_{1},\ldots,e_{m})$. The \emph{dual hypergraph} of $H$, denoted
$H^{*}=(E^{*},V^{*})$, has vertex set
$E^{*}=\{e_{1}^{*},\ldots,e_{m}^{*}\}$ and family of
hyperedges $V^{*}=(v_{1}^{*},\ldots,v_{n}^{*})$, where
$v_{i}^{*} := \{e_{k}^{*} : v_{i} \in e_{k}\}$.
\end{definition}

Put equivalently, the dual of a hypergraph with incidence matrix $S$ is
the hypergraph associated with the transposed incidence matrix,
$S^{T}$. Clearly, $(H^{*})^{*}=H$. Furthermore, observe that two vertices
belonging to the same set of edges in $H$ correspond to multi-edges in $H^{*}$ and isolated vertices in $H$ correspond to empty edges in
$H^{*}$. Thus, the generality of our Definition~\ref{def:hyp} in permitting
multi-edges, empty edges, and isolated vertices ensures the dual of a hypergraph
is also a hypergraph. Indeed, as a formal matter, one could go so far as
to always consider that hypergraphs present in dual \emph{pairs}.\footnote{Note,
however, this is not necessarily true when restricted to the graph case:
for a graph $G$, its dual $G^{*}$ is 2-uniform (and hence still a graph)
if and only if $G$ is 2-regular, in which case $G$ is a cycle or disjoint
union of cycles.}

Continuing this line of observation, in the complex networks literature,
one of the most oft-used tools for studying hypergraph data is its \textit{line
graph}. In the line graph of a hypergraph, each vertex represents a hyperedge,
and each edge represents an intersection between a pair of hyperedges.
More formally:

\begin{definition}
\label{def:lineGraph}
Let $H=(V,E)$ be a hypergraph with vertex set
$V=\{v_{1},\ldots,v_{n}\}$ and family of hyperedges
$E=(e_{1},\ldots,e_{m})$. The \emph{line graph of $H$}, denoted
$L(H)$, is the graph on vertex set $\{e_{1}^{*},\ldots,e_{m}^{*}\}$ and
edge set
$\{\{e_{i}^{*}, e_{j}^{*}\}: e_{i} \cap e_{j} \neq\varnothing
\text{ for } i \neq j \}$.
\end{definition}

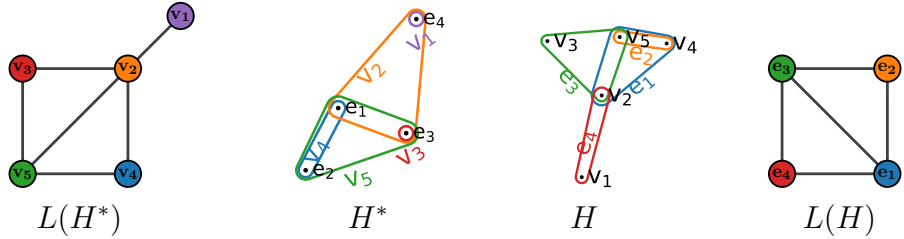
\begin{figure}[t!]
\captionsetup[subfigure]{justification=centering}
\centering
\begin{subfigure}[b]{0.2\textwidth}
\centering
  \scalebox{0.7}{
\begin{tikzpicture}[baseline={(Z.base)}, scale=1]
   	\Vertex[x=0 ,y=0, label=$\mathlarger{\mathlarger{\mathlarger{\bf v_5}}}$,size=0.5, color=myGreen]{5}
   	\Vertex[x=0 ,y=2, label=$\mathlarger{\mathlarger{\mathlarger{\bf v_3}}}$,size=0.5, color=myRed]{3}
   	\Vertex[x=2,y=0, label=$\mathlarger{\mathlarger{\mathlarger{\bf v_4}}}$,size=0.5, color=myBlue]{4}
   	\Vertex[x=2,y=2, label=$\mathlarger{\mathlarger{\mathlarger{\bf v_2}}}$,size=0.5, color=myOrange]{2}
	\Vertex[x=3,y=3, label=$\mathlarger{\mathlarger{\mathlarger{\bf v_1}}}$, size=0.5, color=myPurple]{1}
    \node (Z) at (1.5,0) {} ; 
    \Edge(4)(5)
   \Edge(4)(2)
   \Edge(5)(2)
   \Edge(1)(2)
   \Edge(3)(2)
   \Edge(3)(5)
  \end{tikzpicture}  
}
\caption*{ \hspace{-7mm}  \large $L(H^*)$} \label{fig:line_dual} 
\end{subfigure}
\qquad
\begin{subfigure}[b]{0.13\textwidth}
\includegraphics[width = \linewidth]{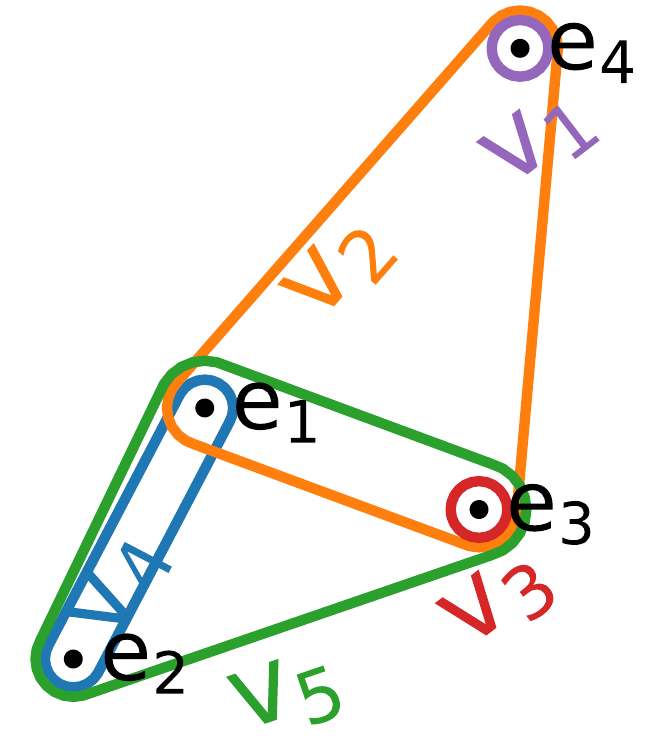} 
\caption*{\large $H^*$} \label{fig:dual}
\end{subfigure}
\qquad \quad
\begin{subfigure}[b]{0.13\textwidth}
\includegraphics[width = \linewidth]{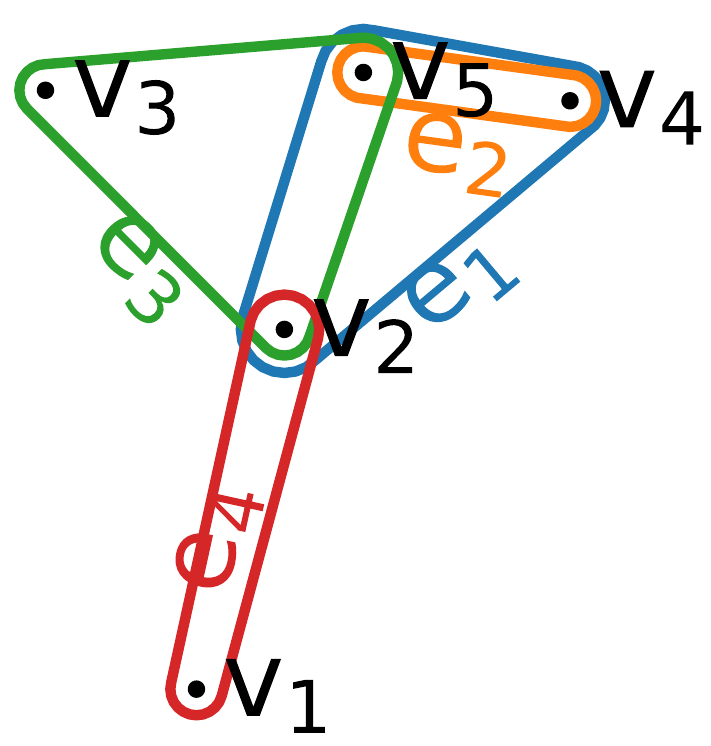}
\caption*{\hspace{-10mm} \large$H$} \label{fig:forward}
\end{subfigure}
\qquad
\begin{subfigure}[b]{0.2\textwidth}
  \scalebox{0.7}{
\begin{tikzpicture}[baseline={(Z.base)}, scale=1]
   	\Vertex[x=0 ,y=0, label=$\mathlarger{\mathlarger{\mathlarger{\bf e_4}}}$,size=0.5, color=myRed]{4}
   	\Vertex[x=0 ,y=2, label=$\mathlarger{\mathlarger{\mathlarger{\bf e_3}}}$,size=0.5, color=myGreen]{3}
   	\Vertex[x=2,y=0, label=$\mathlarger{\mathlarger{\mathlarger{\bf e_1}}}$,size=0.5, color=myBlue]{1}
   	\Vertex[x=2,y=2, label=$\mathlarger{\mathlarger{\mathlarger{\bf e_2}}}$,size=0.5, color=myOrange]{2}
    \node (Z) at (1,0) {} ; 
   \Edge(4)(1)
   \Edge(1)(2)
   \Edge(2)(3)
   \Edge(3)(4)
   \Edge(3)(1)
  \end{tikzpicture}  
}
\caption*{\hspace{-13mm} \large$L(H)$} \label{fig:line_forward} 
\end{subfigure}
\caption{From left to right: the line graph of $H^*$, the hypergraphs $H^*$ and $H$, and the line graph of $H$.} \label{fig:dualAndLine}
\end{figure}

In order to additionally capture information about the size of hyperedge
edge intersections, line graphs of hypergraphs may be defined with additional
edge weights where $\{e_{i}^{*},e_{j}^{*}\}$ has weight
$ \llvert  e_{i} \cap e_{j} \rrvert  $. By definition of matrix multiplication, it is easy
to see the line graph of a hypergraph with incidence matrix $S$ has edge-weighted adjacency matrix $S^{T}S$ with diagonal entries all converted
to zero. Figure~\ref{fig:dualAndLine} gives an example of a hypergraph,
its dual, and their respective line graphs. All hypergraph visualizations
in this paper were created using HyperNetX (HNX) \cite{hnxRef}, a recently
released\footnote{\url{https://github.com/pnnl/HyperNetX}} Python library
echoing NetworkX \cite{hagberg2008exploring} for exploratory hypergraph data analytics.

Hypergraph line graphs are also referred to by a plethora of other names.
Berge \cite{Berge1984} refers to $L(H)$ as both the ``line graph'' and ``representative
graph'' of $H$; Naik \cite{naik2018recent,Naik1982} refers to $L(H)$ as the
``intersection graph'' of~$H$. In the complex networks literature on bipartite
graphs or ``2-mode'' graphs, the oft-mentioned ``one-mode projections''
are equivalent to hypergraph line graphs. For instance, $L(H)$ and
$L(H^{*})$ are referred to as the ``top and bottom'' projections in
\cite{Latapy2008}, similarly, \cite{Everett2013} dubs these the ``column
and row'' projections. Moreover, $L(H^{*})$ is commonly referred to as the
``2-section'', ``clique graph'', or ``clique expansion'' of the hypergraph
$H$, since the edge set of $L(H^{*})$ is generated by taking all $2$-element
subsets of each edge in $H$, hence vertices within each hyperedge
$H$ form a clique in $L(H^{*})$. Consequently, if $G$ is a graph,
$L(G^{*})$ is identical to~$G$.

Line graphs play an important role in hypernetwork science. Due to the
relative dearth of hypergraph analytic tools, line graphs are often analyzed
in place of the hypergraphs they were derived from so that classical network
science techniques can be applied. However, hypergraph
line graphs are fundamentally limited in several ways. First, line graphs
are lossy representations of hypergraphs in the sense that distinct hypergraphs
can have identical line graphs. We note such structural loss does
not occur for \textit{graphs}, as Whitney's theorem \cite{Whitney1932} states,
apart from the triangle and 4-vertex star graph, any pair of connected
graphs with isomorphic line graphs must be isomorphic. In the case of hypergraph
line graphs, Kirkland \cite{Kirkland2017} recently illustrated the structural
loss in a severe sense by giving an example of two distinct
$19 \times 19$ incidence matrices $S$ and $R$ respectively, such that both
\begin{gather*}
S^{T}S = R^{T}R,
\\
SS^{T} = RR^{T}.
\end{gather*}
Put equivalently in the language of hypergraphs: although non-isomorphic,
the weighted line graphs of the hypergraphs represented by $S$ and
$R$, \textit{as well as those of their dual hypergraphs}, are both identical.
Kirkland also constructed
infinite familities of such pairs of hypergraphs and showed they constitute
a vanishingly small proportion of hypergraphs. Accordingly, while one isn't
likely to encounter such pairs of hypergraphs in empirical data, Kirkland's
work illustrates structural properties of hypergraphs may be lost even
when simultaneously accounting for hypergraph duality and using weighted
line graphs. Consequently, depending on the properties under consideration,
the extent to which line graphs faithfully represent hypergraphs may be
unclear. Nonetheless, researchers have offered preliminary evidence that
some meaningful, albeit incomplete, hypergraph structure can be extracted
from their line graphs \cite{Everett2013}.

Lastly, as noted in \cite{Latapy2008,Sariyuece2018},
another important limitation of hypergraph line graphs is computational:
sparse hypergraphs can still yield relatively dense line graphs that may
be difficult to analyze or store in computer memory. This can be
easily seen by observing that $k$-way edge intersections (guaranteed by
a vertex of degree~$k$) in the hypergraph yield $\binom{k }{2}$ edges in
its line graph. Particularly if the hypergraph is large and its vertex
degree and edge cardinality distributions are heavily skewed (common features
in real world network data), its line graphs may be too dense to analyze
computationally or even construct at all.

\section{From Graph Walks to Hypergraph Walks}\label{sec:swalk}

One of the most fundamental concepts in graph theory, underpinning a myriad
of areas including Hamiltonian and Eulerian graphs, distance and centrality
measures, stochastic processes on graphs and PageRank, is that of a walk.
For a graph $G=(V,E)$, a \textit{walk of length $k$} is a sequence of vertices
$v_{0},v_{1},\ldots,v_{k}$, such that each pair of successive vertices
are adjacent. By definition of a (simple) graph, two adjacent vertices
belong to exactly one edge, and conversely, two incident edges intersect
in exactly one vertex. Consequently, any valid graph walk can be equivalently
expressed as either a sequence of adjacent vertices or as a sequence of
incident edges, i.e.
\begin{equation*}
\underbrace{v_{0}}_{e_{0} \setminus e_{1}},\underbrace{v_{1}}_{e_{0}
\cap e_{1}},
\ldots,\underbrace{v_{k-1}}_{e_{k-1} \cap e_{k}}, \underbrace{v_{k}}_{e_{k}\setminus e_{k-1}}
\quad \longleftrightarrow \quad \underbrace{e_{1}}_{\{v_{0},v_{1}\}},
\ldots,\underbrace{e_{k}}_{\{v_{k-1},v_{k}
\}}.
\end{equation*}
In the setting of hypergraphs, this simple observation no longer holds.
Hypergraph edge incidence and vertex adjacency is \textit{set-valued} and
\textit{quantitative} in the sense that two hyperedges can intersect at any
number of vertices, and two vertices can belong to any number of shared
hyperedges. This motivates two walk concepts for hypergraphs that are dual
but distinct: walks on the vertex level (consisting of successively
adjacent vertices), and walks on the edge level (consisting of successively
intersecting edges). For ease of presentation, and to be consistent with related prior work, we limit
our exposition to edge-level hypergraph walks. Nonetheless, both notions are captured when duality is considered, as a vertex-based
walk on a hypergraph $H$ is simply an edge-walk on the dual hypergraph
$H^{*}$. We define a hypergraph walk as an ``$s$-walk'' on a hypergraph,
where $s$ controls for the size of edge intersection, as follows:

\begin{definition}
For a positive integer $s$, an \emph{$s$-walk} of length $k$ between hyperedges
$f$ and $g$ is a sequence of hyperedges,
\begin{equation*}
f=e_{i_{0}},e_{i_{1}},\ldots,e_{i_{k}}=g,
\end{equation*}
where for $j=1,\ldots,k$, we have
$s \le  \llvert  e_{i_{j-1}} \cap e_{i_{j}} \rrvert  $ and $i_{j-1} \neq i_{j}$.
\end{definition}

\label{def:swalk}

When interpreted on the dual hypergraph $H^{*}$, an $s$-walk corresponds
to a sequence of adjacent vertices in which each successive pair of vertices
belong to at least $s$ shared hyperedges.
Since in a graph a pair of vertices can belong to at most
1 edge, the usual graph walk between vertices $x$, $y$ on a graph $G$ is equivalent
to a $1$-walk between hyperedges $x^{*}$, $y^{*}$ on the dual, $G^{*}$. Consequently,
the $s=1$ case recovers the usual graph walk and $s$-walks for $s>1$ are
only possible on hypergraphs.

As will become apparent in subsequent sections, a number of basic yet
important properties of walks in graphs immediately extend to $s$-walks
on hypergraphs. For instance, just as any graph walk ending at vertex
$v_{k}$ can be concatenated with any walk starting at vertex $v_{k}$ to
form another walk, any $s$-walk ending at a particular edge can be concatenated
to any other $s$-walk starting at the edge. Consequently, the existence
of an $s$-walk between hyperedges defines an equivalence relation under
which hyperedges can be partitioned into \textit{$s$-connected components},
which we explore in Sect.~\ref{sec:sComp}. Furthermore, this also ensures
the length of the shortest $s$-walk between edges, called \textit{$s$-distance}
(Sect.~\ref{sec:sDist}), satisfies the triangle inequality and defines
a bona-fide distance metric on the hypergraph. Finally, in Sect.~\ref{sec:sPaths} we explore how one may distinguish between different kinds
of $s$-walks in a hierarchical way, and how the subsequent notions of
$s$-traces, $s$-meanders, $s$-paths, and $s$-cycles lend themselves to
discerning substructures native to hypergraphs, such as $s$-triangles. For readers interested in random walks on hypergraphs, Appendix A includes a brief discussion of recent literature and its relationship to $s$-walks.

\paragraph*{Prior work} Many researchers have considered different
notions of ``high-order walks'' on hypergraphs, abstract simplicial complexes,
and related set systems. Concepts closely related to
$s$-walks have for long appeared in the mathematics literature. Bermond,
Heydemann, and Sotteau \cite{bermond1977line} introduced and analyzed \textit{$k$-line graphs} of uniform hypergraphs, which are derived from hypergraphs
by representing each hyperedge as a vertex, and linking two such vertices
if their corresponding hyperedges intersect in at least $k$ vertices. In
this way, a (graph) walk on their line graphs corresponds to an $s$-walk
on a hypergraph. In \cite{lu2011high}, Lu and Peng define higher order
walks on hypergraphs for $k$-uniform hypergraphs as sequences of edges intersecting in \textit{exactly} $s$ vertices, where the vertices within each edge are \textit{ordered}. Their work
is related to a rich literature on Hamiltonian cycles in $k$-uniform hypergraphs
(e.g. \cite{han2010dirac,katona1999hamiltonian}) and takes a spectral
approach: these generalized walks are used to define a so-called $s$-Laplacian matrix. Wang and Lee \cite{wang1999paths} define
hypergraph paths as edge sequences in which no successive intersection
is a subset of any other. Their motivation is to prove enumeration formulas
for certain cycle structures in hypergraphs. In a series of three recent
papers
\cite{cooley2018subcritical,Cooley2015,Cooley2016ThresholdAH}, Kang,
Cooley, Koch, and others consider a notion of $s$-walk between $s$-tuples
of vertices. They conduct a rigorous mathematical analysis of the asymptotic
$s$-walk properties of binomial random $k$-uniform hypergraphs, considering
hitting times, the evolution of high-order $s$-components, and high-order
``hypertree'' structures. Lastly, in
\cite{joslynHICSS,purvine2018topological}, authors of the present work
briefly considered the $s$-walk based notion of $s$-distance as applied
to Domain Name System (DNS) cyber data, and the Enron email dataset, respectively.

\paragraph*{Contributions}
The main contributions of the present work are:
\begin{itemize}
\item developing hypergraph generalizations of graph network science measures
using the $s$-walk framework,
\item applying these new measures to real
hypernetworks, analyzing and comparing the results, discussing structural
insights they reveal, and
\item experimentally testing the degree to which
existing generative hypergraph null models are able to replicate the properties
seen in real data captured by these measures.
\end{itemize}
To make clear how these new hypergraph metrics generalize their graph counterparts,
when appropriate we include a definition subheading called ``Graph case
\& equivalence''. This subheading addresses two distinct questions: first,
it describes how the definition(s) in question reduce to the graph case
when the hypergraph is a graph. As we will see, most of the proposed
hypergraph measures reduce to a graph analog by taking $s=1$ and examining
the dual of the graph, $G^{*}$ (taking the dual converts our edge-based
exposition to match the vertex-based notions common in network science).

Second, this heading describes whether the hypergraph
measure is equivalent to a graph measure on the hypergraph's $s$-line graphs
(Definition~\ref{def:s-line}), which are generalizations of the aforementioned
line graph. In the case of $s$-connected components and $s$-distance measures
(Sects.~\ref{sec:sComp}--\ref{sec:sDist}), these have natural equivalences
on, and thus may be obtained via, $s$-line graphs. However, the $s$-path,
$s$-cycle, and $s$-clustering coefficients (Sect.~\ref{sec:sPaths}) rely
on subset information not encoded in, and hence not discernible from, the
$s$-line graphs. Furthermore, properties of the hypergraph generative models
we consider, such as hypergraph degree distributions and metamorphosis
coefficients, also cannot be determined from the $s$-line graphs. A~takeaway
from this is that our study of $s$-walks and hypernetwork science includes,
but extends beyond, the study of $s$-line graphs.

Lastly, we briefly compare our
approach with that of the related research surveyed in ``Prior work'' above:
while the present work is similarly based around the concept of high-order
hypergraph walks, we utilize them for different ends. In particular, we
use this framework to develop network science concepts that are aimed at
messy, real hypergraph data. In contrast to all the work mentioned above,
apart from that of Wang and Lee \cite{wang1999paths}, we do not assume
$k$-uniformity, as real hypernetworks are frequently non-uniform. Furthermore,
our methods apply to disconnected hypergraphs and permit duplicate hyperedges---both of which are also common in real data. Additionally,
we make design choices to ensure our methods are more computationally tractable
in light of the combinatorial explosion inherent in hypergraphs. For instance,
as opposed to Lu and Peng who define $s$-walks between arbitrary \textit{ordered
$s$-tuples} of vertices,\footnote{Consequently, the $s$-Laplacian matrix
they study is $n^{\underline{s}} \times n^{\underline{s}}$, where
$n$ denotes the number of vertices and
$x^{\underline{k}}=\binom{x }{k} k!$ denotes the falling factorial. Even
for a modestly sized hypergraph on $n=10^{4}$ vertices with $s=20$, this
matrix has size $n^{\underline{s}} \approx 10^{80}$, approximately the
number of atoms in the known universe.} defining our notion of $s$-walk
between pairs of unordered hyperedges (or when working with the dual, pairs
of single vertices) permits the development of methods more tractable in
application to real data.

\section{Hypergraph Walk Framework}\label{sec:sMetrics}

In this section, we explore how analytic tools from network science extend
to hypergraphs in the hypergraph walk framework. Within each subsection,
we focus on one topic (e.g. $s$-distance, a hypergraph
geodesic), and introduce relevant methods in the ``Methods'' section. In
the ``Application to Data'' section, we apply these methods to real hypernetworks
and analyze the results. Our goal is not to
explain why the observed structure exists using domain-specific analyses.
Rather, we identify abstract structural properties revealed by these measures,
and highlight how these properties differ within each dataset (as we vary
the walk order~$s$), as well as across datasets. 
This illustrates
particular properties these measures capture, as well as new
insights revealed by considering $s$-walk based metrics. While we take
a broader, methods-based viewpoint here, we believe such an approach may
be more useful in guiding future application-specific studies of these
methods across multiple domains. To that end, we consider three datasets
from three domains: corporate governance, biology, and text analysis.

\subsection{Test Data Sets}

\begin{figure}[t]
\centering
\begin{subfigure}[b]{0.25\textwidth}
\includegraphics[width = \linewidth]{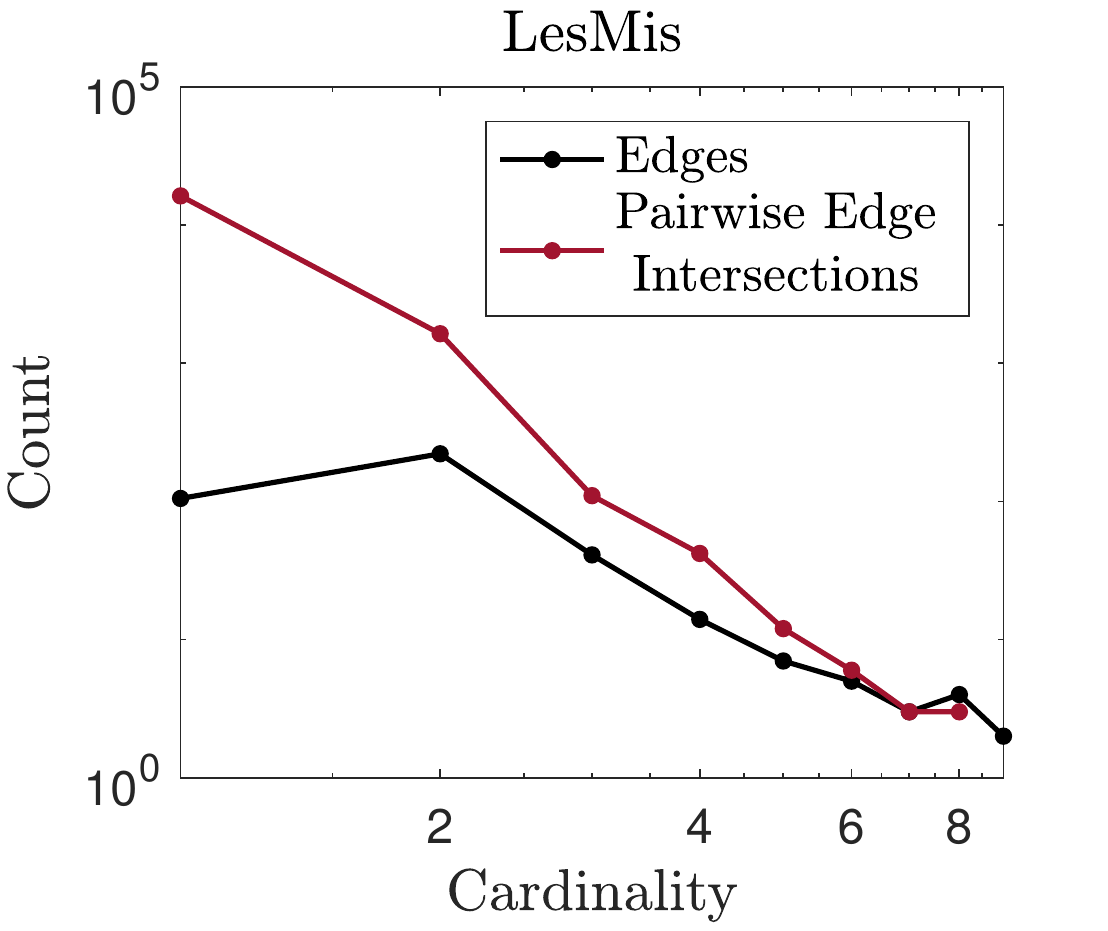} \\
\includegraphics[width = \linewidth]{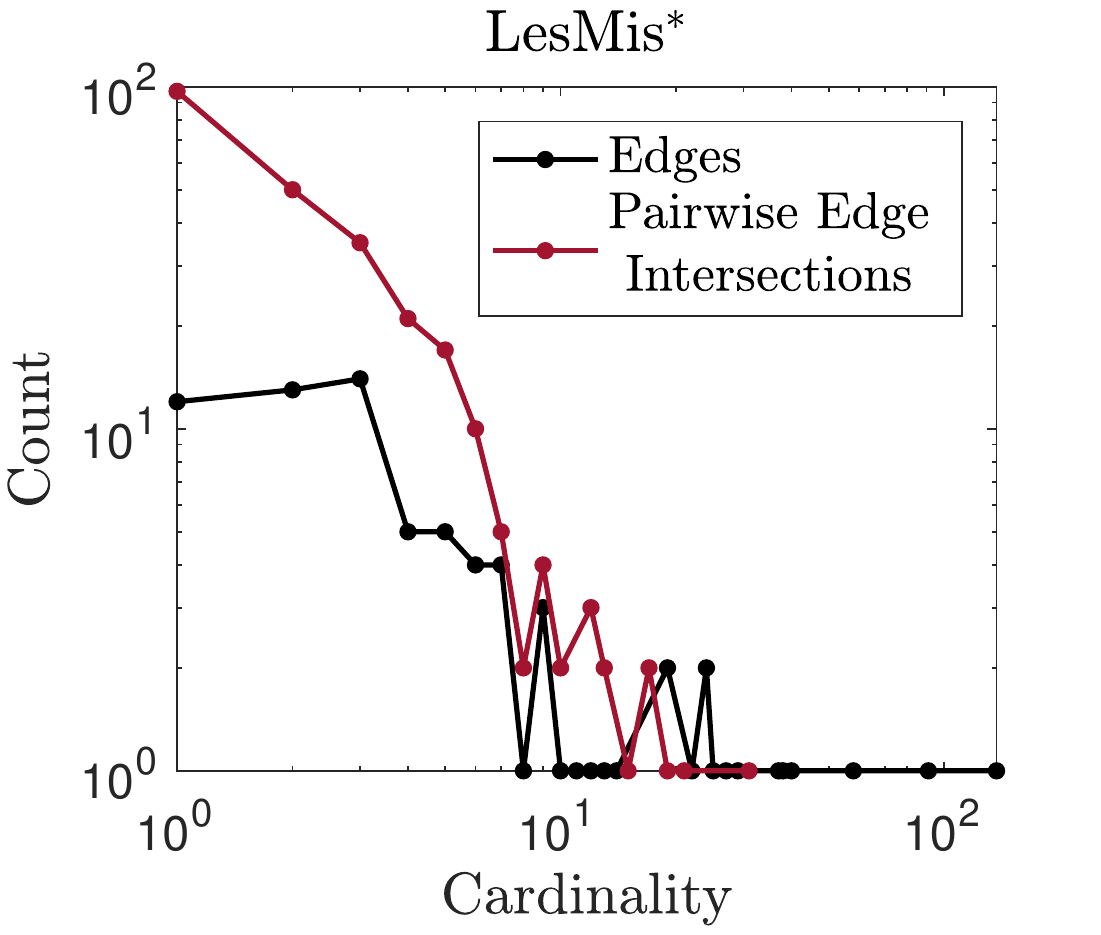} \\
\end{subfigure}
\qquad
\begin{subfigure}[b]{0.25\textwidth}
\includegraphics[width = \linewidth]{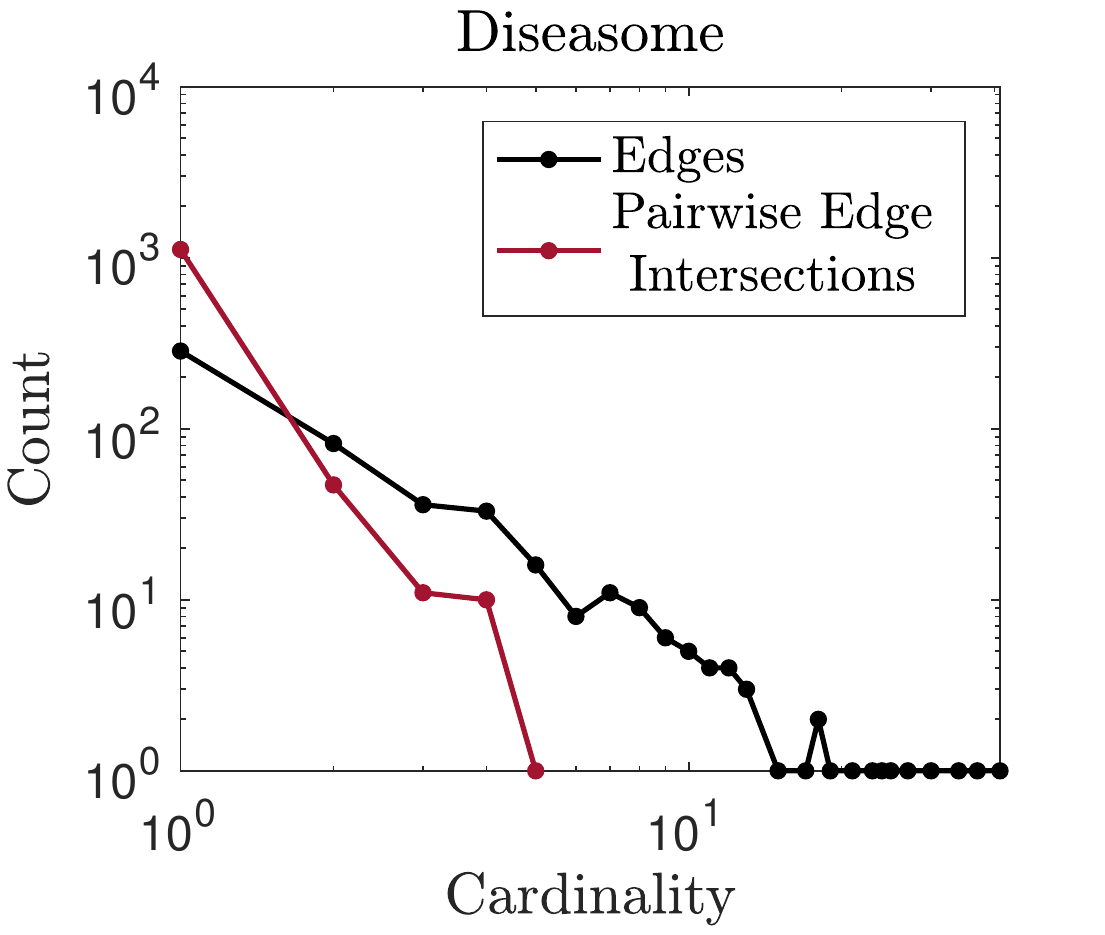} \\
\includegraphics[width = \linewidth]{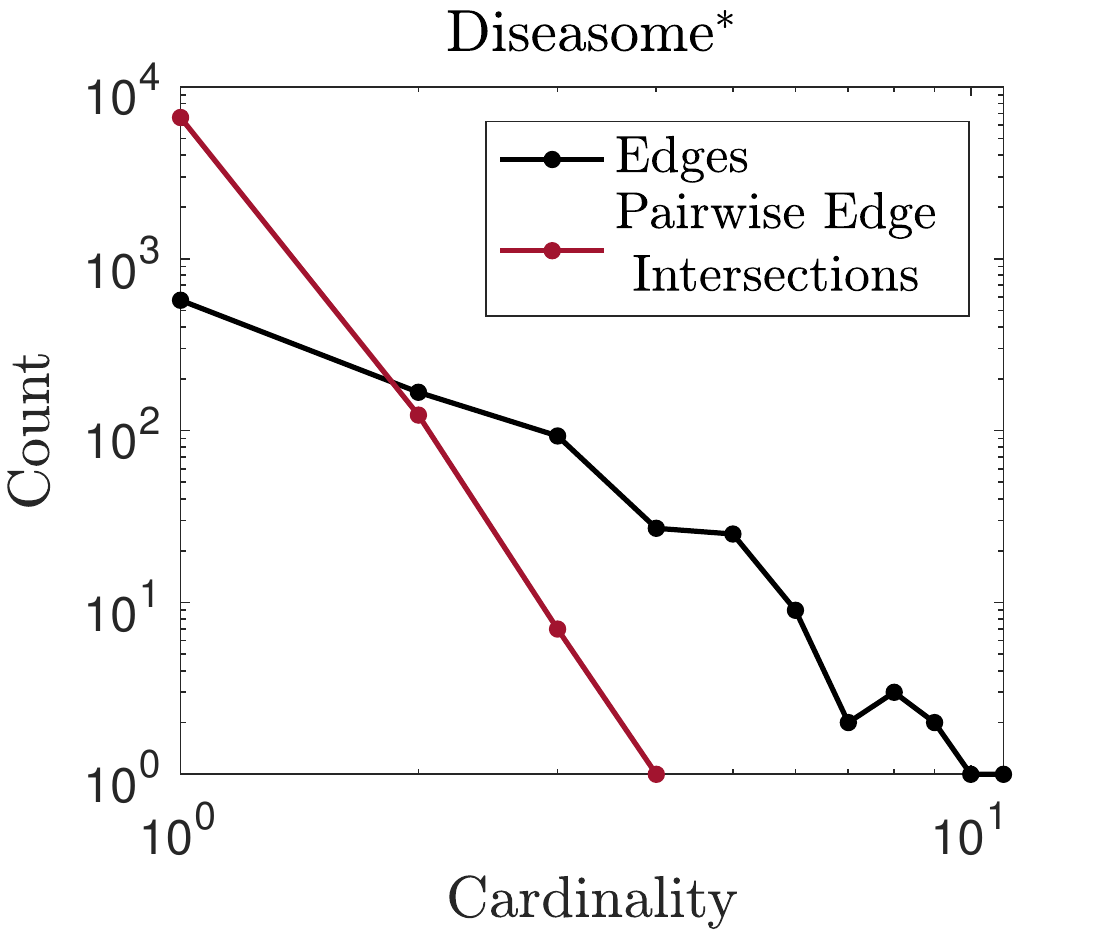} \\
\end{subfigure}
\qquad
\begin{subfigure}[b]{0.25\textwidth}
\includegraphics[width = \linewidth]{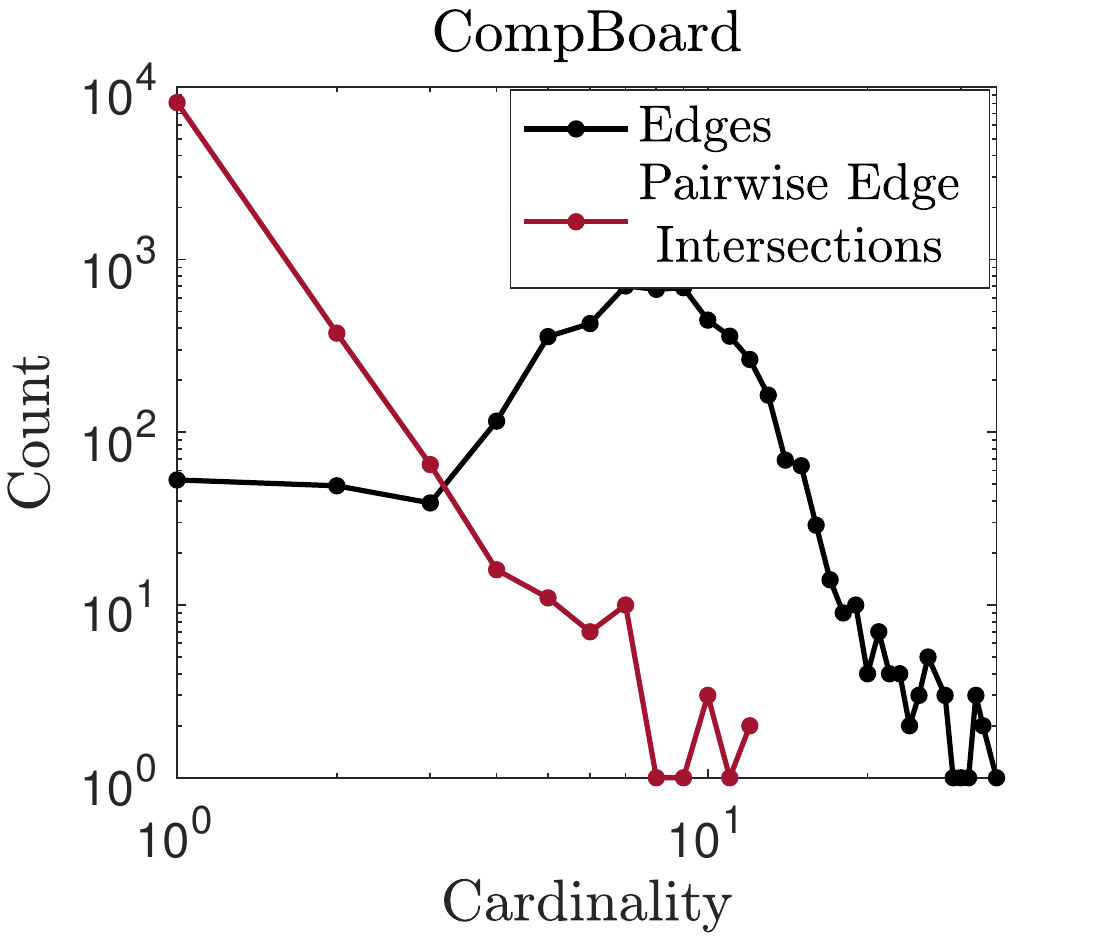} \\
\includegraphics[width = \linewidth]{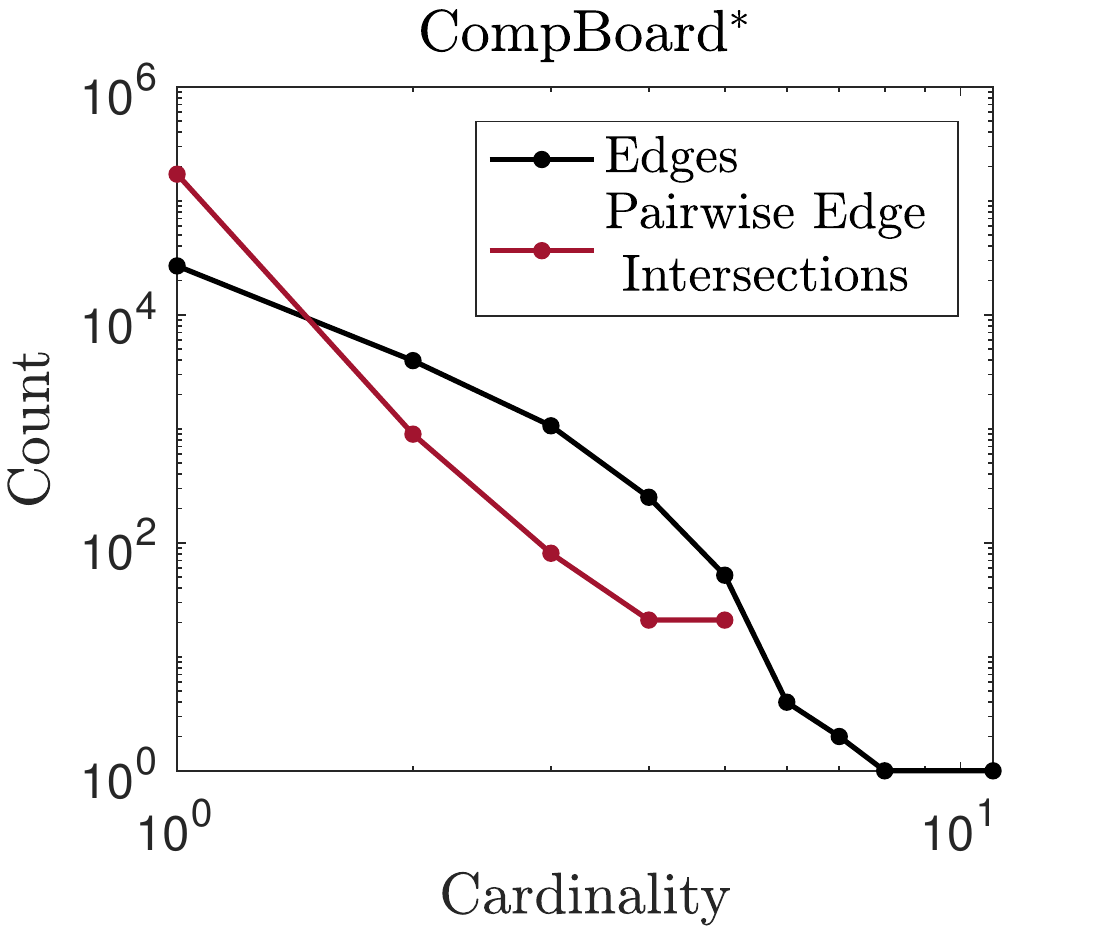} \\
\end{subfigure}
\begingroup
\small
\begin{tabular}{l||cc|cc|cc}
\toprule
 & \multicolumn{1}{c}{\footnotesize LesMis} & \multicolumn{1}{c}{\footnotesize LesMis$^*$}  & \multicolumn{1}{c}{\footnotesize Diseasome} & \multicolumn{1}{c}{\footnotesize Diseasome$^*$} & \multicolumn{1}{c}{\footnotesize CompBoard} & \multicolumn{1}{c}{\footnotesize CompBoard$^*$}  \\ \midrule
\mbox{\footnotesize \# Edges} & 402  & 80 & 516 & 903 & 4,573 & 32,189  \\
\mbox{\footnotesize \# Multi-edges} &211 & 5 & 108 & 422 & 0 & 22,537   \\
\mbox{\footnotesize Max edge size} & 9 & 137 & 41 & 11 & 35  & 11  \\
\mbox{\footnotesize Max edge inter.} & 8 & 31 & 5 & 4 & 12 & 5   \\
\mbox{\footnotesize Density} & 2.68e--2 & 2.68e--2 & 3.33e--3 & 3.33e--3 & 2.67e--4 & 2.67e--4 \\
\bottomrule
\end{tabular}
\endgroup
\caption{{\it Plots}: The edge and pairwise edge intersection distributions for LesMis, Diseasome, and CompBoard (top row) and their dual hypergraphs (bottom row). {\it Table:} Basic statistics on edge and multi-edge counts, the maximum edge and pairwise edge intersection cardinality, and density for each dataset and their dual hypergraphs. } \label{fig:dd}
\end{figure}

For each dataset, we define the associated hypergraph,
review prior graph or hypergraph analyses of these (or closely related)
datasets, and discuss basic properties which are summarized in Fig.~\ref{fig:dd}. In this figure and throughout, we use the same notation as
in Definition~\ref{def:lineGraph} to refer to the dual hypergraph associated
with a data set (e.g. LesMis$^{*}$ refers to the dual hypergraph of LesMis,
in which the roles of the vertices and edges are swapped relative to how
they are defined below). Figure~\ref{fig:dd} plots the edge
cardinality distribution and pairwise edge intersection cardinality distribution.
For instance, the point $(x,y)=(3,100)$ on the ``edges'' distribution means
there are $100$ edges which consist of exactly 3 vertices in that hypergraph;
the same point on the ``pairwise edge intersection'' distribution means
there are $100$ distinct, unordered pairs of hyperedges whose intersection
contains exactly 3 vertices. We remind the reader the edge cardinality
distribution of the dual hypergraph $H^{*}$ is the same as the vertex degree
distribution of~$H$.

The table in Fig.~\ref{fig:dd} highlights basic statistics for each
hypergraph. The number of ``multi-edges'' is the number of edges that duplicate
(in the sense of set equality) another edge, i.e.
$ \llvert  \{i: e_{i}=e_{j} \text{ for } j<i\} \rrvert  $. The maximum edge size and edge
intersection sizes, reported below the multi-edge counts, are particularly
pertinent because they determine the range of interest for our measures:
the former determines the largest value of $s$ for which $s$-walk based
measures are \textit{defined} while the latter determines the maximum value
of $s$ for which $s$-walk based measures are \textit{non-trivial}. Finally,
``density'' measures the number of vertex-hyperedge memberships relative
to the number of possible vertex-hyperedge memberships. Put equivalently,
this is the number of nonzero entries in the incidence matrix $S$ divided
by the product $ \llvert  V \rrvert   \cdot  \llvert  E \rrvert  $. By definition, density is always
the same for the hypergraph and its dual, whereas the other reported values
are edge-based and may differ.

\paragraph*{CompBoard}
\begin{itemize}
\item[] \emph{Data set.} A~company-board network. Vertices represent people and hyperedges represent company
boards. A~vertex belongs to a hyperedge if that person sits on the company
board. The data consists of 4573 companies and 32{,}189 people. Companies are identified by ticker symbols,
excluding any location or exchange code suffixes (e.g. Vodafone group
is represented solely by VOD, not VOD.L or VOD.O) taken from the
NYSE, AMEX, and NASDAQ stock exchange listings\footnote{List of companies
on these exchanges obtained from
\url{https://www.nasdaq.com/screening/company-list.aspx}} on 10/1/2018.
The data was collected from publicly available\footnote{\url{https://www.reuters.com/finance/}}
board director information listed on Reuters. Board director names were
cross referenced against age data to better distinguish different people with the same full name.
\item[] \emph{Prior work.} Company-board network studies are historically
rooted in corporate elite theory, focusing on companies which share a common
board member called \textit{interlocking directorates}. Many such studies
focus on line graph representations of the network, linking companies
whose boards interlock. For instance, Conyon and Muldoon
\cite{Conyon2004} studied the small-world properties of company-board networks
from the US, UK, and Germany, focusing on the clustering coefficient and
average path length of the line graphs. In \cite{Newman2001}, Newman compares
the clustering coefficient of a company-board network line graph to that
of a random model.\footnote{Measuring hypergraph clustering on line graphs has been noted to be potentially misleading
(see, for instance, \cite{Nacher2011,Opsahl2013}) since the cliques generated
heavily skew the number of triangles.} Levine and Roy
\cite{levine1979study} appear to be among the first to analyze bipartite
representations of company-board networks directly, rather than solely
line graphs. They considered topics such as the average path length, connected
component sizes, and proposed a ``rubber-band model''\footnote{Described
as a physical device consisting of two horizontal bars that support ``hooks''
representing companies and board member nodes, with rubber bands that ``join
the appropriate hooks and physically represent the inclusion between persons
and boards''} to cluster the bipartite network. Later, Robins and Alexander
devised a bipartite global clustering coefficient, based on the ratio of
bipartite 4-cycles to 3-paths, to measure ``the extent to which
directors re-meet one another on two or more boards''
\cite{Robins2004}. In Sect.~\ref{sec:sPaths}, we propose a new notion
of hypergraph clustering coefficients and explain how it compares to that
of Robins and Alexander, as well as graph clustering coefficients measured
on the line graph. More generally, since an ``interlocking directorate''
is represented by a hyperedge intersection, our methods can be interpreted
in this context as not only based on the existence of interlocks (i.e. a
pure line graph analysis) but also their size and relative set relationships.
\item[] \emph{Basic properties.} The edge size distribution shows the
sizes of company boards are tightly concentrated around 7--10 members and
drop off sharply at either end: only about 3\% of companies have fewer
than 4 members, and 3\% have more than 14. In contrast, the edge size distribution
of the dual hypergraph is monotonically decreasing, showing more than 99\%
of board members belong to between 1--3 company boards. The pairwise edge
intersection distribution for the hypergraph and its dual similarly exhibit
a sharp decrease, and the range of these distributions imply different
companies share up to a maximum of 12 board members, while different members
serve on up to 5 common company boards. Among the three datasets,
CompBoard is the sparsest: it contains about 0.03\% of possible vertex-hyperedge
memberships, as opposed to 0.33\% and 2.68\% for Diseasome and LesMis,
respectively.
\end{itemize}

\paragraph*{Diseasome}
\begin{itemize}
\item[] \emph{Data set.} A~human gene-disease network from
\cite{Goh2007}. Vertices represent genes and hyperedges represent genetic
disorders. A~vertex belongs to a hyperedge if mutations in that gene are
implicated by that disease. The data consists of 903 genes and 516 diseases.
\item[] \emph{Prior work.} Goh et al. \cite{Goh2007} collected the list
of genes, disorders, and their associations from the Online Mendelian Inheritance
in Man (OMIM) \cite{hamosh2005online} compendium in 2005. Their study considered
the line graphs of the hypergraph and its dual, which they dubbed the Human
Disease Network and Disease Gene Network. They show the size of the largest connected component in these networks
differed with those generated by random models. In Sect.~\ref{sec:sComp} we study a generalized notion of high-order connected components
and compare these against those of random hypergraph models in Sect.~\ref{sec:models}. For a broader discussion of the potential applications
of hypergraphs and hypergraph statistics in biology and genomics, see
\cite{Klamt2009}.
\item[] \emph{Basic properties.} The edge size distributions of Diseasome
and its dual show the most genes implicated by a disease is 41 while
the most diseases implicated by a gene is 11. The pairwise edge intersection
size distribution show
94\% of pairs of diseases implicating common genes share exactly 1 gene; conversely, examining this distribution for the dual hypergraph reveals 98\% of gene pairs associated with a common disease share exacltly 1 disease. 
Among the three datasets, Diseasome and its dual features
the narrowest range of pairwise edge intersection sizes, with maximum edge
sizes of 5 and 4, respectively.
\end{itemize}

\paragraph*{LesMis}
\begin{itemize}
\item[] \emph{Data set.} A~character-scene network from
\cite{knuth1993stanford}. Vertices represent characters and hyperedges
represent scenes from Victor Hugo's novel, Les Mis\'{e}rables. There are
80 characters and 402 scenes.
\item[] \emph{Prior work.} This dataset was collected by Donald Knuth
\cite{knuth1993stanford} and can be structured according to different granularities
in which hyperedges represent the scenes, chapters, books, or volumes of
the novel. The line graph of the LesMis hypergraph, often dubbed the Les
Mis co-appearence network, has appeared frequently in network science literature
for the purpose of demonstrating clustering or modularity methods
\cite{Garriga2010,Newman2004}, or centrality and ranking methods
\cite{Alvarez-Socorro2015}. With regard to the latter, we apply our proposed
hypergraph centrality measure to rank LesMis characters in Sect.~\ref{sec:sDist}.
\item[] \emph{Basic properties.} In LesMis and its dual, the largest hyperedge
features 9 characters and 137 scenes, respectively, with the latter hyperedge
(unsurprisingly) corresponding to the protagonist, Jean Valjean. Compared
against other datasets, the edge intersection size distributions are particularly
distinct. LesMis$^{*}$ features the largest range of edge intersection
sizes across all datasets. Both LesMis and its dual are also notable for
featuring the most edge intersections relative to the number of possible
edge intersections: respectively, 22\% and 8\% of all pairs of edges in
LesMis and its dual intersect, whereas this ratio is an order of magnitude
smaller for Diseasome and its dual and two orders of magnitude smaller
for CompBoard and its dual.
\end{itemize}

\subsection{Connected Components}\label{sec:sComp}

\begin{table}[t!]
\caption{The $s$-line graphs for two hypergraphs.}
\label{fig:sLineGraphs}
\setlength{\tabcolsep}{5mm} 
\def\arraystretch{1.25} 
\centering
\scalebox{0.9}
{
\begin{tabular}{|c||c|c|c|c|c|}
  \hline
  {\bf Hypergraph $H$}    &   {\bf $L_1(H)$}    &   {\bf $L_2(H)$} &   {\bf $L_3(H)$} &   {\bf $L_4(H)$} &   {\bf $L_5(H)$}
  \\ \hline
   \begin{minipage}{.2\textwidth}
      \includegraphics[width=\linewidth]{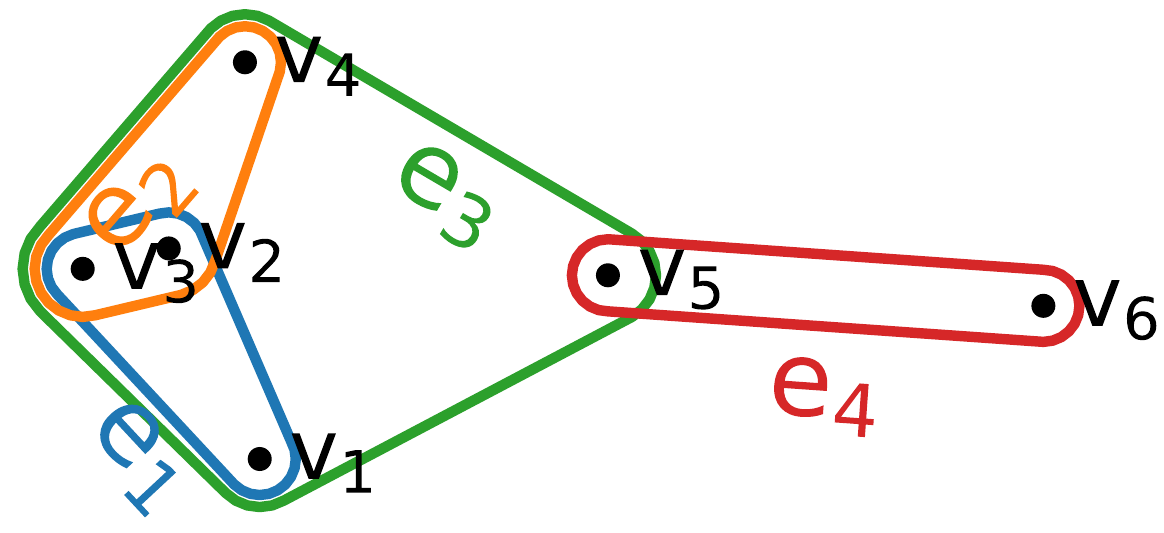}
    \end{minipage}
  &
  \scalebox{0.7}{
\begin{tikzpicture}[baseline={(Z.base)}, scale=1]
   	\Vertex[x=0.5 ,y=.134,label=$\mathlarger{\mathlarger{\mathlarger{\bf e_3}}}$,size=0.5, color=myGreen]{3}
   	\Vertex[x=0 ,y=1,label=$\mathlarger{\mathlarger{\mathlarger{\bf e_1}}}$,size=0.5,color=myBlue]{1}
   	\Vertex[x=1,y=1,label=$\mathlarger{\mathlarger{\mathlarger{\bf e_2}}}$,size=0.5,color=myOrange]{2}
   	\Vertex[x=0.5,y=-0.866,label=$\mathlarger{\mathlarger{\mathlarger{\bf e_4}}}$,size=0.5,color=myRed]{4}
    \node (Z) at (0.5,0.067) {} ; 
   \Edge(1)(2)
   \Edge(1)(3)
   \Edge(3)(4)
   \Edge(2)(3)
    \addvmargin{1mm}
  \end{tikzpicture}  
}
   &
   \scalebox{0.7}{
\begin{tikzpicture}[baseline={(Z.base)}, scale=1]
   	\Vertex[x=0.5 ,y=.134,label=$\mathlarger{\mathlarger{\mathlarger{\bf e_3}}}$,size=0.5,  color=myGreen]{3}
   	\Vertex[x=0 ,y=1,label=$\mathlarger{\mathlarger{\mathlarger{\bf e_1}}}$,size=0.5,color=myBlue]{1}
   	\Vertex[x=1,y=1,label=$\mathlarger{\mathlarger{\mathlarger{\bf e_2}}}$,size=0.5, color=myOrange]{2}
   	\Vertex[x=0.5,y=1.7,label=$\mathlarger{\mathlarger{\mathlarger{\bf e_4}}}$,size=0.5, color=myRed]{4}
    \node (Z) at (0.5,0.917) {} ; 
   \Edge(1)(2)
   \Edge(1)(3)
   \Edge(2)(3)
    \addvmargin{1mm}
  \end{tikzpicture}  
}
   &
   \scalebox{0.7}{
\begin{tikzpicture}[baseline={(Z.base)}, scale=1]
   	\Vertex[x=0.5 ,y=.134,label=$\mathlarger{\mathlarger{\mathlarger{\bf e_3}}}$,size=0.5, color=myGreen]{3}
   	\Vertex[x=0 ,y=1,label=$\mathlarger{\mathlarger{\mathlarger{\bf e_1}}}$,size=0.5,color=myBlue]{1}
   	\Vertex[x=1,y=1,label=$\mathlarger{\mathlarger{\mathlarger{\bf e_2}}}$,size=0.5, color=myOrange]{2}
    \node (Z) at (0.5,0.567) {} ; 
   \Edge(1)(3)
   \Edge(2)(3)
    \addvmargin{1mm}
  \end{tikzpicture}  
}
 & 
\scalebox{0.7}{
\begin{tikzpicture}[baseline={(Z.base)}, scale=1]
   	\Vertex[x=0.5 ,y=0.5,label=$\mathlarger{\mathlarger{\mathlarger{\bf e_3}}}$,size=0.5,  color=myGreen]{3}
    \node (Z) at (0.5,0.5) {} ; 
    \addvmargin{1mm}
  \end{tikzpicture}  
}
 & 
\scalebox{0.7}{
\begin{tikzpicture}[baseline={(Z.base)}, scale=1]
   	\Vertex[x=0.5 ,y=0.5,label=$\mathlarger{\mathlarger{\mathlarger{\bf e_3}}}$,size=0.5,  color=myGreen]{3}
    \node (Z) at (0.5,0.5) {} ; 
    \addvmargin{1mm}
  \end{tikzpicture}  
}
  \\ \hline
   \begin{minipage}{.2\textwidth}
      \includegraphics[width=\linewidth]{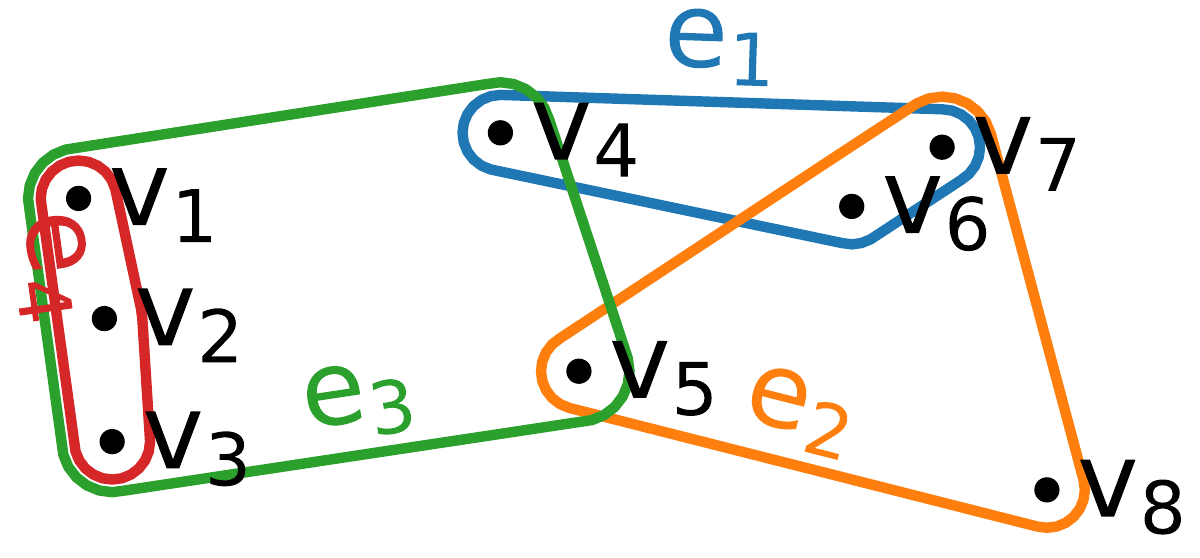}
    \end{minipage}
&
\scalebox{0.7}{
\begin{tikzpicture}[baseline={(Z.base)}, scale=1]
   	\Vertex[x=0.5 ,y=.134,label=$\mathlarger{\mathlarger{\mathlarger{\bf e_3}}}$,size=0.5, color=myGreen]{3}
   	\Vertex[x=0 ,y=1,label=$\mathlarger{\mathlarger{\mathlarger{\bf e_1}}}$,size=0.5, color=myBlue]{1}
   	\Vertex[x=1,y=1,label=$\mathlarger{\mathlarger{\mathlarger{\bf e_2}}}$,size=0.5, color=myOrange]{2}
   	\Vertex[x=0.5,y=-0.866,label=$\mathlarger{\mathlarger{\mathlarger{\bf e_4}}}$,size=0.5, color=myRed]{4}
    \node (Z) at (0.5,0.067) {} ; 
   \Edge(1)(2)
   \Edge(1)(3)
   \Edge(3)(4)
   \Edge(2)(3)
    \addvmargin{1mm}
  \end{tikzpicture}  
}   
&
\scalebox{0.7}{
\begin{tikzpicture}[baseline={(Z.base)}, scale=1]
        \Vertex[x=0 ,y=1,label=$\mathlarger{\mathlarger{\mathlarger{\bf e_1}}}$,size=0.5, color=myBlue]{1}
        \Vertex[x=1,y=1,label=$\mathlarger{\mathlarger{\mathlarger{\bf e_2}}}$,size=0.5, color=myOrange]{2}
   	\Vertex[x=0 ,y=.134,label=$\mathlarger{\mathlarger{\mathlarger{\bf e_3}}}$,size=0.5,  color=myGreen]{3}
   	\Vertex[x=1, y=.134,label=$\mathlarger{\mathlarger{\mathlarger{\bf e_4}}}$,size=0.5, color=myRed]{4}
    \node (Z) at (0.5,0.433) {} ; 
   \Edge(1)(2)
   \Edge(3)(4)
    \addvmargin{1mm}
  \end{tikzpicture}  
}
  & 
\scalebox{0.7}{
\begin{tikzpicture}[baseline={(Z.base)}, scale=1]
   	\Vertex[x=0 ,y=1,label=$\mathlarger{\mathlarger{\mathlarger{\bf e_1}}}$,size=0.5, color=myBlue]{1}
   	\Vertex[x=1,y=1,label=$\mathlarger{\mathlarger{\mathlarger{\bf e_2}}}$,size=0.5, color=myOrange]{2}
   	\Vertex[x=0 ,y=.134,label=$\mathlarger{\mathlarger{\mathlarger{\bf e_3}}}$,size=0.5,  color=myGreen]{3}
   	\Vertex[x=1, y=.134,label=$\mathlarger{\mathlarger{\mathlarger{\bf e_4}}}$,size=0.5, color=myRed]{4}
    \node (Z) at (0.5,0.433) {} ; 
   \Edge(3)(4)
    \addvmargin{1mm}
  \end{tikzpicture}  
}
 &
\scalebox{0.7}{
\begin{tikzpicture}[baseline={(Z.base)}, scale=1]
   	\Vertex[x=0.8,y=0.9,label=$\mathlarger{\mathlarger{\mathlarger{\bf e_2}}}$,size=0.5, color=myOrange]{2}
   	\Vertex[x=0.2 ,y=.234,label=$\mathlarger{\mathlarger{\mathlarger{\bf e_3}}}$,size=0.5,  color=myGreen]{3}
    \node (Z) at (0.5,0.433) {} ; 
    \addvmargin{1mm}
  \end{tikzpicture}  
}
&
\scalebox{0.7}{
\begin{tikzpicture}[baseline={(Z.base)}, scale=1]
   	\Vertex[x=0.5 ,y=0.5,label=$\mathlarger{\mathlarger{\mathlarger{\bf e_3}}}$,size=0.5,  color=myGreen]{3}
    \node (Z) at (0.5,0.5) {} ; 
    \addvmargin{1mm}
  \end{tikzpicture}  
}
  \\ \hline
  \end{tabular}
  }
\end{table}

\paragraph*{Methods} Under Definition~\ref{def:hyp}, the graph notions of connectedness
and connected components extend naturally to the $s$-walk framework.

\begin{definition}
For a hypergraph $H=(V,E)$, a subset of hyperedges $C \subseteq E$ is called
\emph{$s$-connected} if there exists an $s$-walk between all
$f,g \in C$ and is further called an \emph{$s$-connected component} if there
is no $s$-connected $J \subseteq E$ such that $C \subsetneq J$.
\end{definition}

Since for any $e \in E$, there can be no $s$-walk from $e$ to any other
hyperedges if $ \llvert  e \rrvert  <s$, the order of an $s$-connected component is bounded
above by $ \llvert  E_{s} \rrvert  $, where $E_{s}=\{e \in E:  \llvert  e \rrvert  \geq s\}$. More precisely,
for any positive integer $s$ and hypergraph $H$, the edges in
$E_{s}$ can always be partitioned into $s$-connected components. We call
a hypergraph $s$-connected if $E_{s}$ is $s$-connected.

While an $s$-connected component of a hypergraph
$H$ is an equivalence class of \textit{edges}, a vertex-based notion of
$s$-connected components is obtained by simply applying the above definition
to the dual hypergraph $H^{*}$. In comparing these edge and vertex-based
notions, note the number of $1$-connected components for
$H$ and $H^{*}$ are always the same: it is straightforward to see,
in either case, the number of such components is the same as the number
of nontrivial connected components (i.e. excluding isolated vertices)
in the bipartite graph representation of the hypergraph. In this sense,
edge and vertex-based connectedness are equivalent for $s=1$ and whenever
$H$ is a graph. However, for $s\geq 2$, the number of $s$-connected components
for $H$ and $H^{*}$ may differ. Hence, $s$-connectedness in hypergraphs
is richer and more varied for high-orders, yielding dual but distinct vertex
and edge-based notions.

An effective way of visualizing and studying basic properties of $s$-connected
components is via its $s$-line graph. As previously mentioned, $s$-line
graphs were studied for $k$-uniform hypergraphs by Bermond, Heydemann,
and Sotteau \cite{bermond1977line} as early as 1977. A~definition for the
general case may be stated as follows:

\begin{definition}
\label{def:s-line}
Let $H=(V,E)$ be a hypergraph with vertex set
$V=\{v_{1},\ldots,v_{n}\}$ and edge set $E \supseteq E_{s}$ where
$E_{s}=\{e \in E:  \llvert  e \rrvert   \geq s\}=\{e_{1},\ldots,e_{k}\}$ for an integer
$s \ge 1$. The \emph{$s$-line graph of $H$}, denoted $L_{s}(H)$, is the
graph on vertex set $\{e_{1}^{*},\ldots,e_{k}^{*}\}$ and edge set
$\{\{e_{i}^{*}, e_{j}^{*}\}:  \llvert  e_{i} \cap e_{j} \rrvert   \geq s \text{ for } i
\neq j \}$.
\end{definition}

In other words, each vertex in the $s$-line graph represents a hyperedge
with at least $s$ vertices in the hypergraph, and two vertices are linked
in the $s$-line graph if their corresponding hyperedges intersect in at
least $s$ vertices in the hypergraph. In this way, the $1$-line graph is simply
the line graph from Definition~\ref{def:lineGraph} and the connected components
of the $s$-line graph represent the $s$-connected components of the hypergraph.
Hence, we have: \\

\noindent {\bf Graph case \& equivalence}:
If $H$ is a graph,
$H$ is connected iff $H^{*}$ is $1$-connected. A~hypergraph $H$ is
$s$-connected iff $L_{s}(H)$ is connected. \\

Table~\ref{fig:sLineGraphs} presents examples of two hypergraphs and
their associated $s$-line graphs. Observe that both hypergraphs have identical
$1$-line graphs. Nonetheless, comparing their $s$-line graphs for
$s=2,3,4$ suggests differences otherwise lost when solely considering the
(usual) line graph,
which $s$-line graphs generalize.

Although more general, $s$-line graphs are still subject to a limitation
underlying (the usual) hypergraph line graphs: they do not uniquely identify
a hypergraph, up to isomorphism.
For instance, while we previously observed the two author-paper
networks in Fig.~\ref{fig:authPap} to be different, the corresponding
hypergraphs formed by letting hyperedges denote authors have the same
$s$-line graphs for $s=1,2$ and either trivial or empty $s$-line graphs
for $s>2$. More generally, Kirkland's aforementioned work
\cite{Kirkland2017} shows even when duality is considered, $s$-line graphs
may not uniquely identify a hypergraph. Nevertheless, $s$-line graphs
can be utilized to determine a number of informative $s$-walk properties,
including $s$-distance, which we explore in the next section. It is worth
repeating, however, the study of high-order hypergraph $s$-walks is not limited to $s$-line graphs. As we will see in Sect.~\ref{sec:sPaths}, $s$-line graphs cannot distinguish between
finer classes of $s$-walks, such as $s$-meanders and $s$-paths, and consequently
cannot be used to compute $s$-clustering coefficients, for example.
Returning again to the examples in Fig.~\ref{fig:authPap}, we will see that while these hypergraphs have
identical nontrivial $s$-line graphs, they are distinguished by their
$s$-paths.

\paragraph*{Application to data} Table~\ref{fig:sCompViz2} presents the
$s$-components of LesMis, Diseasome, and CompBoard
$s=1,\ldots,5$ by visualizing their $s$-line graphs. A~page-sized version
of this table is included in the \hyperref[app]{Appendix}.

\begin{table}[t]
\caption{The $s$-connected components of the datasets. For a full-sized image, see Appendix B.}
\label{fig:sCompViz2}
\centering
\scalebox{0.8}
{
\begin{tabular}{|c||c|c|c|c|c|}
  \hline
    &   {$1$-components}    &   {$2$-components} &   {$3$-components} &   {$4$-components} &   {$5$-components}
  \\ \hline
  \rotatebox[origin=c]{90}{LesMis$^*$}
  &
   \begin{minipage}{.2\textwidth}
     \includegraphics[width=\linewidth, height=0.7\linewidth]{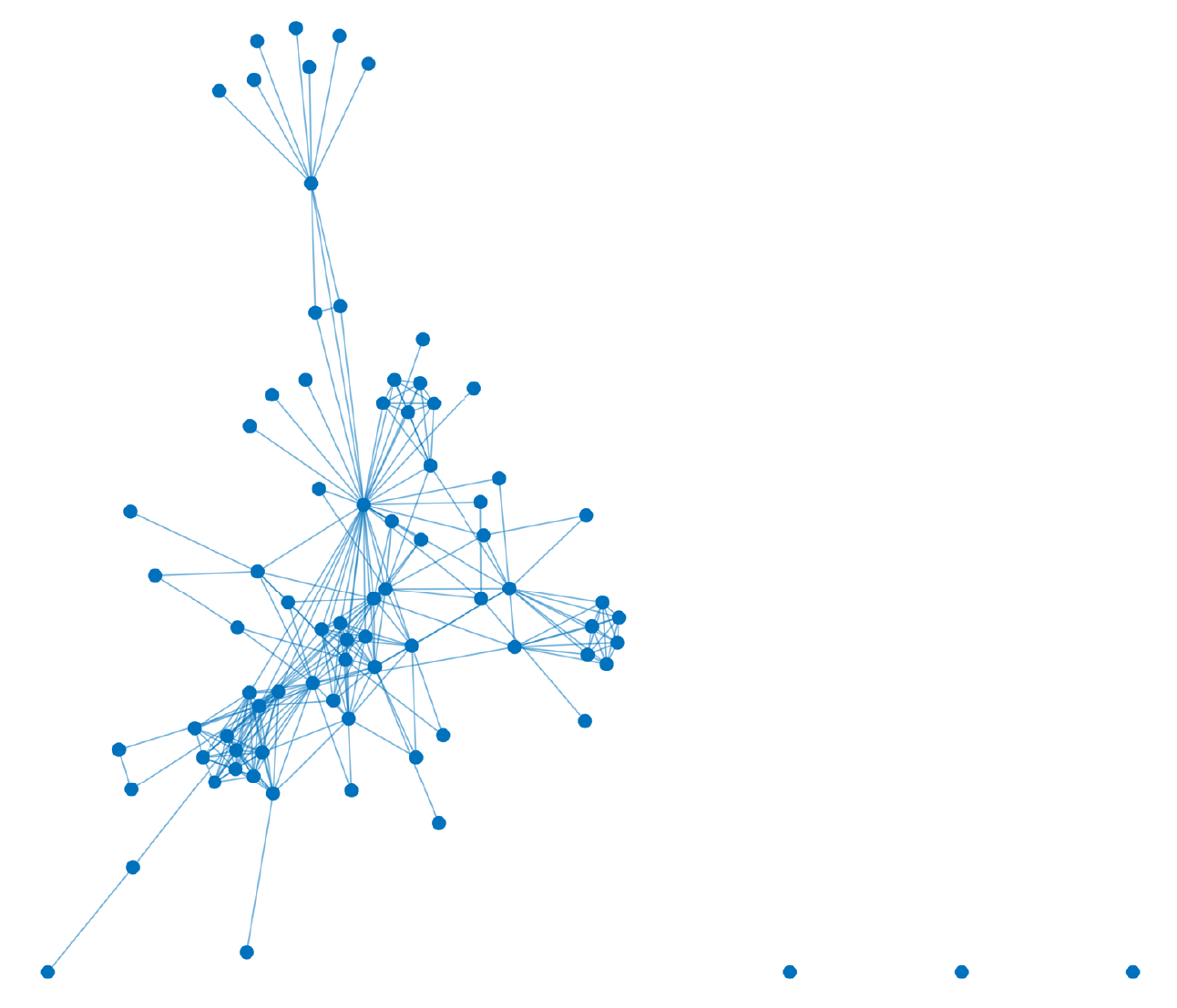}
    \end{minipage}
   &
   \begin{minipage}{.2\textwidth}
      \includegraphics[width=\linewidth, height=0.7\linewidth]{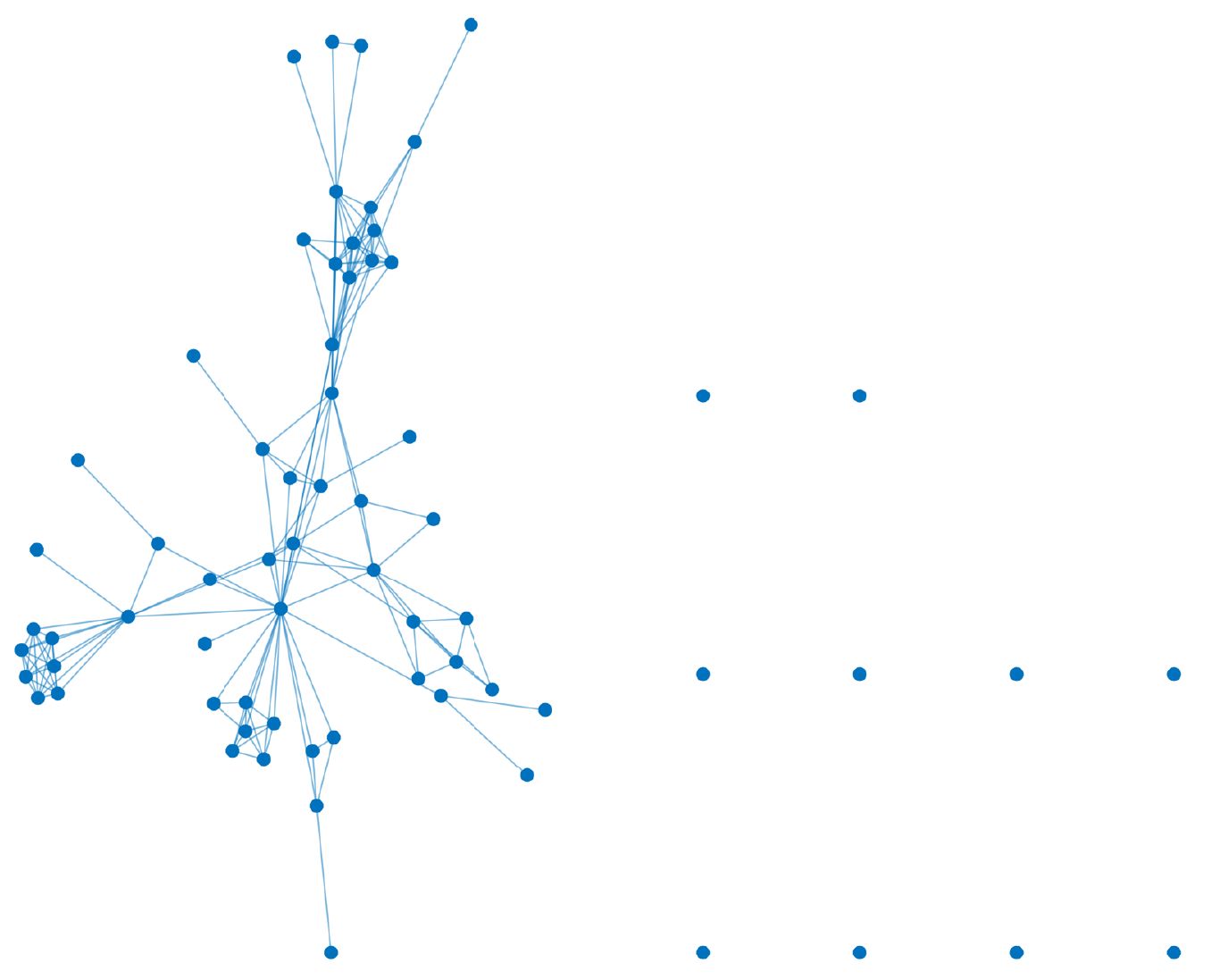}
    \end{minipage}
   &
   \begin{minipage}{.2\textwidth}
      \includegraphics[width=\linewidth, height=0.7\linewidth]{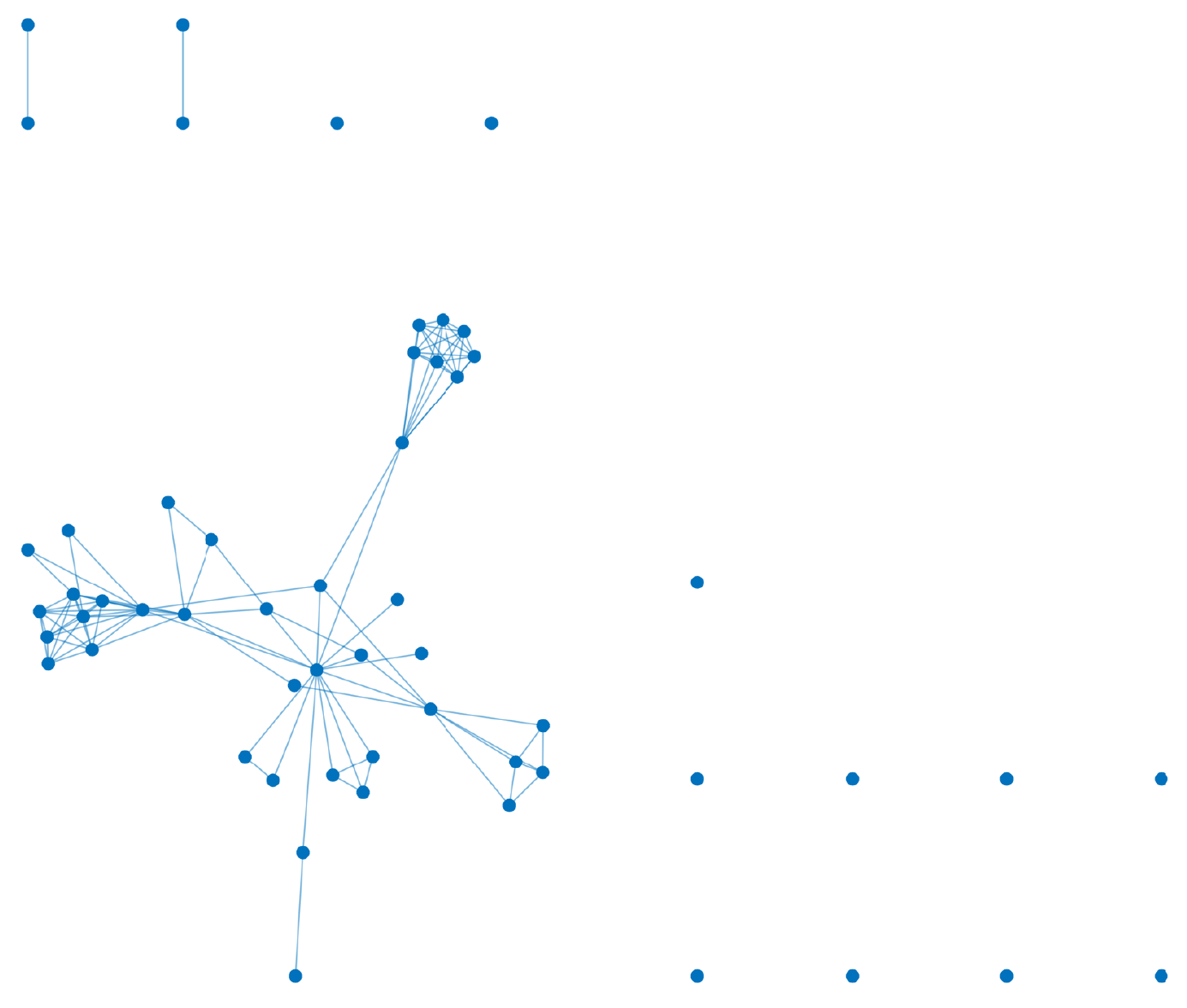}
    \end{minipage}
 & 
   \begin{minipage}{.2\textwidth}
      \includegraphics[width=\linewidth, height=0.7\linewidth]{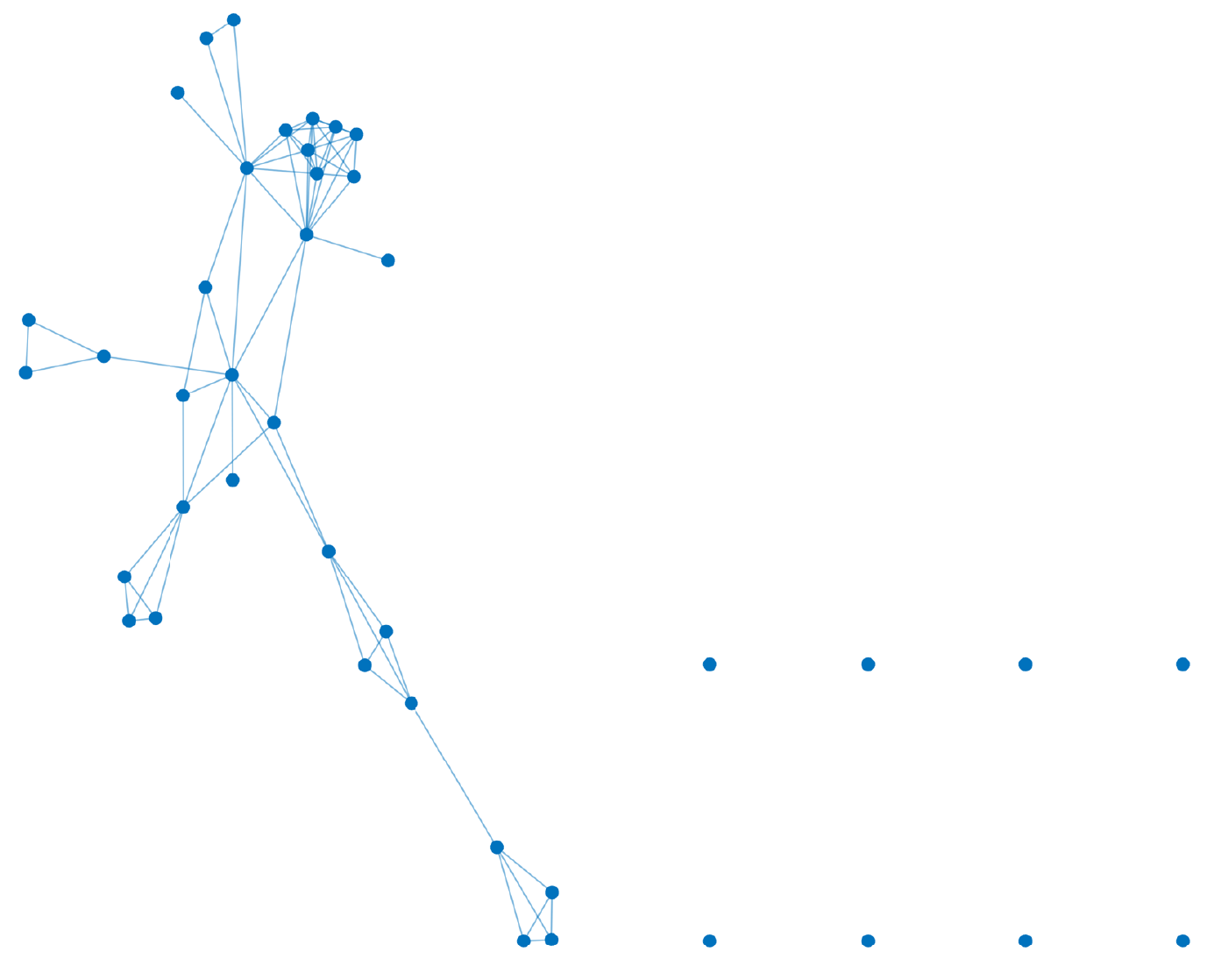}
    \end{minipage}
 & 
   \begin{minipage}{.2\textwidth}
   \vspace{0.5mm}
      \includegraphics[width=\linewidth, height=0.75\linewidth]{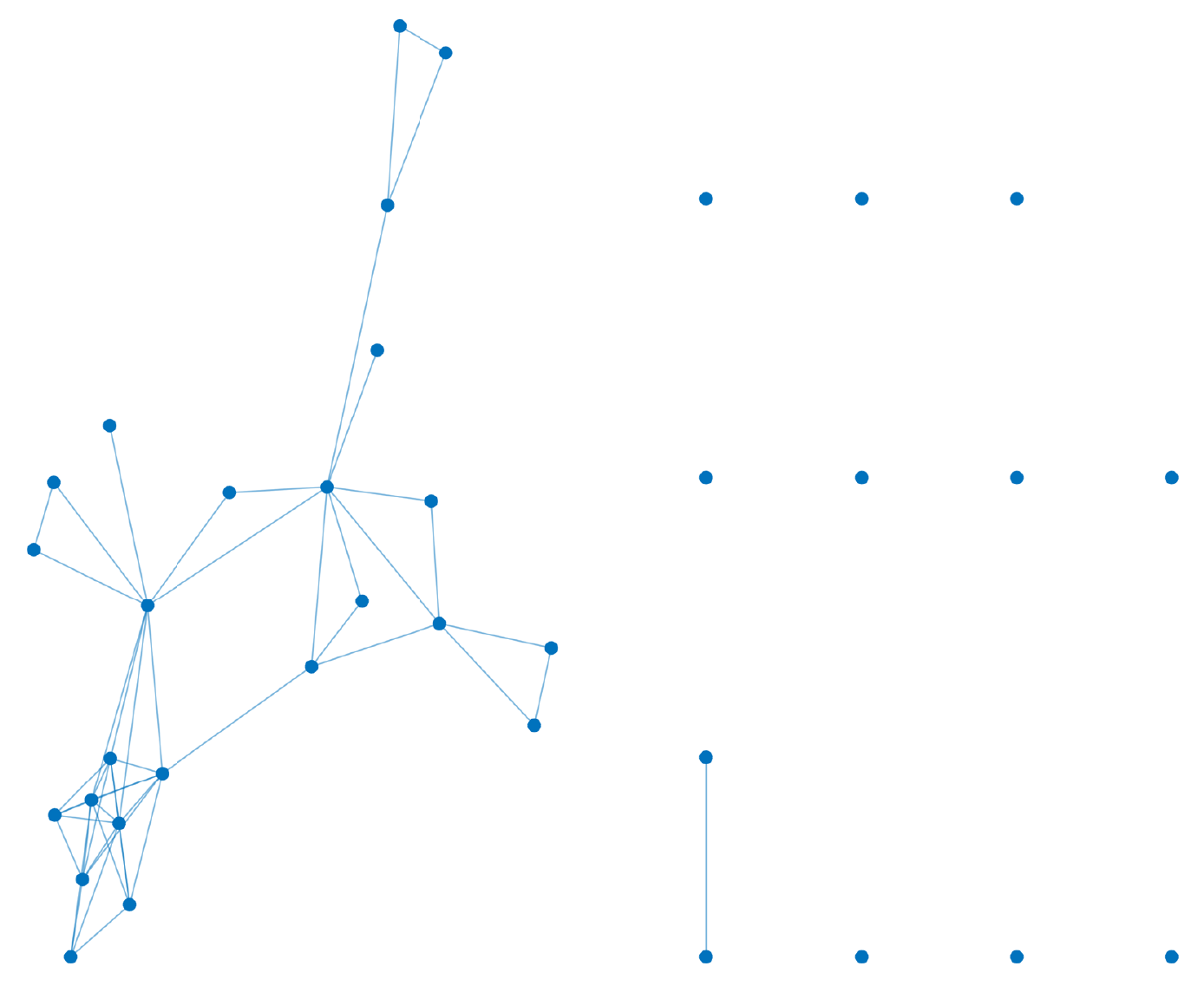}
    \end{minipage}
  \\ \hline
  \rotatebox[origin=c]{90}{Diseasome}
&
   \begin{minipage}{.2\textwidth}
      \includegraphics[width=\linewidth]{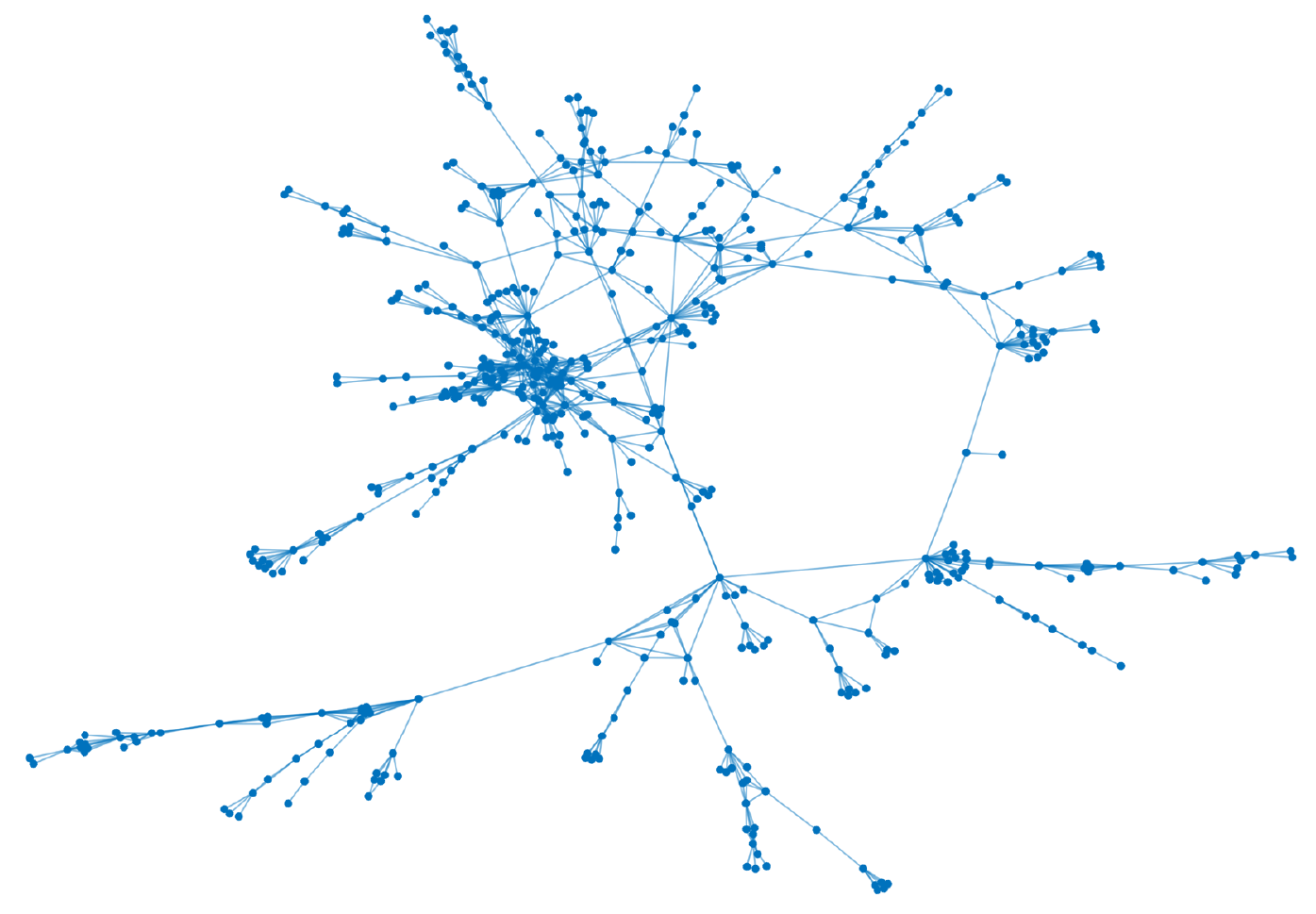}
    \end{minipage}
&
   \begin{minipage}{.2\textwidth}
      \includegraphics[width=\linewidth]{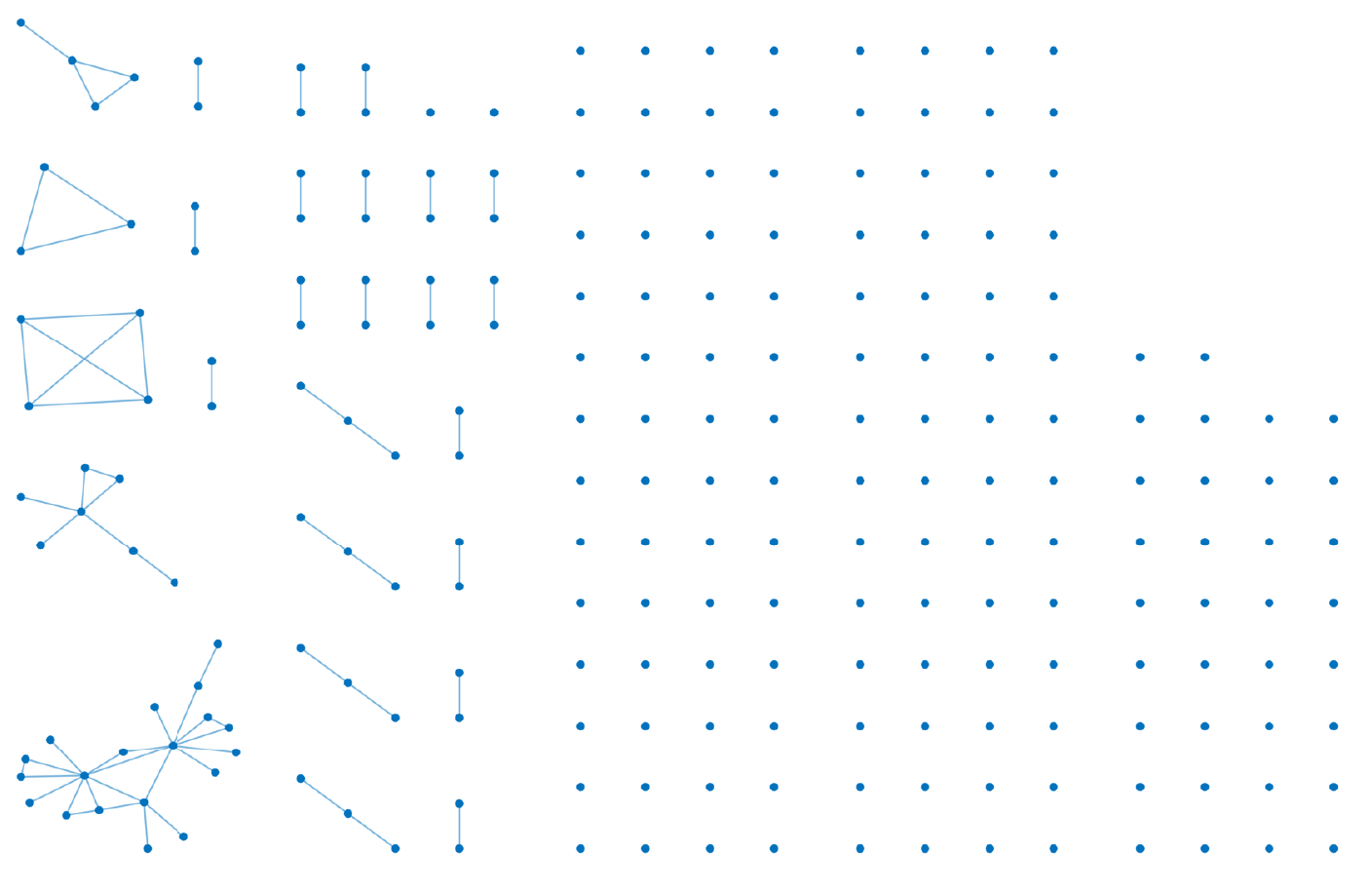}
    \end{minipage}
  & 
   \begin{minipage}{.2\textwidth}
      \includegraphics[width=\linewidth, height=0.7\linewidth]{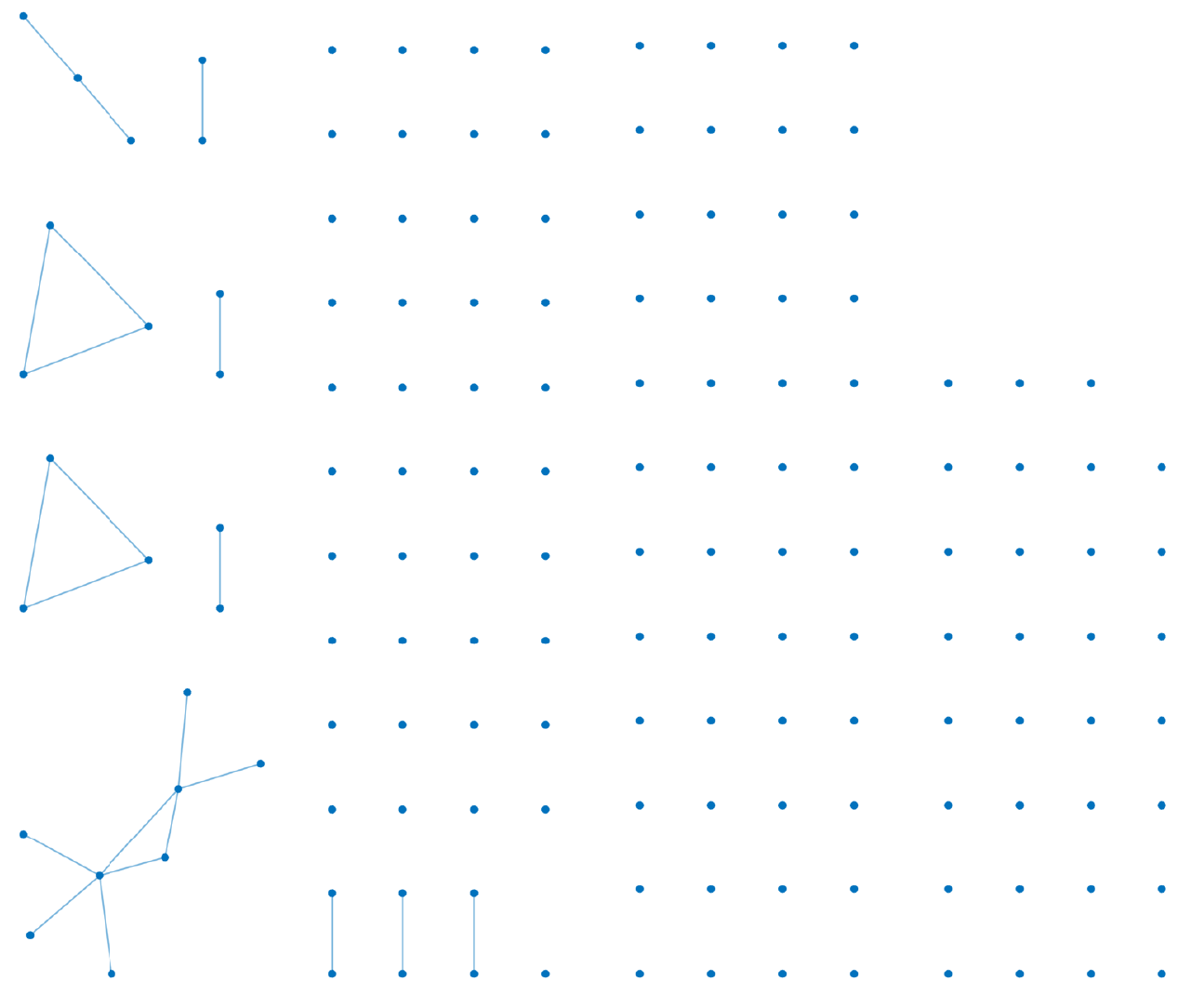}
    \end{minipage}
 &
   \begin{minipage}{.2\textwidth}
   \vspace{0.5mm}
      \includegraphics[width=\linewidth]{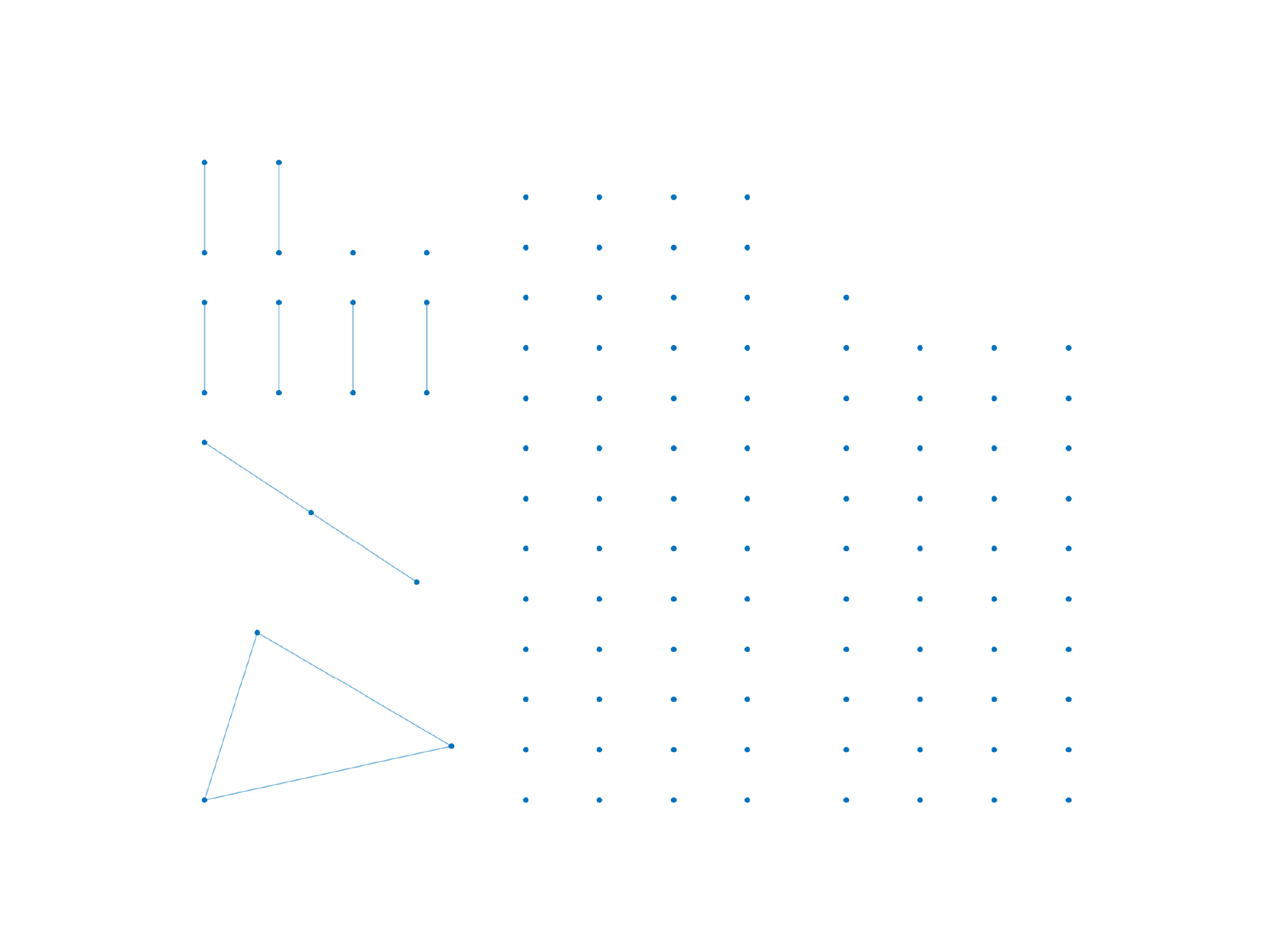}
    \end{minipage}
&
   \begin{minipage}{.2\textwidth}
      \includegraphics[width=\linewidth]{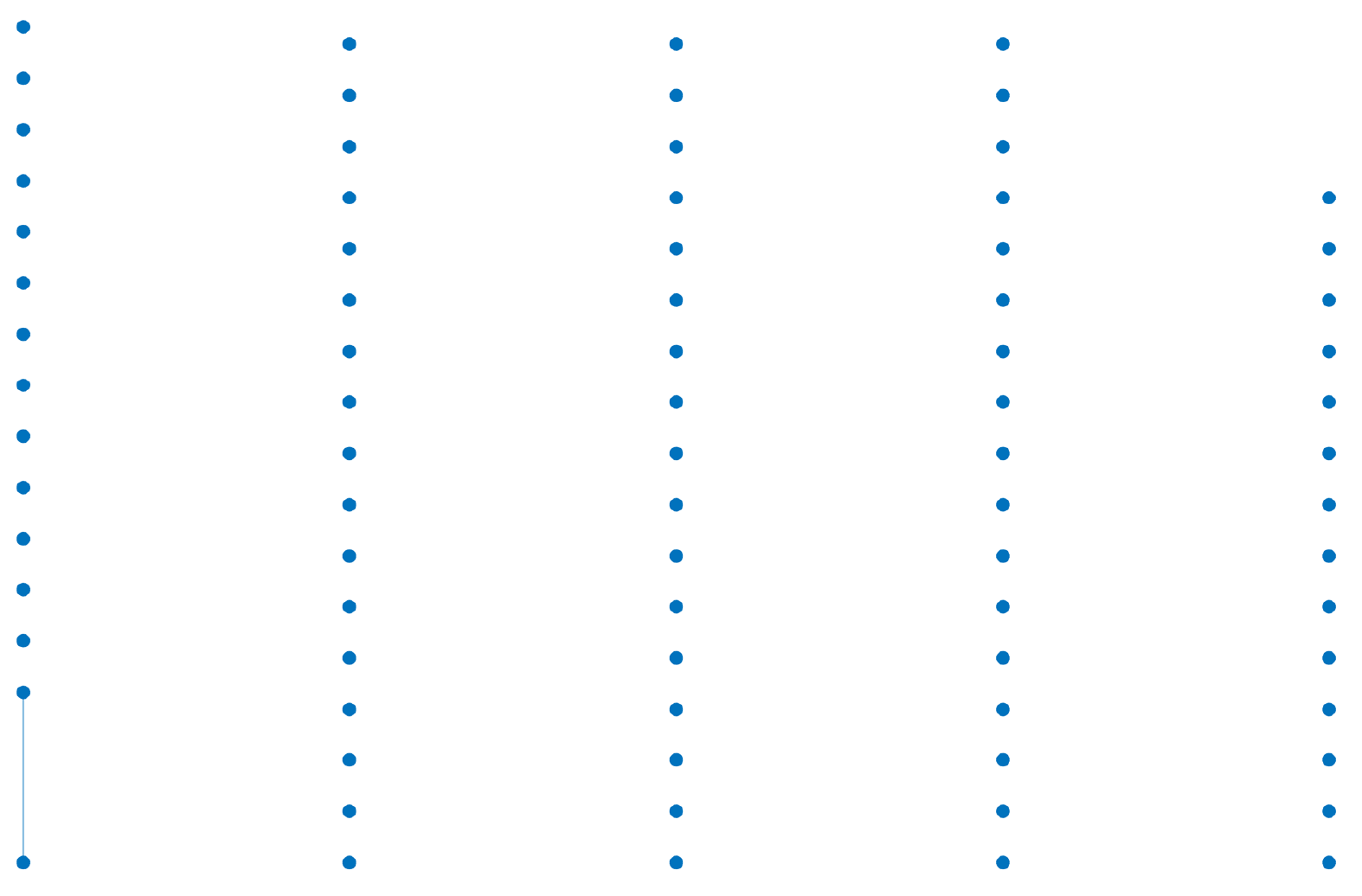}
    \end{minipage}
  \\ \hline
  \rotatebox[origin=c]{90}{CompBoard}
&
   \begin{minipage}{.2\textwidth}
   \centering
      \includegraphics[width=\linewidth]{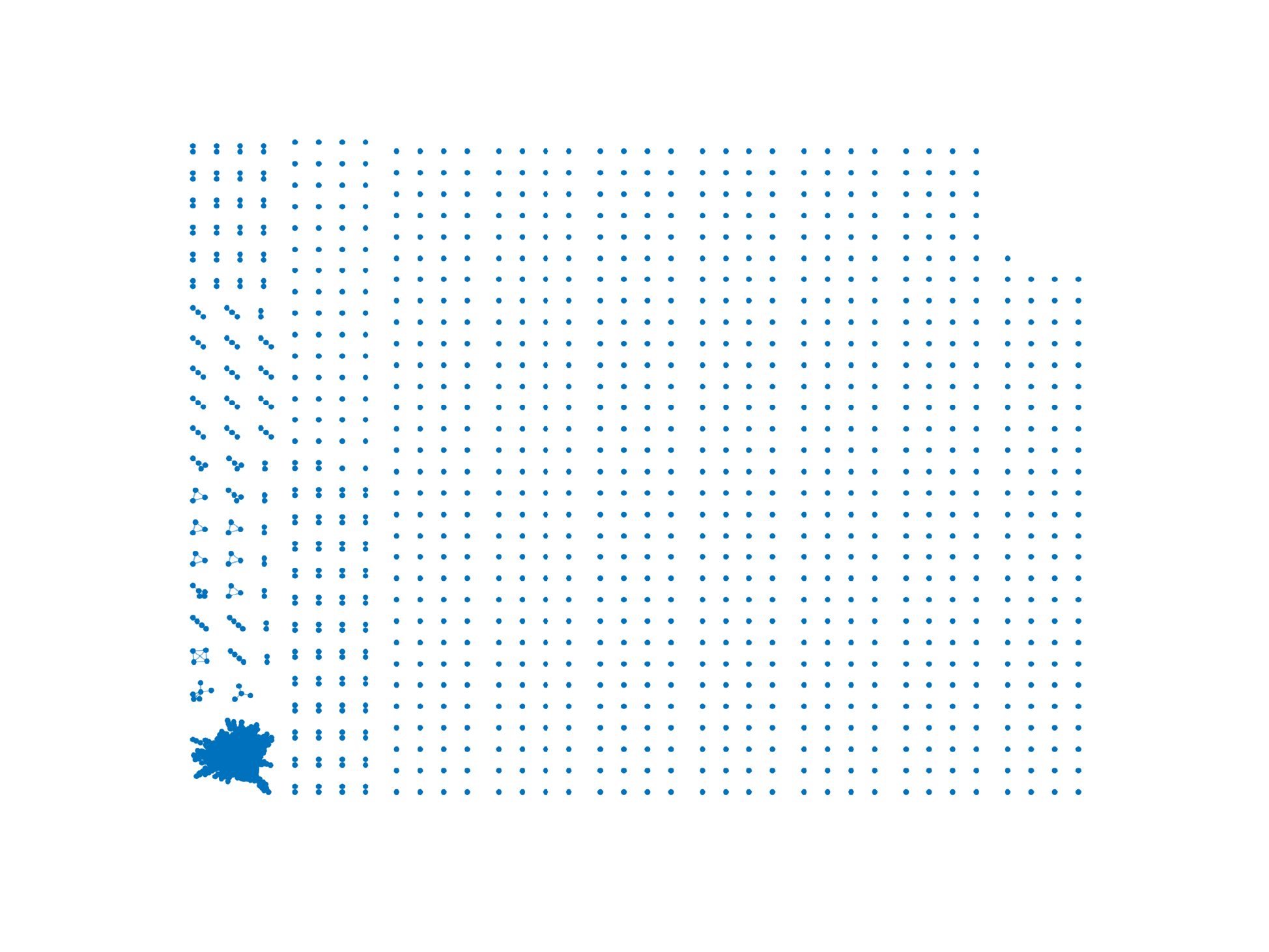}
    \end{minipage}
&
   \begin{minipage}{.2\textwidth}
   \centering
    \includegraphics[width=\linewidth, height=0.75\linewidth]{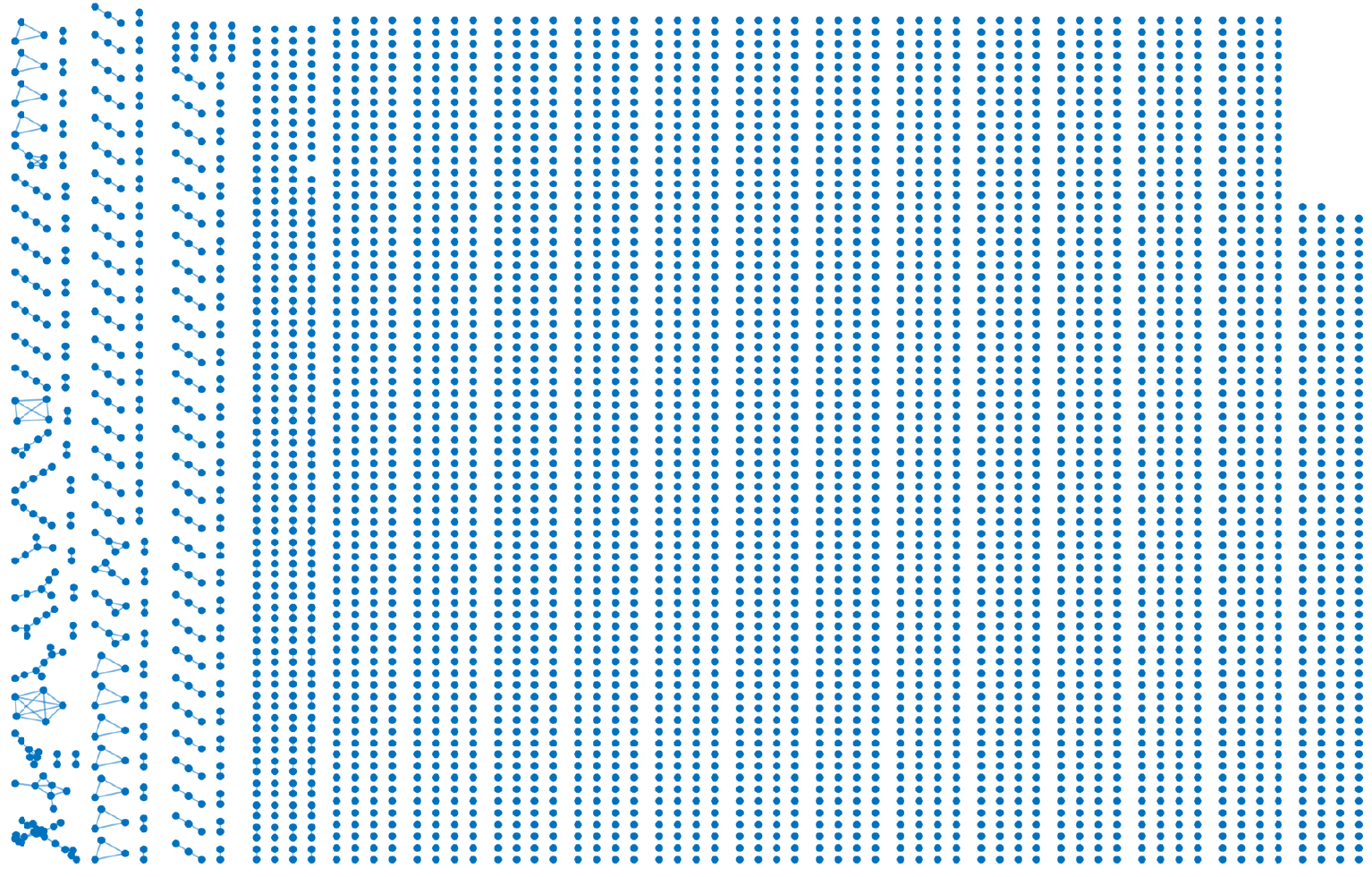}
    \end{minipage}
  & 
   \begin{minipage}{.2\textwidth}
   \centering
      \includegraphics[width=\linewidth,height=0.75\linewidth]{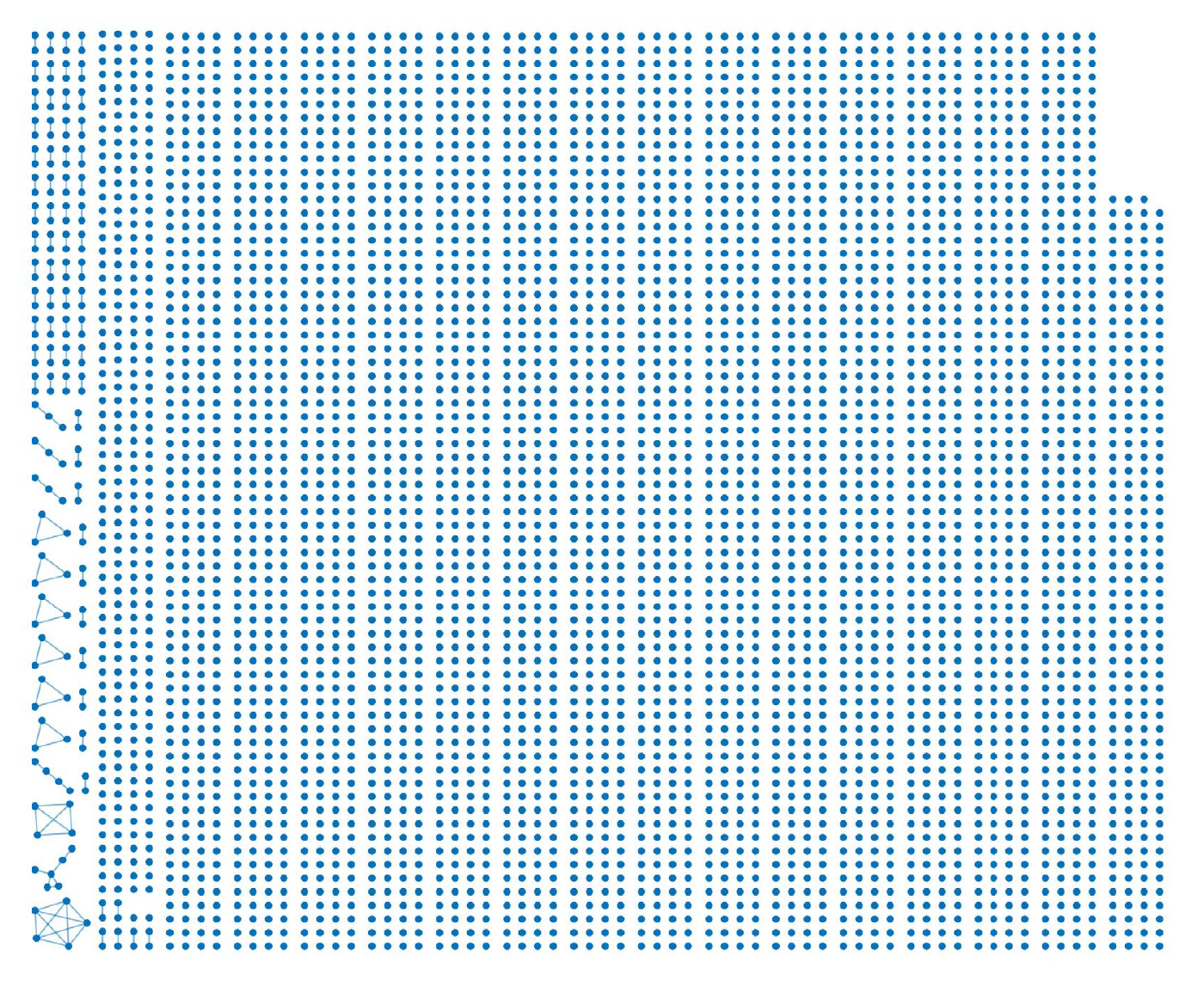}
    \end{minipage}
 &
   \begin{minipage}{.2\textwidth}
   \centering
      \includegraphics[width=\linewidth, height=0.75\linewidth]{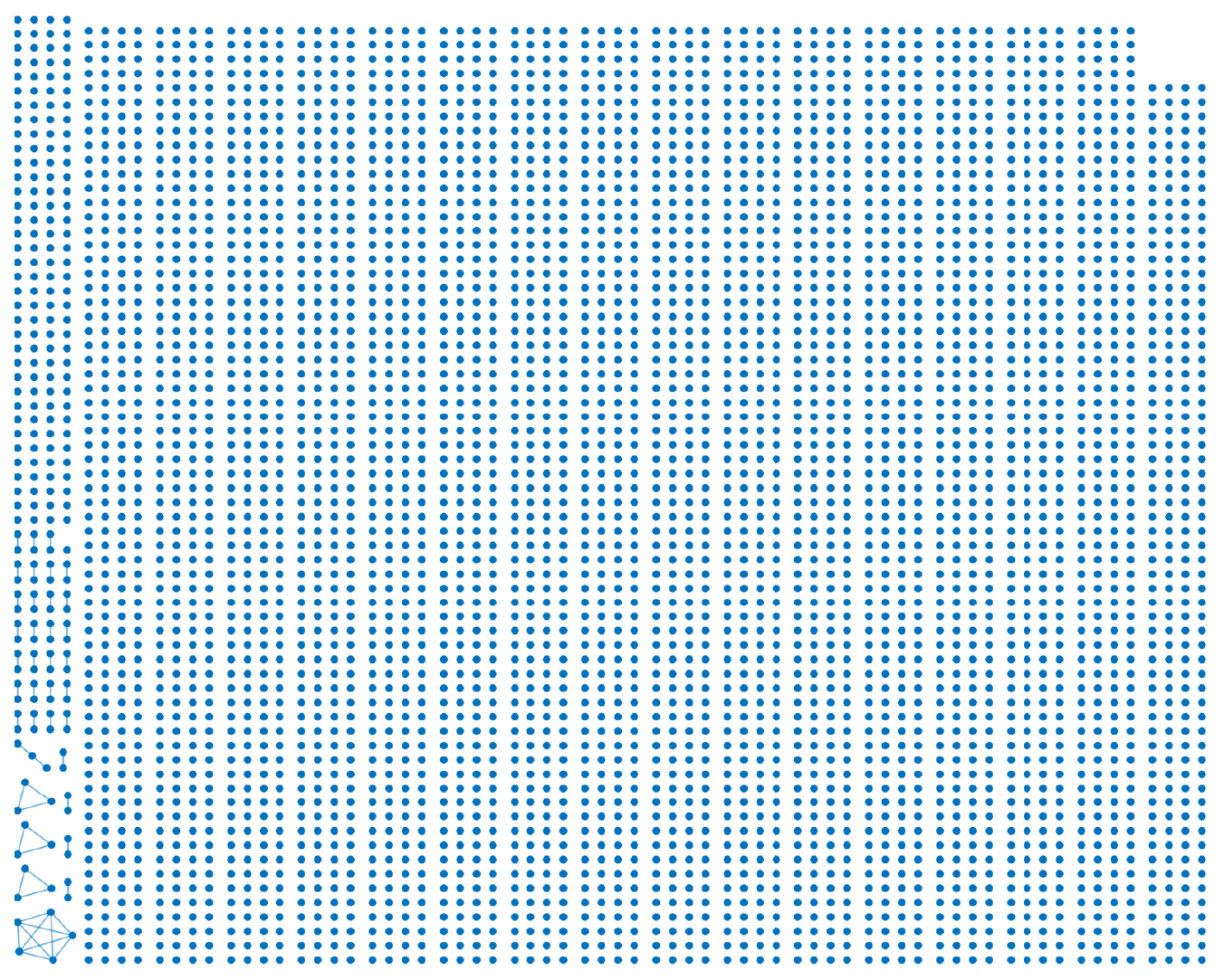}
    \end{minipage}
&
   \begin{minipage}{.2\textwidth}
   \centering
   \vspace{0.5mm}
      \includegraphics[width=\linewidth]{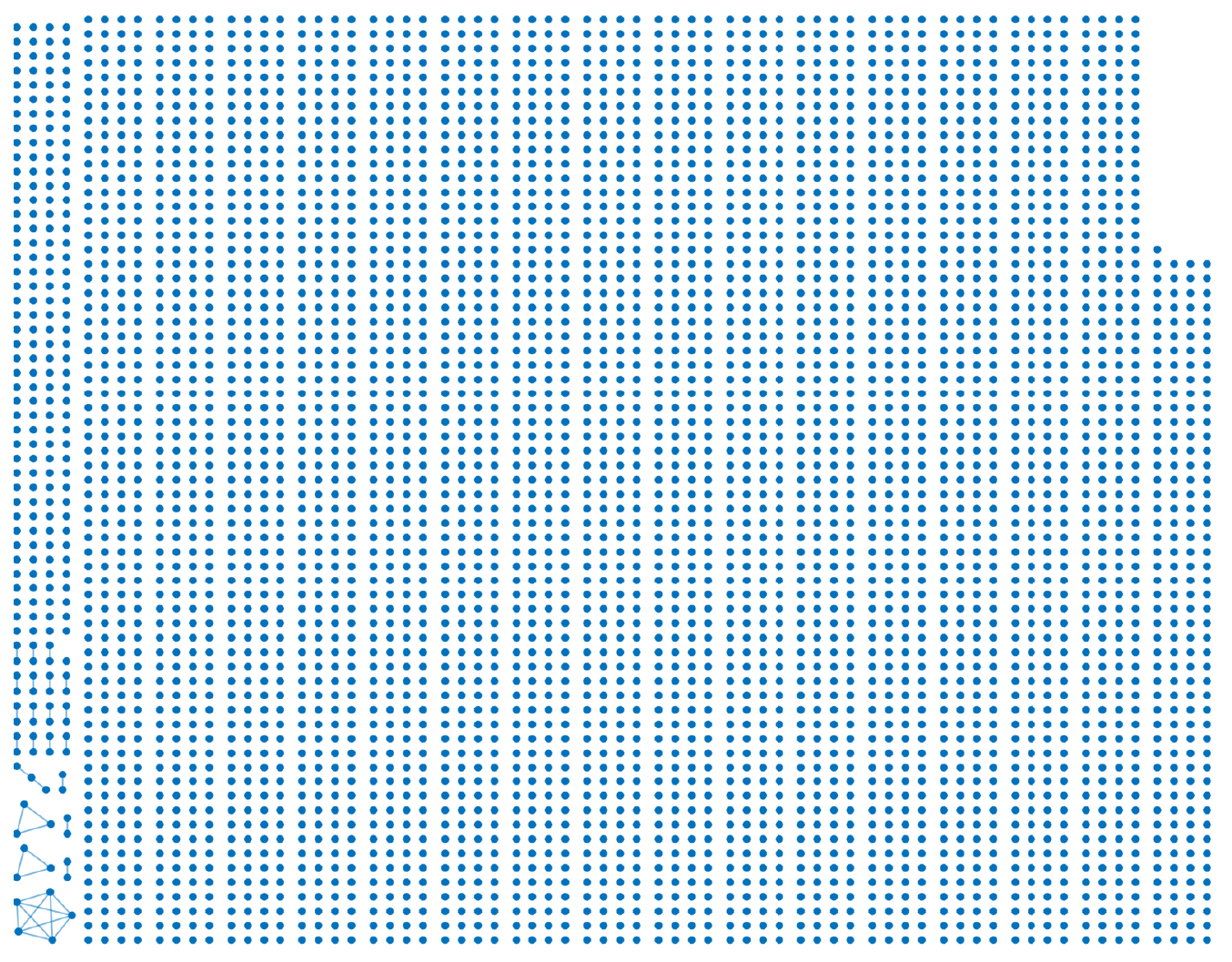}
    \end{minipage}
  \\ \hline
  \end{tabular}
  }
\end{table}

Qualitative differences are readily apparent from the visualization.
For LesMis, the majority of hyperedges are contained
within a giant component for $s=1,\ldots,5$. This means one can link
most characters with each other via a pathway of characters
co-occurring in at least one to five scenes together. For $s=1$ in Diseasome, we similarly observe a giant component; however, for
$s\geq 2$, this giant component fragments into small, roughly equally sized
components. Here, as $s$ increases from one to five, many of the
shared-gene pathways linking diseases for $s=1$ break down. By $s=5$, the
$s$-components consist almost entirely of isolated hyperedges (apart from
a single pair of closely related diseases,``Diabetes Mellitus'' and ``Mature
onset diabetes of the young (MODY)'') implying diseases associated
with 5 or more genes do not share 5 or more of those genes with other disorders.
The most dramatic fragmentation occurs for CompBoard. For
$s=1$, $74\%$ of the companies are contained within the giant
component (pictured in the lower-left hand corner), while for $s=2$, this
drops to $0.5\%$. This affirms shared board-member pathways linking
companies almost always rely on \textit{single} shared board members.

To quantify these changes in $s$-connected
component sizes more rigorously, we compute several entropy-based
measures on the $s$-connected component size probability distribution,
$\boldsymbol{p}_{s}=\langle p^{s}_{1},\ldots,p^{s}_{k} \rangle $, defined by
taking $p^{s}_{j}$ to be the fraction of hyperedges in $E_{s}$ that are
in the $j$'th $s$-component.
For a discrete probability distribution $\boldsymbol{p}$, its Shannon entropy is
given by $H(\boldsymbol{p})=-\sum_{i=1}^{k} p_{i} \log _{2}(p_{i})$. However,
direct comparisons of Shannon entropy on our data may be problematic, as the number of hyperedges in $E_{s}$ and number of $s$-components
varies not only between datasets, but also as $s$ varies within each dataset,
thereby complicating cross-comparisons of the (unitless) Shannon entropy.
To facilitate more meaningful entropy comparisons, we consider
the normalized entropy
\begin{equation*}
\widetilde{H}(\boldsymbol{p})=\frac{H(\boldsymbol{p})}{\log _{2}(k)} \in [0,1],
\end{equation*}
called $p$-smoothness by the authors \cite{Joslyn2016}.
This normalization derives from the fact that ${H}$ achieves
its maximum value of $\log _{2}(k)$ for the uniform distribution,
$\langle \frac{1}{k},\ldots,\frac{1}{k} \rangle $, and a minimum
value of $0$ for a fully skewed distribution, e.g. $\langle 0,\ldots, 0, 1\rangle $. For the special case in which $k=1$, one takes the
limiting definition as $k \to 1$, and defines
$\widetilde{H}(\boldsymbol{p}):= 1$.

We consider the $p$-smoothness of the $s$-component size distribution
$\boldsymbol{p}_{s}$, which we denote
$\widetilde{H}_{s} = \widetilde{H}(\boldsymbol{p}_{s})$. If all
$s$-components are equally sized, then $\widetilde{H}_{s}$ is 1, whereas
if the disparity between component sizes is maximal (e.g. $ \llvert  E_{s} \rrvert  -1$
hyperedges in one $s$-component, and $1$ hyperedge in the other), then
$\widetilde{H}_{s}$ approaches~0. In this sense, $p$-smoothness reflects
how smooth or uniform the $s$-component sizes are, but may not reflect
how dispersedly the hyperedges are distributed among $s$-connected components.
For that purpose, we consider an additional measure, again from
\cite{Joslyn2016}, aptly called dispersion. The dispersion
of the $s$-component size distribution compares the number of
$s$-components to the number of \textit{possible} $s$-components on a logarithmic
scale, i.e.
\begin{equation*}
D_{s}=\frac{\log _{2}( \llvert  C_{s} \rrvert  )}{\log _{2}( \llvert  E_{s} \rrvert  )} \in [0,1],
\end{equation*}
where $C_{s}$ denotes the set of $s$-connected components and
$E_{s}$ denotes the set of $s$-hyperedges.

\begin{figure}[t]
\centering
\begin{subfigure}[b]{0.27\textwidth}
\includegraphics[width = \linewidth]{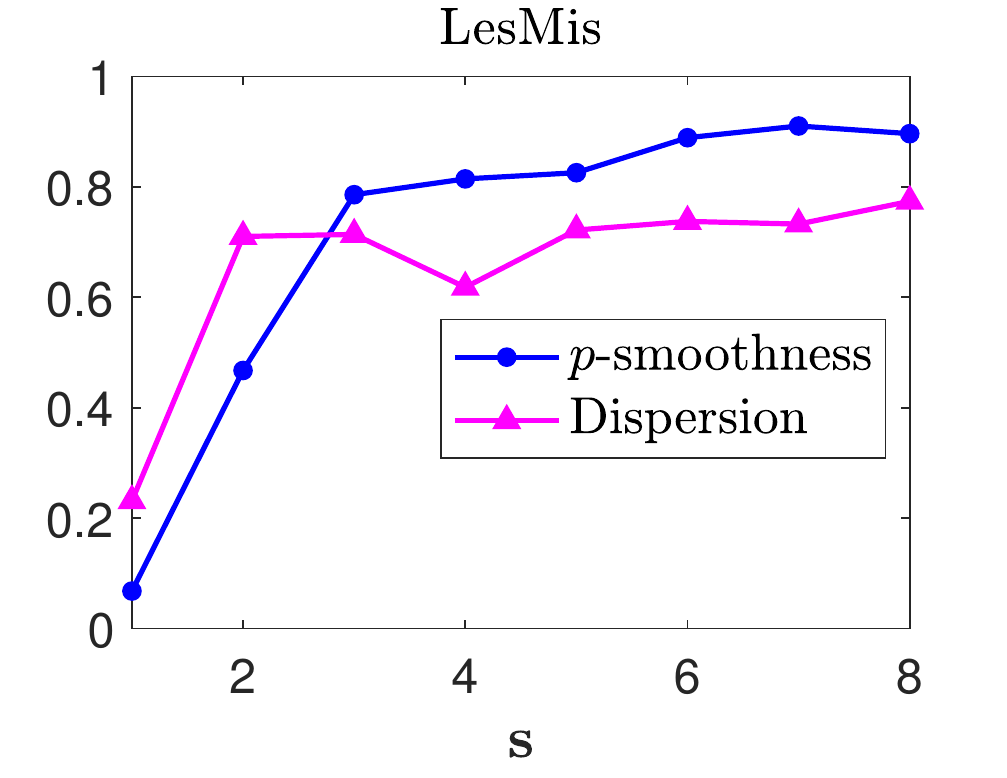} \\
\includegraphics[width = \linewidth]{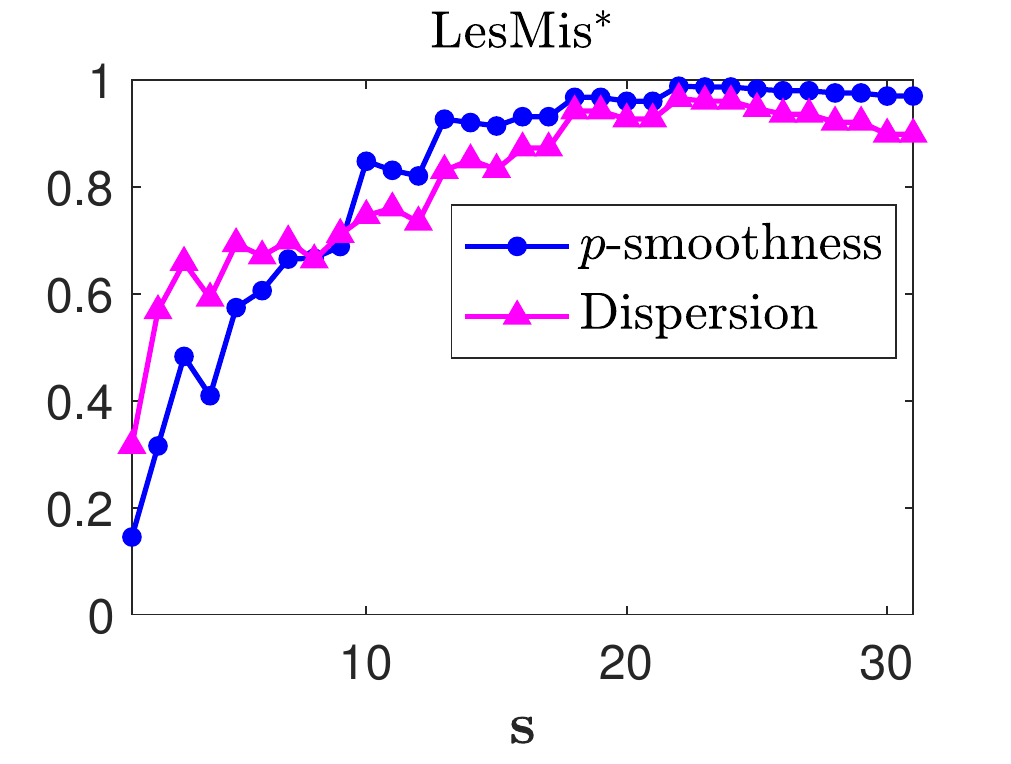} \\
\end{subfigure}
\qquad
\begin{subfigure}[b]{0.27\textwidth}
\includegraphics[width = \linewidth]{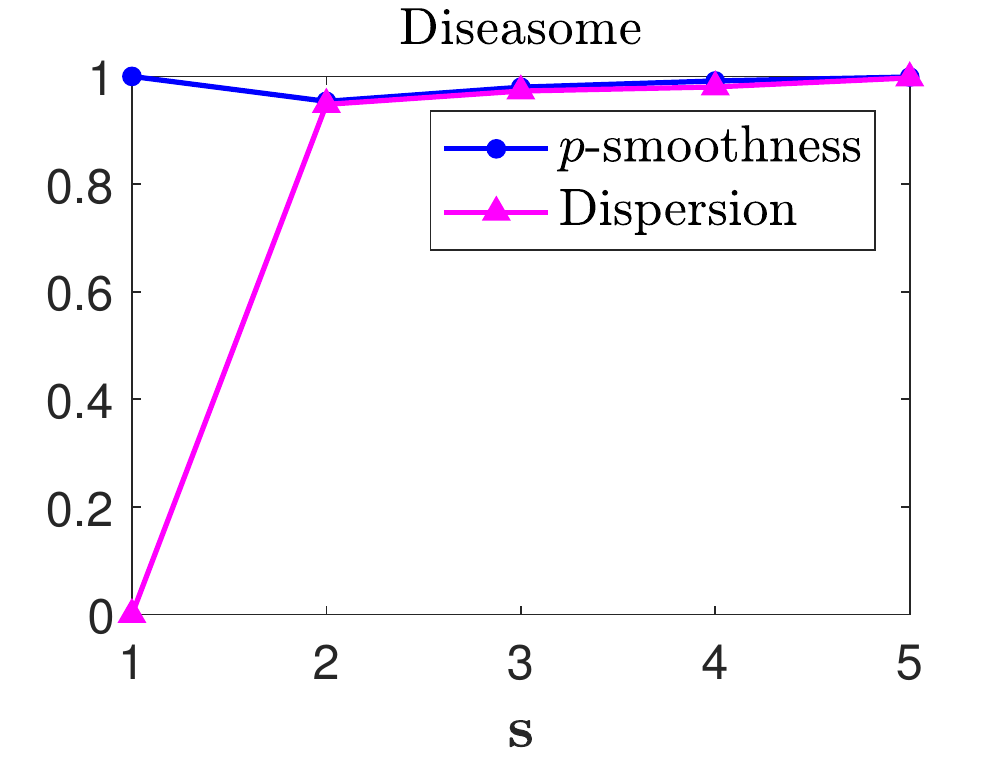} \\
\includegraphics[width = \linewidth]{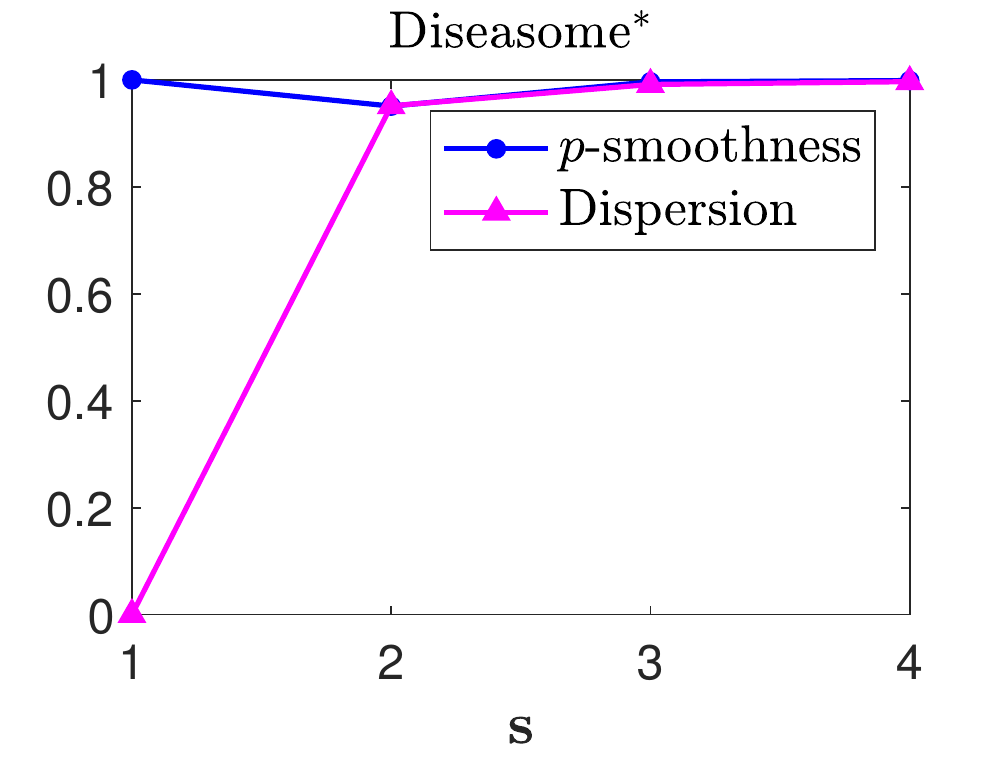} \\
\end{subfigure}
\qquad
\begin{subfigure}[b]{0.27\textwidth}
\includegraphics[width = \linewidth]{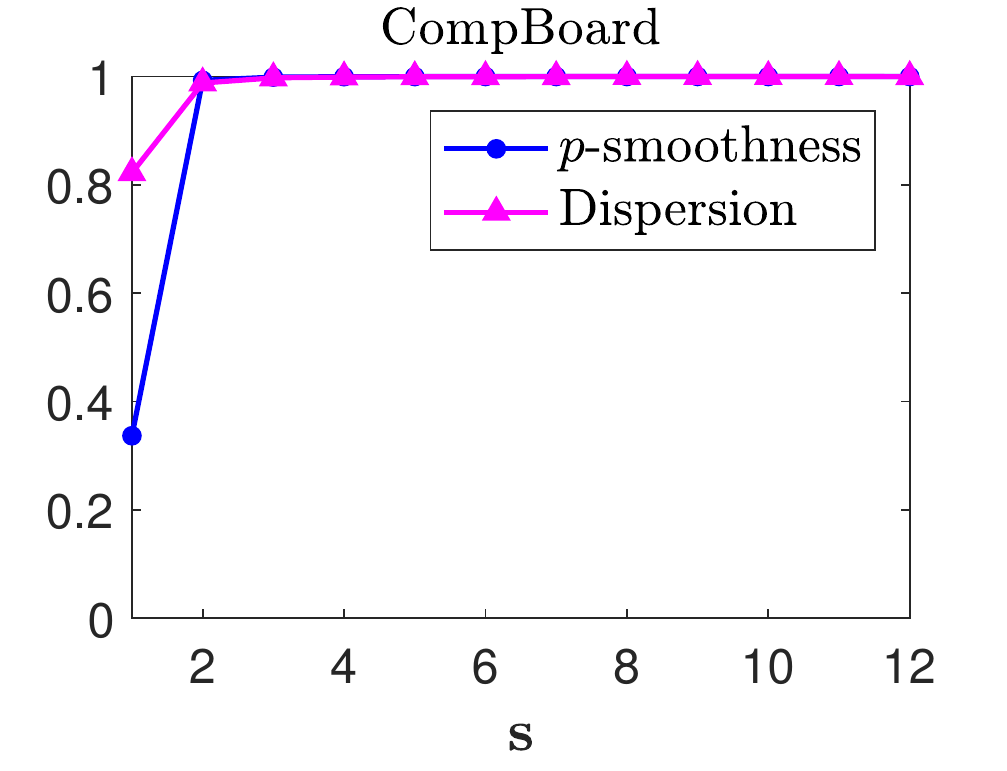} \\
\includegraphics[width = \linewidth]{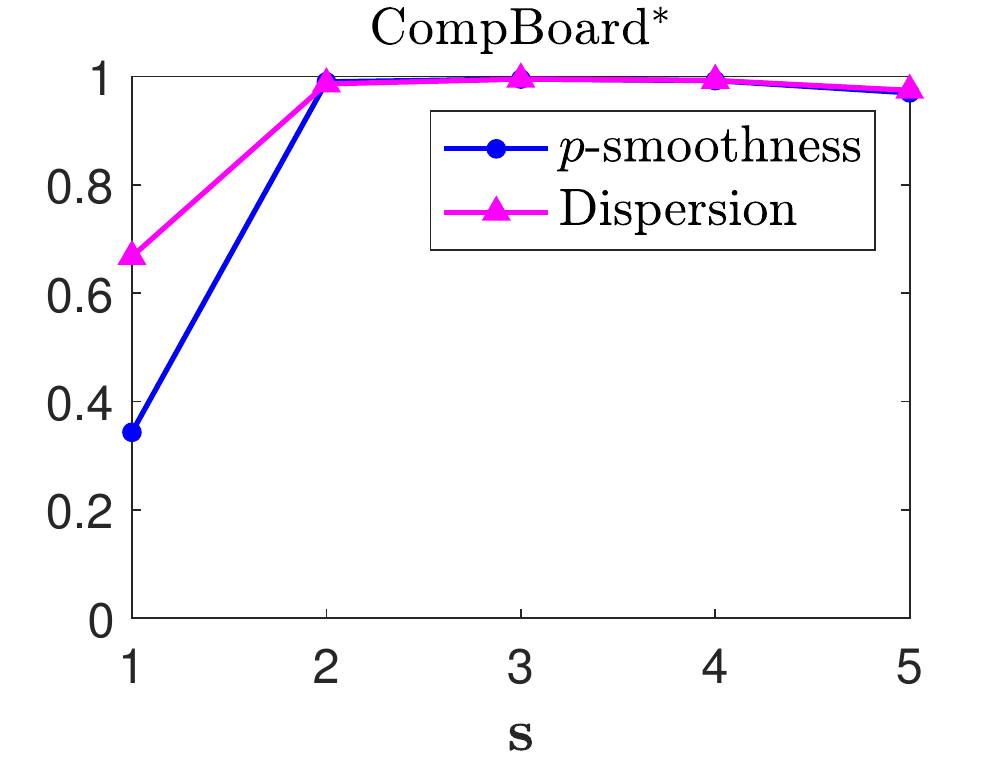} \\
\end{subfigure}
\vspace{-5mm}
\caption{The $p$-smoothness (normalized entropy) and dispersion values of the $s$-connected components for LesMis, Diseasome, and CompBoard (top row) and their dual hypergraphs (bottom row).} \label{fig:entropy}
\end{figure}

Fig.~\ref{fig:entropy} plots the $p$-smoothness and dispersion
for $s=1,\ldots, s_{\max }$, where
$s_{\max }=\max_{f,g \in E}  \llvert  f \cap g \rrvert  $. For $s>s_{\max }$, the $s$-components
are either all isolated hyperedges, or non-existent. In all datasets, both dispersion and $p$-smoothness tend to increase in
$s$, although, as evident from LesMis, this increase is not always monotonic.
LesMis$^{*}$ exhibits lower values of $p$-smoothness
for each of $s=1,\ldots,5$ relative to those for corresponding values of
$s$ in the other datasets, consistent with the highly skewed distribution
reflecting the large component we observed in the visualization. CompBoard exhibits a large separation between $p$-smoothness
and dispersion for $s=1$. In this case, while the component size distribution
is still skewed---and hence has low $p$-smoothness---the remaining
$s$-components consist of many isolated hyperedges, reflected in the high
dispersion value. Lastly, for Diseasome, $p$-smoothness is
maximal while dispersion is minimal for $s=1$, and for $s \geq 2$ both
$p$-smoothness and dispersion closely coincide at values near~1. This reflects
the fragmentation of a single giant component into many $s$-components
(hence the high dispersion) that are equally sized (hence the high
$p$-smoothness).

\subsection{Distance and Centrality}\label{sec:sDist}

\paragraph*{Methods}
Under Definition~\ref{def:hyp}, it is straightforward to
show the length of the shortest $s$-walk serves as a distance metric function
over a set of hyperedges. More precisely

\begin{prop}
Let $H=(V,E)$ be a hypergraph and $E_{s}=\{e\in E:  \llvert  e \rrvert   \geq s \}$. Define
the \emph{$s$-distance} function
$d_{s}: E_{s} \times E_{s} \to \mathbb{Z}_{\geq 0}$ by
\begin{equation*}
d_{s}(f,g)= %
\begin{cases}
\textit{length of the shortest }s\textit{-walk} &
\textit{if an }s\textit{-walk between }f,g\textit{ exists},
\\
\infty & \textit{otherwise}.
\end{cases} %
\end{equation*}
Then $(E_{s},d_{s})$ is a metric space.
\end{prop}

We omit the proof, as the triangle inequality can be proved constructively,
and the other metric space axioms follow immediately from Definition~\ref{def:swalk}. \\

\noindent {\bf Graph case \& equivalence}:
If $H$ is a graph, then
the graph distance between vertices $x$ and $y$ in $H$ is equivalent to
the $1$-distance between hyperedges $x^{*}$ and $y^{*}$ in $H^{*}$.
For a hypergraph $H$, the $s$-distance between $x$ and
$y$ is equivalent to the graph distance between $x^{*}$ and $y^{*}$ in
$L_{s}(H)$. Consequently, the forthcoming $s$-distance based measures in
Definitions~\ref{def:distanceMetrics}--\ref{def:harmDist} are equivalent
to their graph counterparts on $L_{s}(H)$ and, whenever $H$ is a graph,
reduce to their graph counterparts on $H^{*}$ for $s=1$. \\

With $s$-distance serving as hypergraph geodesic distance, hypergraph
$s$-analogs of local and global distance-based graph invariants easily
extend.

\begin{definition} \label{def:distanceMetrics}
Let $H=(V,E)$ be a hypergraph. 
\item[$(i)$] The {\bf $s$-eccentricity} of a hyperedge $f$ is $\max\limits_{g \in E_s} d_s(f,g)$.
\begin{itemize}
\item[$-$] The {\bf $s$-diameter} is the maximum $s$-eccentricity over all edges in $E_s$, while the {\bf $s$-radius} is the minimum. 
\end{itemize}
\item[$(ii)$] The {\bf average $s$-distance} of $H$ is $\displaystyle{|E_s| \choose 2}^{-1} \displaystyle\sum\limits_{f,g \in E_s} d_s(f,g)$.
\item[$(iii)$] The {\bf $s$-closeness centrality} of a hyperedge $f$ is $\displaystyle\frac{|E_s|-1}{\displaystyle\sum\limits_{g \in E_s} d_s(f,g)}$.
\end{definition}

Important caveats arise when applying Definition~\ref{def:distanceMetrics} to real data. As we've observed, $H$ may contain more than one
$s$-component for some values of $s$, in which case the $s$-distance between
some pairs of edges is infinite. Consequently, the $s$-eccentricity of
every edge (and hence $s$-diameter and $s$-radius) and mean $s$-distance
are all infinite; similarly, the $s$-closeness centrality of every edge
is trivially~0. Similar to how these issues are sometimes addressed for
graphs, one alternative is to compute these measures on only the largest
$s$-component. Depending on the analyst's aims, this approach might
be satisfactory, particularly if the majority of hyperedges in
$E_{s}$ are contained within the largest $s$-component, as was seemingly
the case in LesMis$^*$.

However, restricting to the largest component may be unsatisfactory in
cases where the largest $s$-component does not constitute the overwhelming
majority of edges in $E_{s}$, as in CompBoard
for $s\geq 2$. In such cases, one may wish to compute $s$-eccentricity
on a per-component basis, taking the extrema over all $s$-components as
the $s$-diameter and $s$-radius. One may similarly compute mean $s$-distance
or $s$-closeness per-component, however, it is unclear how to properly
synthesize these values in order to obtain (in the former case) a \textit{single}
global numerical measure or (in the latter case) a ranking over \textit{all}
hyperedges in the entire network. Instead of a per-component approach,
an elegant alternative for averaging graph distances in disconnected graphs,
advocated by Newman \cite{Newman2003}, is to use the harmonic mean instead
of the arithmetic. This approach was adopted by Latora and Marchiori
\cite{Latora2001} to define network \textit{efficiency} as the reciprocal
of the harmonic mean path length, proposed as a quantitative measure of
small-worldness. Latora and Marchiori termed this measure ``efficiency''
in reference to how efficiently information might be exchanged over the
network. Later, a similar approach was adopted by Rochat
\cite{rochat2009closeness} to define the \textit{harmonic closeness centrality
index} of vertices in a disconnected graph. Extending these notions to
the hypergraph context, a more practical definition of the aforementioned
$s$-distance based notions is given by:

\begin{definition}\label{def:harmDist}
Let $H=(V,E)$ be a hypergraph and let $C_s$ denote the set of its $s$-connected components. 
\item[$(i)$] The {\bf $s$-eccentricity} of a hyperedge $f \in C$ where $C \in C_s$ is $\max\limits_{g \in C} d_s(f,g)$.
\begin{itemize}
\item[$-$] The {\bf $s$-diameter} is the maximum $s$-eccentricity over all edges in $E_s$, while the {\bf $s$-radius} is the minimum. 
\end{itemize}
\item[$(ii)$] The {\bf average $s$-efficiency} of $H$ is $\displaystyle{|E_s| \choose 2}^{-1} \displaystyle\sum\limits_{\substack{f,g \in E_s \\ f\not=g}} \frac{1}{d_s(f,g)}$.
\item[$(iii)$] The {\bf harmonic $s$-closeness centrality index} of a hyperedge $f$ is $\displaystyle\frac{1}{|E_s|-1}\displaystyle\sum\limits_{\substack{g \in E_s \\ f \not=g}} \frac{1}{d_s(f,g)}$.
\end{definition} 

We take the limiting value of 0 for the summand in
(\textup{ii}) and (\textup{iii}) when $f$ and $g$ are in different $s$-components. Both
(\textup{ii}) and (\textup{iii}) are numerical quantities between 0 and 1, with larger
values indicative of closer $s$-distances between hyperedges, either globally
(in the former case) or locally (in the latter case).
If $ \llvert  E_{s} \rrvert  =1$, they are undefined. In practice, one may
ignore such isolated hyperedges when computing centrality or assign them
a value of 0 by convention, analogous to how
\cite[p.~221]{Freeman1978} sets the closeness centrality value of an isolated
vertex in a graph to~0. 

\paragraph*{Application to data}

\begin{figure}[t]
\centering
\begin{subfigure}[b]{0.28\textwidth}
\includegraphics[width = \linewidth]{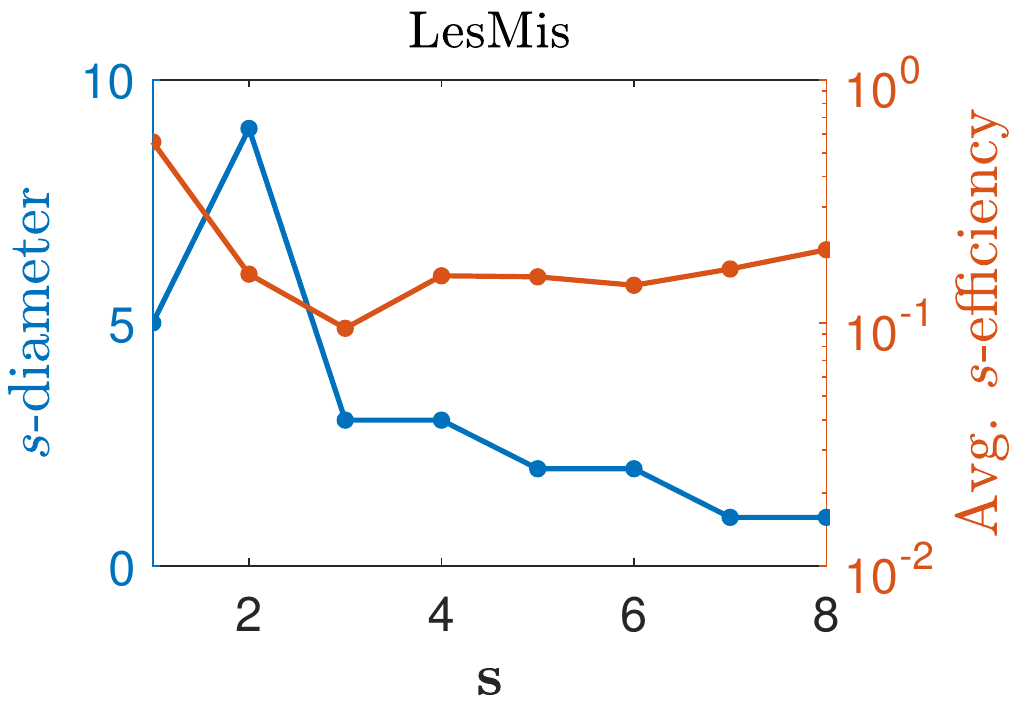} \\
\includegraphics[width = \linewidth]{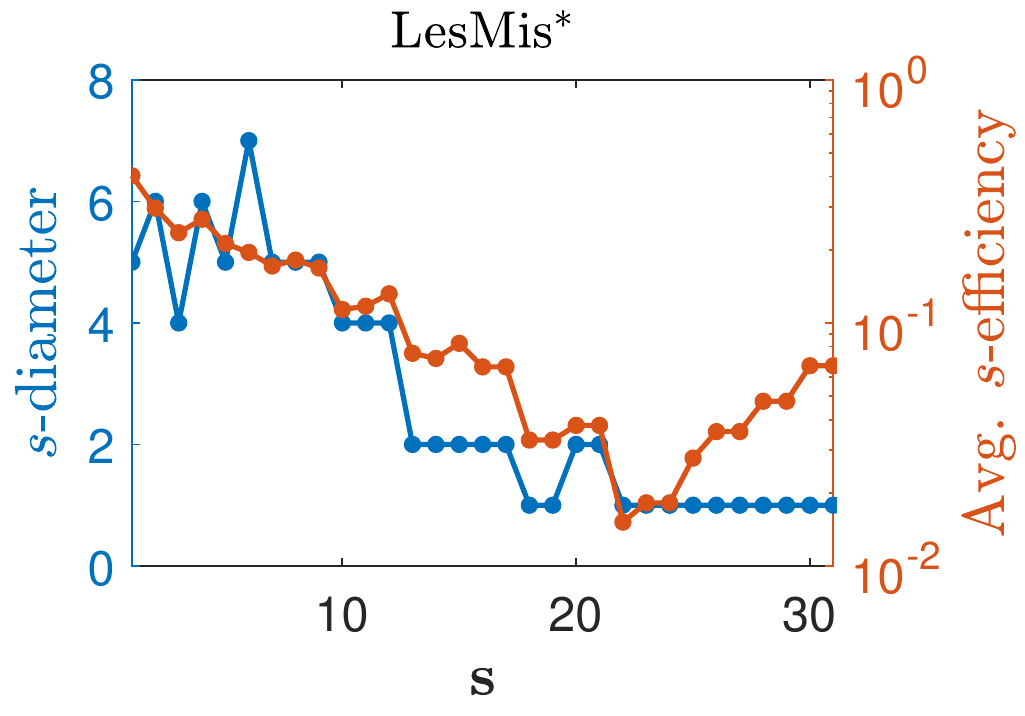} \\
\end{subfigure}
\qquad
\begin{subfigure}[b]{0.28\textwidth}
\includegraphics[width = \linewidth]{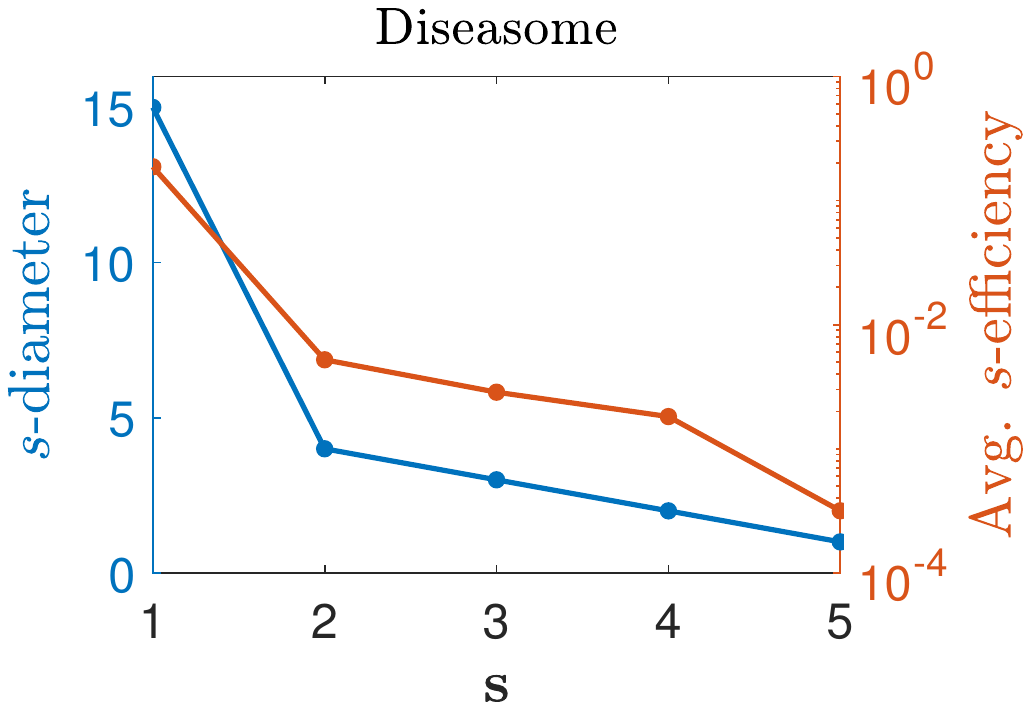} \\
\includegraphics[width = \linewidth]{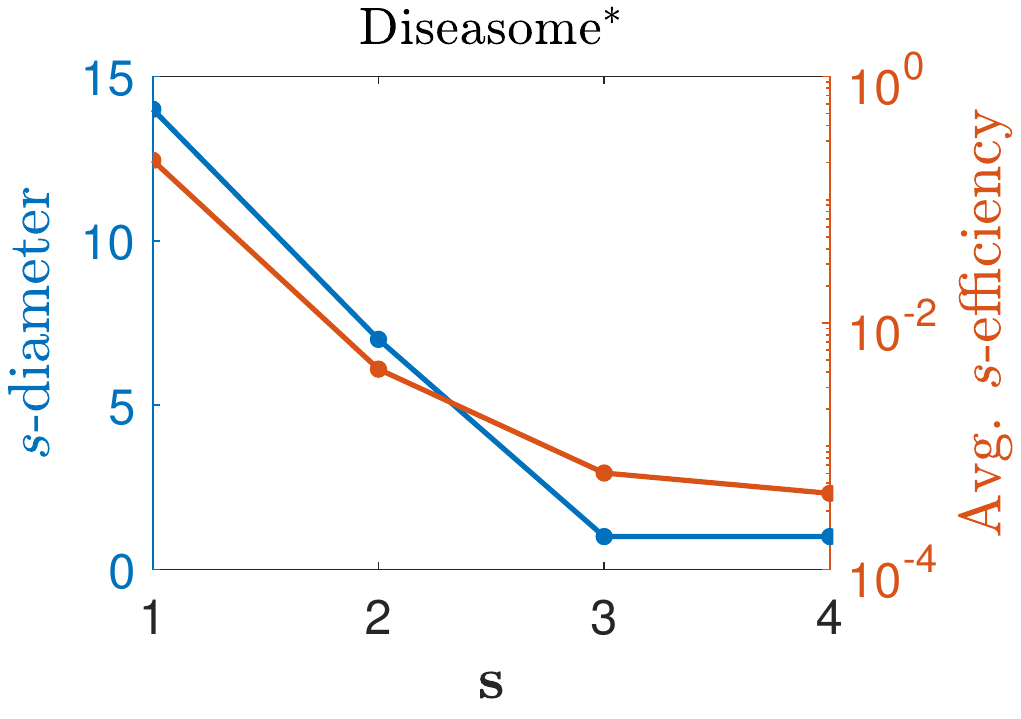} \\
\end{subfigure}
\qquad
\begin{subfigure}[b]{0.28\textwidth}
\includegraphics[width = \linewidth]{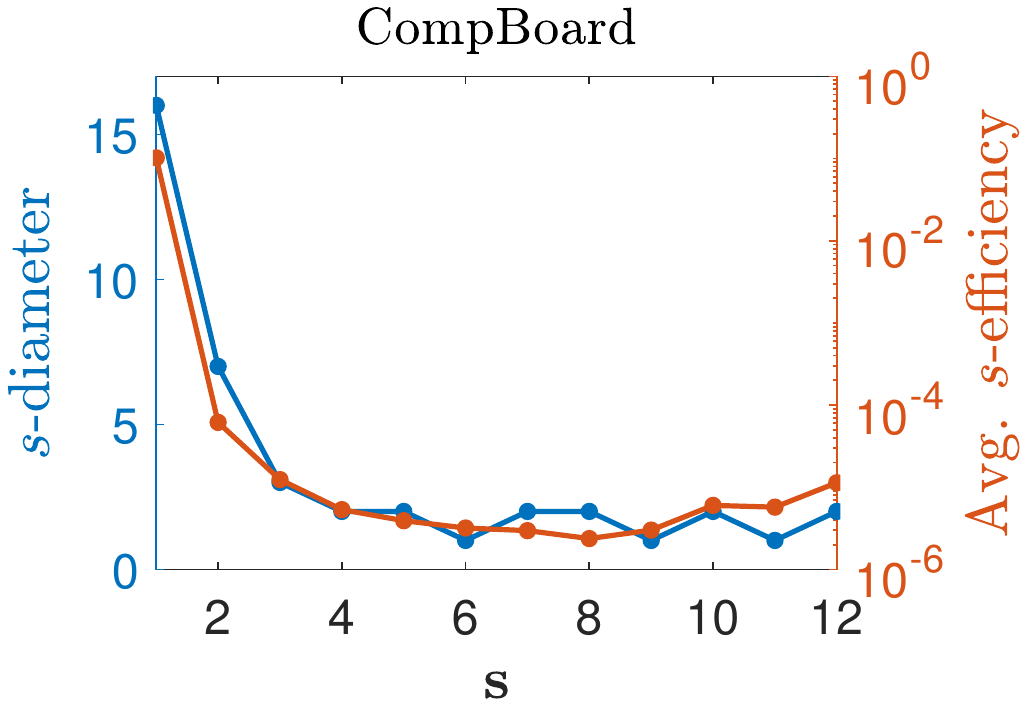} \\
\includegraphics[width = \linewidth]{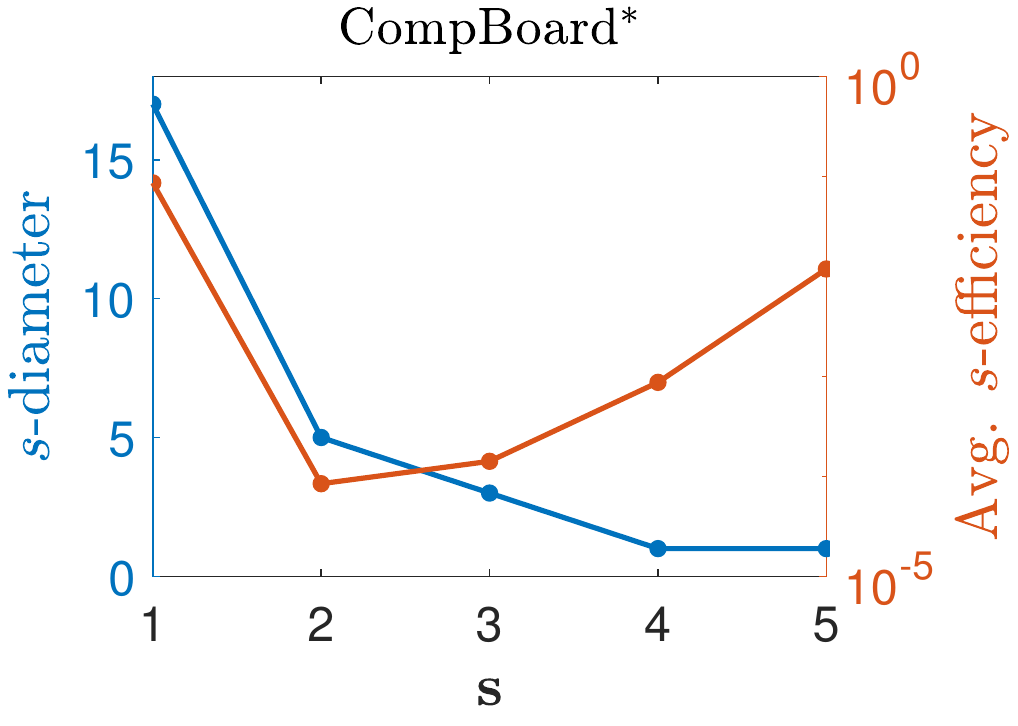} \\
\end{subfigure}
\vspace{-5mm}
\caption{Maximum $s$-diameter over all $s$-components and average $s$-efficiency for LesMis, Diseasome, and CompBoard (top row) and their dual hypergraphs (bottom row). Note the $s$-diameter values are in linear scale whereas the $s$-efficiency values are in logarithmic scale.} \label{fig:dist}
\end{figure}

We compute
three of the aforementioned $s$-distance-based measures: the average
$s$-efficiency index, the $s$-diameter (i.e. the maximum $s$-eccentricity
over all $s$-components) and the harmonic $s$-closeness centrality index.

\begin{figure}[t]
\centering
\scalebox{0.55}{
\setlength{\tabcolsep}{4pt}
\begin{tabular}{l @{\quad} l l l}
\toprule
& \multicolumn{3}{l}{\!\!CompBoard} \\
\cmidrule(l{-4pt}r{18pt}){2-4}
\small{Rank}  & $s=1$ & $s=2$ & $s=3$  \\ 
\midrule
1 & TGT & QRTEB & LBTYK \\ \cline{4-4}
2 & LOW & LSXMK & \multicolumn{1}{|l|}{NUW}  \\ 
3 & MMM & LBTYK & \multicolumn{1}{|l|}{NUM} \\
4 & MDLZ & LBRDK & \multicolumn{1}{|l|}{JRS}  \\
5 & AVY & SIRI & \multicolumn{1}{|l|}{NZF}  \\
6 & DWDP & GLIBP & \multicolumn{1}{|l|}{NIM} \\ \cline{4-4}
7 & CAH & LTRPB & LBRDK  \\ \cline{4-4}
8 & PYPL & ZG &  \multicolumn{1}{|l|}{LSXMK} \\ 
9 & DE & DISCK &  \multicolumn{1}{|l|}{QRTEB} \\ \cline{4-4} \cline{4-4}
10 & TXN & LEXEB &  \multicolumn{1}{|l|}{CRESY} \\ 
11 & R & SJR &  \multicolumn{1}{|l|}{LND}  \\
12 & UTX & TRIP &  \multicolumn{1}{|l|}{IRCP} \\ 
13 & UPS & EXPE &  \multicolumn{1}{|l|}{IRS}  \\ \cline{4-4} \cline{3-3}
14 & GWW &  \multicolumn{1}{|l|}{P} & DISCK  \\ 
15 & CSX & \multicolumn{1}{|l|}{CHTR} & DMF  \\ \cline{3-3}
\bottomrule
\end{tabular}
}
\label{tab:compBoard}
\quad
\scalebox{0.55}{
\setlength{\tabcolsep}{4pt}
\begin{tabular}{l @{\quad} l l l}
\toprule
& \multicolumn{3}{l}{\!\!Diseasome} \\
\cmidrule(l{-4pt}r{18pt}){2-4}
\small{Rank} & $s=1$ & $s=2$ & $s=3$  \\ 
\midrule
1 & Colon cancer & Breast cancer & Colon cancer \\
2 & Diabetes mellitus & Colon cancer & Breast cancer  \\
3 & Breast cancer & Ovarian cancer & Ovarian cancer \\ \cline{4-4}
4 & Glioblastoma & Lymphoma & \multicolumn{1}{|l|}{Turcot syndrome}  \\
5 & Leukemia & Gastric cancer & \multicolumn{1}{|l|}{Lymphoma}  \\
6 & Hepatic adenoma & Pancreatic cancer & \multicolumn{1}{|l|}{Hepatic adenoma} \\  \cline{3-3} \cline{4-4} \cline{4-4}
7 & Gastric cancer & \multicolumn{1}{|l|}{Li-Fraumeni synd.} & \multicolumn{1}{|l|}{Pancreatic cancer}  \\
8 & Lipodystrophy & \multicolumn{1}{|l|}{Osteosarcoma} & \multicolumn{1}{|l|}{Prostate cancer} \\  \cline{3-3}  \cline{3-3} \cline{4-4} \cline{4-4}
9 & Pancreatic cancer & \multicolumn{1}{|l|}{Adenomas} & \multicolumn{1}{|l|}{Cone dystrophy} \\
10 & Ovarian cancer & \multicolumn{1}{|l|}{Cafe-au-lait spots} & \multicolumn{1}{|l|}{Retinitis pigmentosa} \\
11 & Thyroid carcinoma & \multicolumn{1}{|l|}{Muir-Torre synd.} & \multicolumn{1}{|l|}{Cardiomyopathy}  \\  \cline{3-3}  \cline{3-3}
12 & Cardiomyopathy & \multicolumn{1}{|l|}{Prostate cancer} & \multicolumn{1}{|l|}{LCA disease} \\
13 & Neurofibromatosis & \multicolumn{1}{|l|}{Fanconi anemia} & \multicolumn{1}{|l|}{Charcot-Marie-Tooth}  \\
14 & Prostate cancer & \multicolumn{1}{|l|}{Lung cancer} & \multicolumn{1}{|l|}{Dejerine-Sottas synd.}  \\  \cline{3-3}
15 & Lymphoma & Turcot syndrome & \multicolumn{1}{|l|}{Neuropathy}  \\ \cline{4-4}
\bottomrule
\end{tabular}
}
\label{tab:diseasome}
\quad
\scalebox{0.55}{
\setlength{\tabcolsep}{4pt}
\begin{tabular}{l @{\quad} l l l}
\toprule
& \multicolumn{3}{l}{\!\!LesMis$^*$} \\
\cmidrule(l{-4pt}r{18pt}){2-4}
\small{Rank}  & $s=1$ & $s=2$ & $s=3$ \\ 
\midrule 
1 & Jean Valjean & Jean Valjean & Jean Valjean \\ 
2 &  Gavroche & Marius & Enjolras \\
3 & Marius & Enjolras & Marius \\
4 & Javert & Fantine & Fantine \\
5 & M. Th\'{e}nardier & M. Th\'{e}nardier & Javert \\
6 & Enjolras & Javert & M. Th\'{e}nardier \\ \cline{2-2} \cline{4-4}
7 & \multicolumn{1}{|l|}{Lesgle} & Mme Th\'{e}nardier& \multicolumn{1}{|l|}{Lesgle} \\
8 & \multicolumn{1}{|l|}{Fantine} & Courfeyrac & \multicolumn{1}{|l|}{Courfeyrac} \\ \cline{2-2} \cline{4-4} \cline{4-4}
9 & Cosette & Gavroche & \multicolumn{1}{|l|}{Combeferre} \\
10 & Mme. Th\'{e}nardier & Cosette & \multicolumn{1}{|l|}{Cosette} \\ \cline{2-2} \cline{4-4}
11 & \multicolumn{1}{|l|}{Babet} & Combeferre & Gavroche \\ \cline{3-3}
12 & \multicolumn{1}{|l|}{Gueulemer} & \multicolumn{1}{|l|}{Lesgle} & Mme Th\'{e}nardier \\ \cline{2-2} \cline{4-4}
13 & Claquesous & \multicolumn{1}{|l|}{Joly} & \multicolumn{1}{|l|}{Bahorel} \\ \cline{3-3} 
14 & Montparnasse & Mlle Gillenormand & \multicolumn{1}{|l|}{Joly} \\ \cline{4-4}
15 & Bishop Myriel & M. Gillenormand & Feuilly \\
\bottomrule
\end{tabular}
}
\label{tab:lesMis}
\[
\includegraphics[scale=0.41]{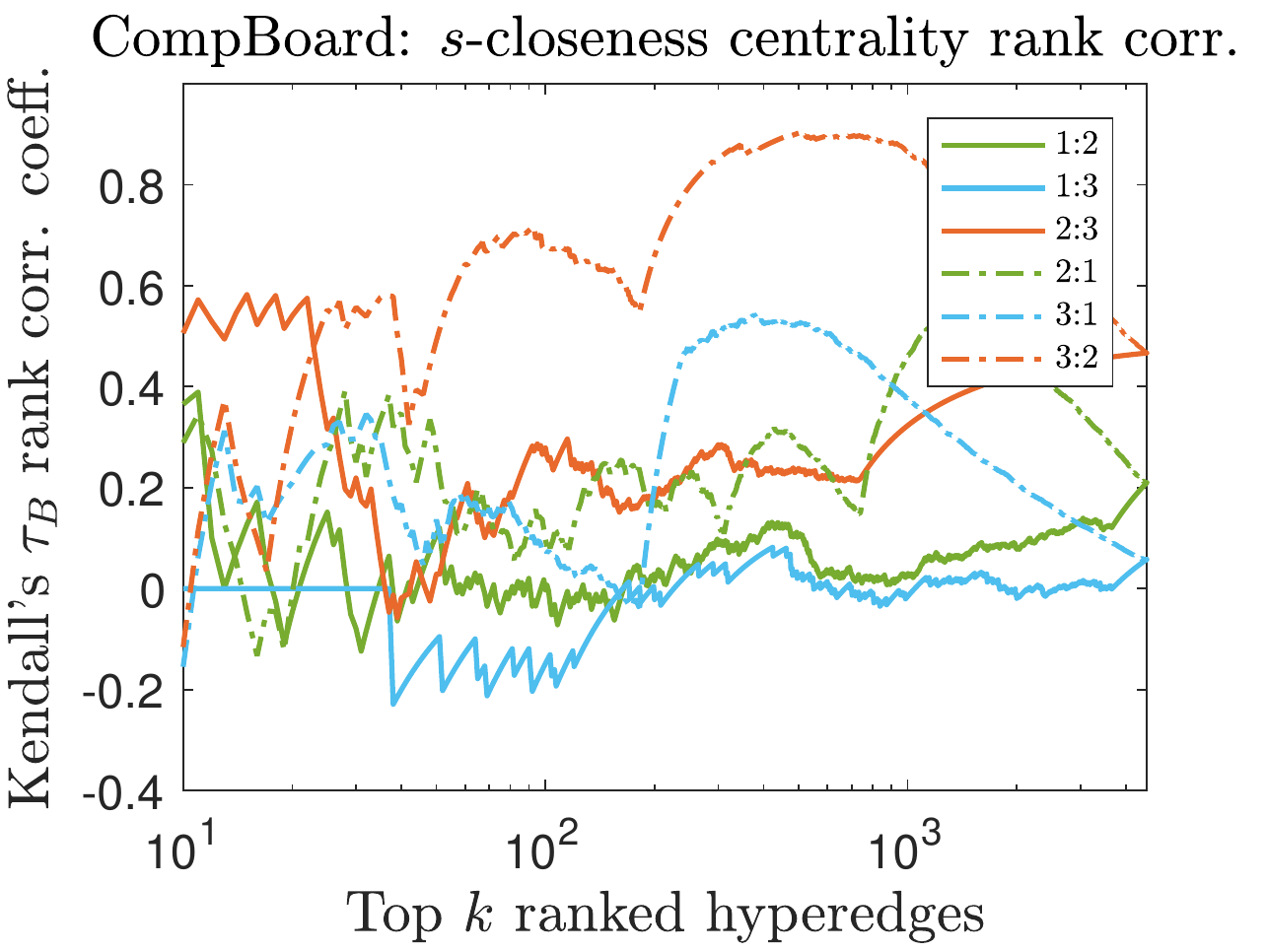}
\quad
\includegraphics[scale=0.41]{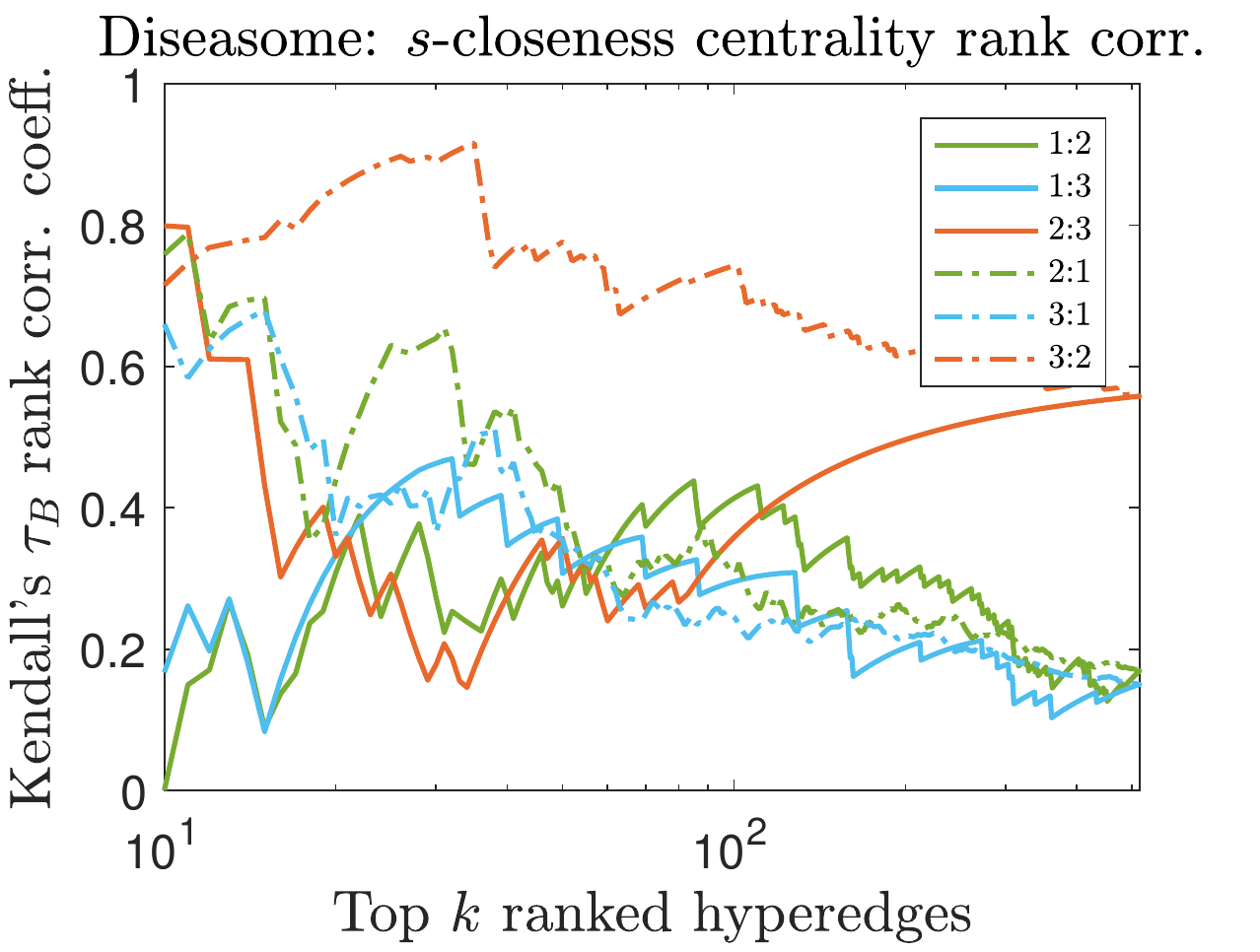}
\quad
\includegraphics[scale=0.41]{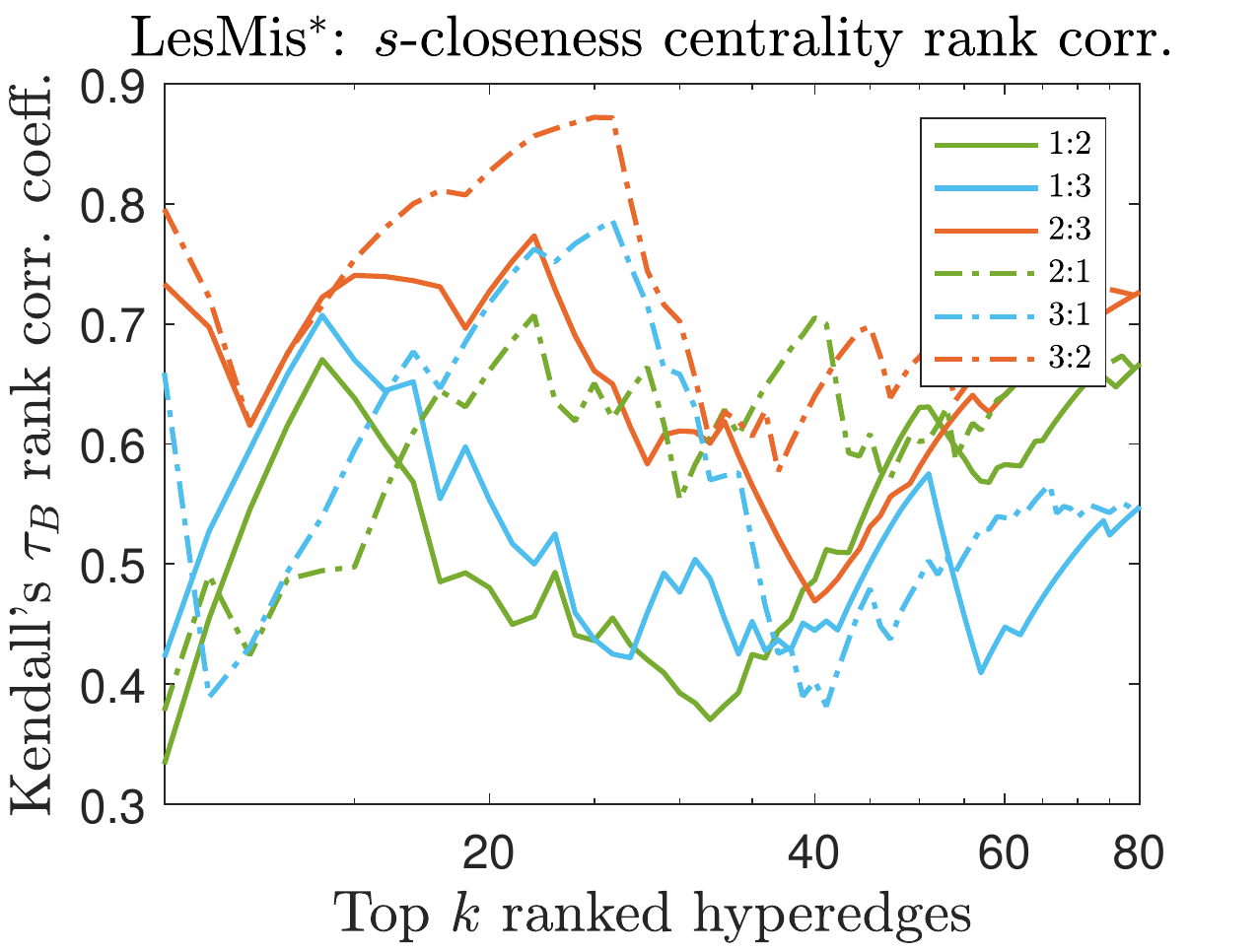}
\]
\caption{{\it Top row}: top 15 ranked hyperedges for $s=1,2,3$ in CompBoard, Diseasome, and LesMis$^*$. Boxes enclosing items in the table indicate those hyperedges are tied in rank.  {\it Bottom row}: Kendall's $\tau_B$ rank correlation coefficient between the top $k$ ranked hyperedges for a value of $s$ (listed first in the legend) compared against those hyperedge's ranking under a different value of $s$ (listed second).} \label{fig:sCen}
\end{figure}

Fig.~\ref{fig:dist} plots the maximum $s$-diameter (over all
$s$-components) and average $s$-efficiency for the hypergraph and its
dual. \textit{Larger}
values of average $s$-efficiency imply \textit{smaller} $s$-distances among
the hyperedges in question. For some of the data (e.g. LesMis,
Diseasome, Diseasome$^*$) average $s$-efficiency and
$s$-diameter tends to decrease as $s$ increases. In these networks, the
shortest $s$-walks linking hyperedges tend to become longer (or infinite)
as $s$ is increased. However, for CompBoard$^*$, average $s$-efficiency \textit{increases} in $s$ for each
$s\geq 2$. This suggests that, among company board members who sit on multiple
boards, those who sit on more boards tend to (on average) be closer to
one another in $s$-distance. LesMis$^*$ exhibits a similar phenomena regarding average
$s$-efficiency, where for characters appearing
in at least $s\geq 22$ scenes, the more scenes they appear in, the closer
they are to each other in $s$-distance.

Turning to $s$-diameter, it is possible for
$s$-diameter (taken as the maximum over all $s$-components) to increase
or decrease in~$s$. In the former scenario, as $s$ is increased, shorter
$s$-walks linking hyperedges may disappear, and those edges may only be
linked via longer $s$-walks, thereby increasing $s$-diameter. LesMis exhibits this most prominently, where $s$-diameter increases from 5 to
9 as $s$ increases from 1 to~2. On the other hand, if increasing $s$ eliminates
\textit{all} $s$-walks between pairs of hyperedges, then these hyperedges
are separated into different $s$-components in which hyperedges may be
closer to each other. In such cases, the $s$-diameter may decrease,
as in Diseasome. Consistent with our intuition from the Diseasome visualization
in Fig.~\ref{fig:sCompViz2}, this $s$-diameter drop reflects the fragmentation
of the network into small components; accordingly average $s$-efficiency
also drops because of the infinite $s$-distances between edges in different
$s$-components.

Lastly, Fig.~\ref{fig:sCen} (top row) lists the top 15 hyperedges
in CompBoard, Diseasome, and LesMis$^*$ for $s=1,2,3$, as
ranked according to their harmonic $s$-closeness centrality. Boxes enclosing
hyperedges indicate a tie in $s$-closeness centrality. Comparing the ordinal
rankings across the datasets, for some data the
the top 15 ranked hyperedges for $s=1$ remain within the top ranked for
$s=2$ (e.g. for LesMis$^*$, 10 remain in the top 15) whereas in other
data, the top ranked hyperedges may change completely (e.g. in CompBoard,
\textit{none} of the top 15 companies with highest $1$-closeness centrality
remain in the top 15 for $2$-closeness centrality).

A~drop in a hyperedge's rank from $s=1$ to 2 may indicate short pathways
linking that hyperedge to others rely on sparse hyperedge intersections.
For example, in Diseasome we observe that while
``Colon cancer'' and ``Breast cancer'' remain in the top 3 ranked hyperedges
for $s=1,2,3$, ``Diabetes mellitus'' drops from having the second largest
$1$-centrality, to having the 34th largest $2$-centrality. ``Diabetes mellitus''
shares genes with 24 other diseases and hence this hyperedge intersects
with 24 other hyperedges. However, of these 24 diseases, ``Diabetes mellitus''
shares at least 2 genes with only two diseases: ``Obesity'' and ``Mature
Onset Diabetes of the Young (MODY)''. Thus, any $2$-walk between ``Diabetes
mellitus'' and another disease can only go through one of these diseases,
which (in this case) results in larger average $2$-distance between diabetes
and other diseases, relative to the average 2-distance between other pairs
of diseases. In contrast, ``Breast cancer'' shares at least 2 genes with
9 other diseases, and (on average) can be linked to other diseases via
a shorter 2-walk than for ``Diabetes mellitus''.

To more rigorously explore these changes in ordinal rankings
by $s$-closeness, we compute Kendall's $\tau _{B}$ rank correlation coefficient
between the top $k$ ranked hyperedges for one value of $s$ and the rankings
of those same hyperedges under another value of~$s$. We compute this coefficient
for each of $k=10,\ldots,  \llvert  E \rrvert  $ and for each ordered pair of $s$-values
from $\{1,2,3\}$. Hyperedges with equal $s$-closeness centrality are considered
tied in rank, and we assign the minimum $s$-closeness centrality score
of 0 to any hyperedge with fewer than $s$ vertices. Kendall's $\tau _{B}$ ranges from $-$1 (if the ordinal rankings
are perfectly inverted) to 1 (if the ordinal rankings are identical), and is explicitly formulated to handle ties
in rank \cite{Agresti2012}. Fig.~\ref{fig:sCen} plots results for CompBoard, Diseasome, and LesMis$^*$.
CompBoard exhibits an absence of correlation for the $1$-closeness rankings
when compared against 2 or 3, and a stronger correlation for the $3$-closeness
rankings compared against the $2$-closeness rankings. When all hyperedges
in the network are considered (i.e. for $k= \llvert  E \rrvert  $, given by the rightmost
points in each plot), the $1$-closeness rankings of LesMis$^*$ exhibits
the strongest correlations between the 2 and 3-closeness rankings.

\subsection{Paths, Cycles, and Clustering Coefficients} \label{sec:sPaths}

\paragraph*{Methods} So far, our methods have centered solely around the base
definition of $s$-walk. However, just as graph walks may be distinguished
into finer classes such as trails, paths, circuits and cycles, $s$-walks
may also be distinguished from each other and organized hierarchically. As we'll show, doing so allows one to define high-order
substructures native to hypergraphs, such as $s$-triangles, that cannot
be determined from their $s$-line graphs.

\begin{definition}
For a hypergraph $H=(V,E)$, let the sequence of hyperedges  ${\omega}=(e_{i_0},e_{i_1},\dots,e_{i_k})$ be an $s$-walk of length $k$. For ease of notation let $I_j=e_{i_{j-1}} \cap e_{i_j}$ be the $j$'th intersection. The $s$-walk $\omega$ may be further defined as:

\begin{enumerate}
\item[$(i)$] An {\bf $s$-trace} if ${i_x} \not = {i_{y}}$ for all $x\not=y$ (all hyperedges are pairwise distinct by label).
\item[$(ii)$] An {\bf $s$-meander} if $\omega$ is an $s$-trace in which $I_x \not= I_{y}$ for all $x\not=y$ (all intersections are pairwise distinct).
\item[$(iii)$]  An {\bf $s$-path} if $\omega$ is an $s$-meander in which $I_x \setminus I_{y} \not= \varnothing$ for all $x\not=y$ (no intersection is included in another). \label{spath}
\end{enumerate} 
\end{definition}

\noindent {\bf Graph case \& equivalence}:
If $H$ is a graph, a
$1$-trace on $H^{*}$ is equivalent to a walk on $H$ in which vertices are
distinct but edges may be repeated. Furthermore, if $H$ is a graph,
$s$-meanders and $s$-paths on $H^{*}$ are both equivalent
to a graph path on~$H$. However, if $H$ is a hypergraph, a path in
$L_{s}(H)$ does not necessarily correspond to an $s$-path in~$H$. Consequently,
the forthcoming $s$-path based triadic notions in Definition~\ref{def:sTriadic} cannot be obtained from $L_{s}(H)$ but reduce to their
usual graph counterparts on $H^{*}$ for $s=1$ whenever $H$ is a graph. \\

We note Wang and Lee \cite{wang1999paths} also define hypergraph paths
using the same subset condition stated in Definition 10 above. The notions of $s$-walk, $s$-trace, $s$-meander, and $s$-path form a nested
hierarchy: every $s$-trace, $s$-meander, or $s$-path is an $s$-walk; every
$s$-meander and $s$-path is an $s$-trace; and every $s$-path is an
$s$-meander. However, in each case, the reverse may not be true (e.g. an
$s$-meander may not be an $s$-path). With regard to $s$-distance (Sect.~\ref{sec:sDist}), it is straightforward to show constructively that if
there exists an $s$-walk (resp. $s$-trace, $s$-meander) of length
$k$ between two hyperedges, there exists an $s$-trace (resp. $s$-meander,
$s$-path) of length \textit{at most}~$k$. This implies the length of the shortest
$s$-walk between two hyperedges is equivalent to the length of the shortest
$s$-path; consequently, $s$-distance as given by the length of the shortest
$s$-walk is equivalent to that given by the length of the shortest
$s$-path.

While not having ramifications for the notion of $s$-distance, the finer
classes of $s$-walks above provide a means, within the $s$-walk framework,
to define high-order substructures or motifs that cannot be determined
from the $s$-line graph. To define an example of these substructures, we
require the notion of a \textit{closed} walk. Analogous to its usage in graph
theory, we call an $s$-walk \textit{closed} if ${i_{0}}={i_{k}}$, and call
a closed $s$-path an \textit{$s$-cycle}. As a point of clarification, closed
$s$-traces, meanders, or paths are still considered valid $s$-traces, meanders
or paths (that is, only the terminal edges are exempt from the $s$-trace
requirement that all edges be distinct by label). Using $s$-cycles, we
define hypergraph $s$-analogs of triadic measures commonly applied to graph
data. Whereas graph triadic notions like the local clustering
coefficient \cite{Watts1998} are defined for \textit{vertices}, the $s$-analogs
below are defined for \textit{hyperedges}, keeping consistent with the rest
of our presentation. We remind the reader vertex-based notions are
obtained by simply applying the below definition to the dual hypergraph,
$H^{*}$.

\begin{definition} \label{def:sTriadic}
 For a hypergraph $H$, an {\bf $s$-triangle} is a closed $s$-path of length $3$ and an {\bf $s$-wedge} is an $s$-path of length 2. For an $s$-wedge $e_0,f,e_2$, we say $f$ is the center of the $s$-wedge. 
\begin{itemize}

\item[$(i)$] The {\bf $s$-local clustering coefficient} of a hyperedge $f \in E_s$ is given by
\[
s\mbox{-LCC}(f)= \begin{cases} \frac{\mbox{number of $s$-triangles containing $f$}}{\mbox{number of $s$-wedges centered at $f$}} & \mbox{if $f$ is the center of an $s$-wedge} \\ 0 & \mbox{otherwise}. \end{cases}
\]  
\item[$(ii)$] The {\bf $s$-global clustering coefficient} of a hypergraph $H$ is given by
\[
s\mbox{-GCC}(H)=\frac{3 \cdot \mbox{total number of $s$-triangles }}{\mbox{total number of $s$-wedges}}.
\]
\end{itemize}
\end{definition}

In the same way as for the LCC of graphs, one may obtain a global measure
for the $s$-LCC of a hypergraph by taking the mean $s$-local clustering
coefficient over all edges in $E_{s}$.

\begin{figure*}[t]
    \centering
    \begin{subfigure}[t]{0.33\textwidth}
        \centering
        \includegraphics[height=1in]{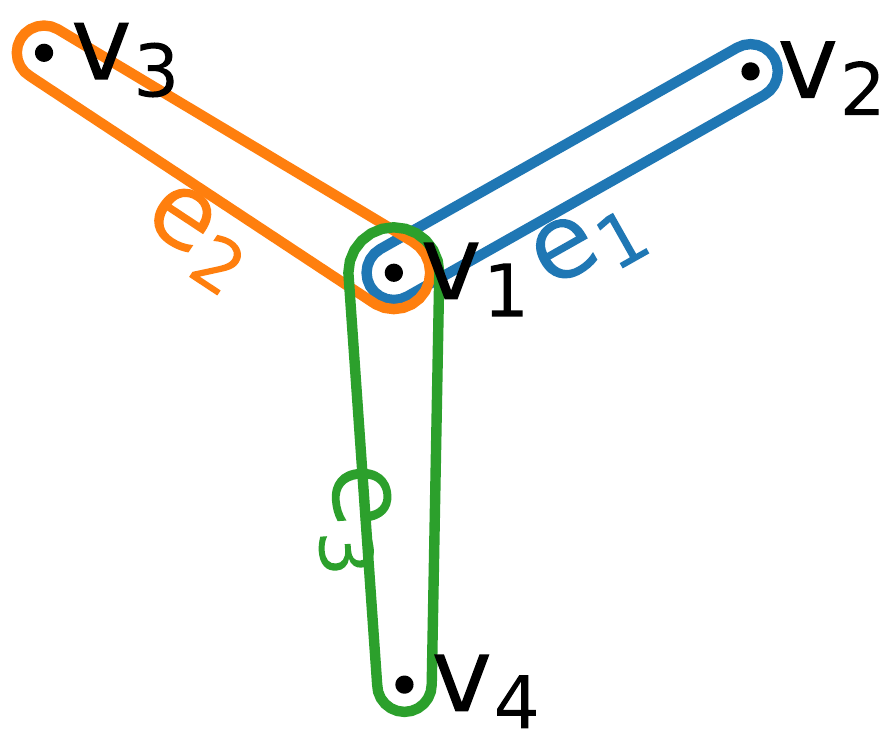}
        \caption{closed $1$-trace of length 3} \label{subfig:trace}
    \end{subfigure}%
    ~ 
    \begin{subfigure}[t]{0.33\textwidth}
        \centering
        \includegraphics[height=1in]{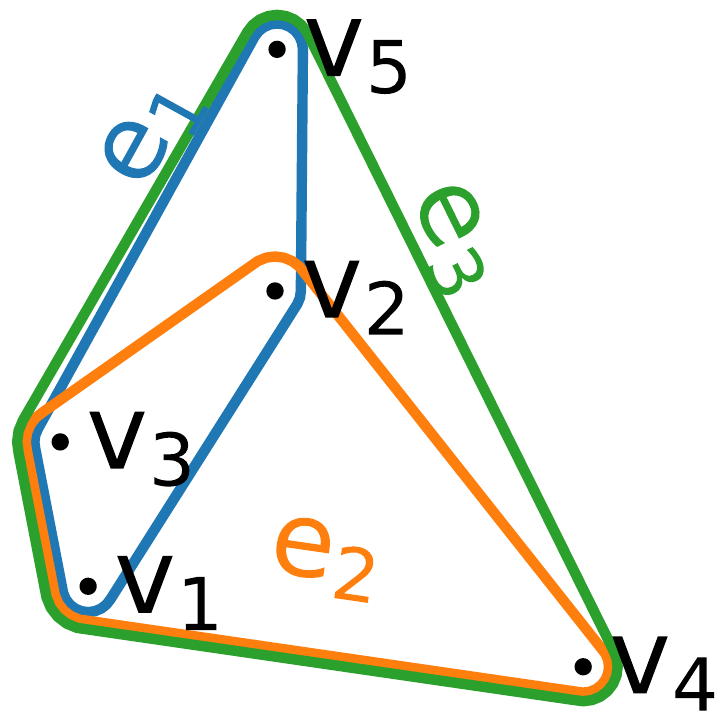}
        \caption{closed $3$-meander of length 3} \label{subfig:meander}
    \end{subfigure}%
    ~
     \begin{subfigure}[t]{0.33\textwidth}
        \centering
        \includegraphics[height=1in]{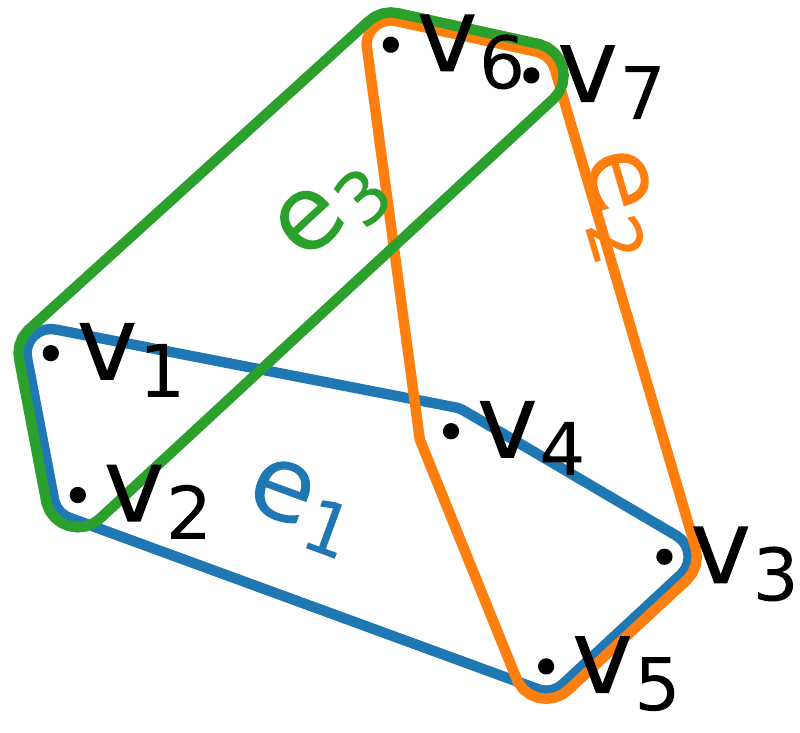}
        \caption{closed $2$-path of length 3 \\  \centering ($2$-triangle)} \label{subfig:tri}
    \end{subfigure}%
    \caption{Examples of different closed $s$-walks of length 3 given by \textcolor{myBlue}{$e_1$},\textcolor{myOrange}{$e_2$},\textcolor{myGreen}{$e_3$},\textcolor{myBlue}{$e_1$}.}\label{fig:cycle}
\end{figure*}

Fig.~\ref{fig:cycle} illustrates examples of three different hypergraphs
induced by a closed $s$-walk $e_{1}$, $e_{2}$, $e_{3}$, $e_{1}$ of length 3; namely,
from left to right: a closed $s$-trace that is not an $s$-meander, a closed
$s$-meander that is not an $s$-path, and a closed $s$-path of length 3
(i.e. an $s$-triangle). Observe the 1-line graphs of all three of
these hypergraphs consists of a single triangle, while only the rightmost
pictured hypergraph in Fig.~\ref{fig:cycle} is a 1-triangle (as well
as a 2-triangle).
Another example may be found by reconsidering the author-paper
networks in Fig.~\ref{fig:authPap}: for the hypergraphs constructed by
letting hyperedges denote authors, the $2$-walk $A$, $B$, $C$, $A$ is a 2-triangle
for the leftmost network, while the same walk is a 2-trace for the rightmost.
These examples illustrate $s$-triangles cannot be determined from
line graphs (a~fact that is unsurprising, since $s$-line graphs
do not encode the subset relationships stipulated for $s$-paths).

Since other definitions of hypergraph clustering coefficients
have appeared in the complex networks literature, it is worth clarifying
how these notions compare to ours. Estrada \cite{Estrada2006} proposes
a global hypergraph clustering coefficient as a ratio of (non $s$-walk
based) hypergraph triangles to hypergraph wedges. More precisely, Estrada
defines a hypertriangle as an alternating vertex-hyperedge sequence with
three distinct vertices and three distinct hyperedges such that for each
subsequence $v_{i}$, $e_{k}$, $v_{j}$, we have that
$v_{i},v_{j} \in e_{k}$ (put equivalently, these are 6-cycles in the bipartite
representation of the hypergraph). Thus, returning to the rightmost hypergraph
pictured in Fig.~\ref{fig:cycle}, the alternating sequence given by interlacing
the pair of vertex and hyperedge triples $(v_{1},v_{3},v_{6})$ and
$(e_{1},e_{2},e_{3})$ constitutes a triangle, as does the same pair with
$v_{1}$ replaced with $v_{2}$. It is easy to see the existence
of an $s$-triangle implies the existence of at least one such hypertriangle
as defined by Estrada; however, the converse is not necessarily true (e.g. while
the hypergraph pictured in the center of Fig.~\ref{fig:cycle} contains
many such hypertriangles, neither this hypergraph nor its dual contain
any $s$-triangles). In this sense, Estrada's notion of clustering and ours
are fundamentally different.

Other proposed notions of hypergraph clustering differ to ours in being
based on averaging various \textit{pairwise} set theoretic measures between
pairs of hyperedges or vertices. For instance, Latapy, Magnien and Vecchio
\cite{Latapy2008} propose a pairwise clustering coefficient between hyperedges
$e_{i}$, $e_{j}$ as $\frac{ \llvert  e_{i} \cap e_{j} \rrvert  }{ \llvert  e_{i} \cup e_{j} \rrvert  }$, which is the Jaccard similarity coefficient between the sets of vertices
constituting the two hyperedges (or, when applied to the dual, the Jaccard
similarity between the sets of hyperedges to which two vertices belong).
They then define a local and global notion of hypergraph clustering by
averaging this quantity. Zhou and Nakhleh \cite{Zhou2011} propose local
and global hypergraph clustering coefficients based on the pairwise \textit{excess
overlap} between hyperedges. As described and studied further by the authors
in \cite{Dewar2018}, excess overlap measures the proportion of the vertices
in exactly one of the edges that are neighbors of vertices in only the
other edge. Lastly, notions of bipartite graph clustering proposed in
the literature, (applicable to hypergraphs via the bicolored graph-hypergraph correspondence
mentioned in Sect.~\ref{sec:Prelim}) are frequently based on bipartite 4-cycles
\cite{Aksoy2017,Robins2004}. In the language of hypergraphs, a bipartite
4-cycle is a subhypergraph on two hyperedges and two vertices. Hence (in
addition to again not being based in high-order $s$-walks) these bipartite
4-cycle based notions of clustering differ from our $s$-triangle based
notions in involving only pairs (rather than triples) of hyperedges.

\begin{figure}[t]
\centering
\begin{subfigure}[b]{0.49\textwidth}
\includegraphics[width = 0.5\linewidth]{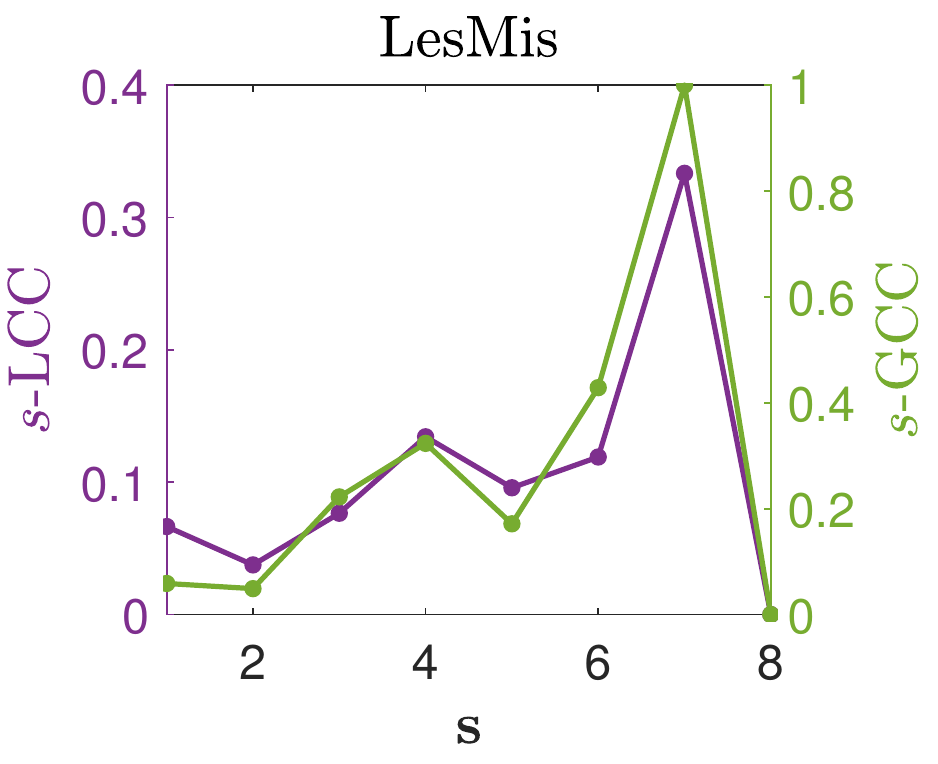}\includegraphics[width = 0.5\linewidth]{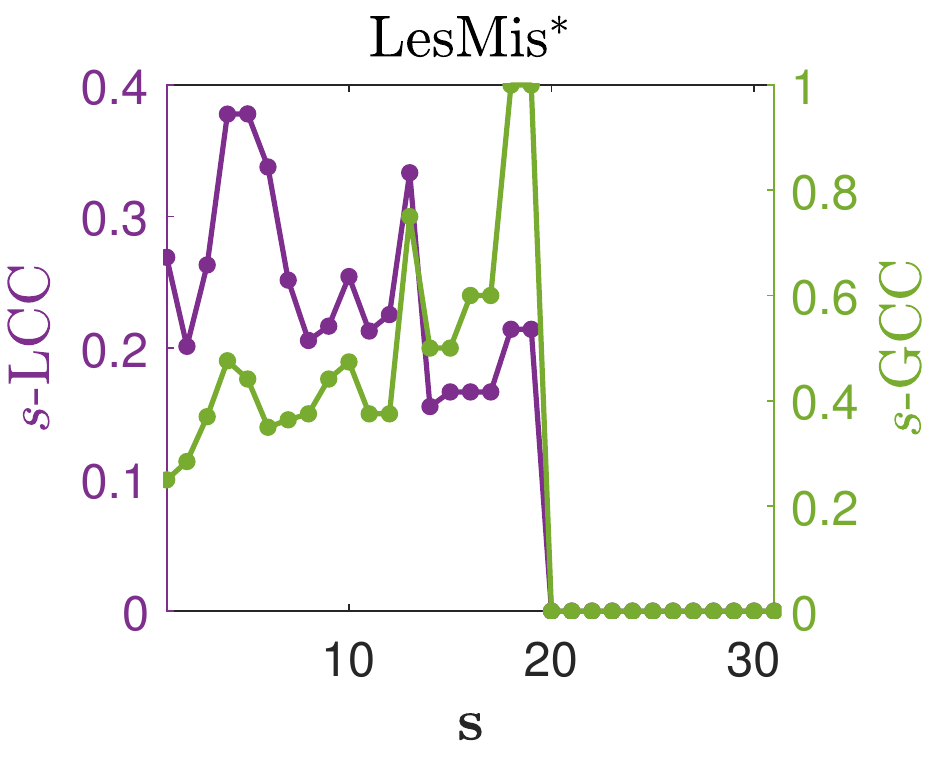} \\
\includegraphics[width = 0.5\linewidth]{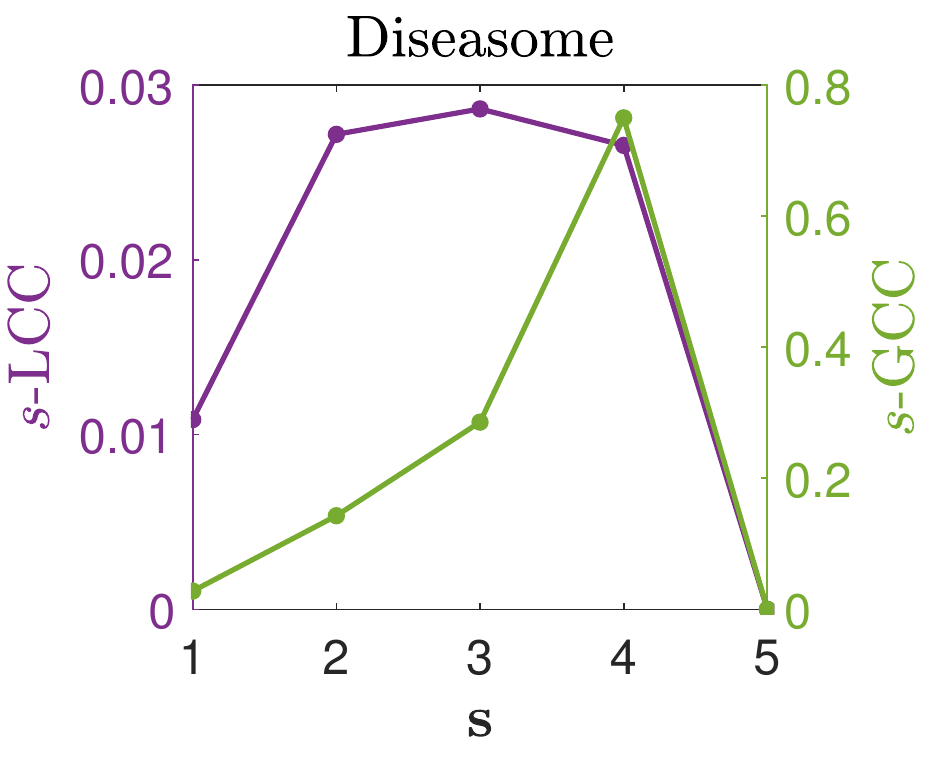}\includegraphics[width = 0.5\linewidth]{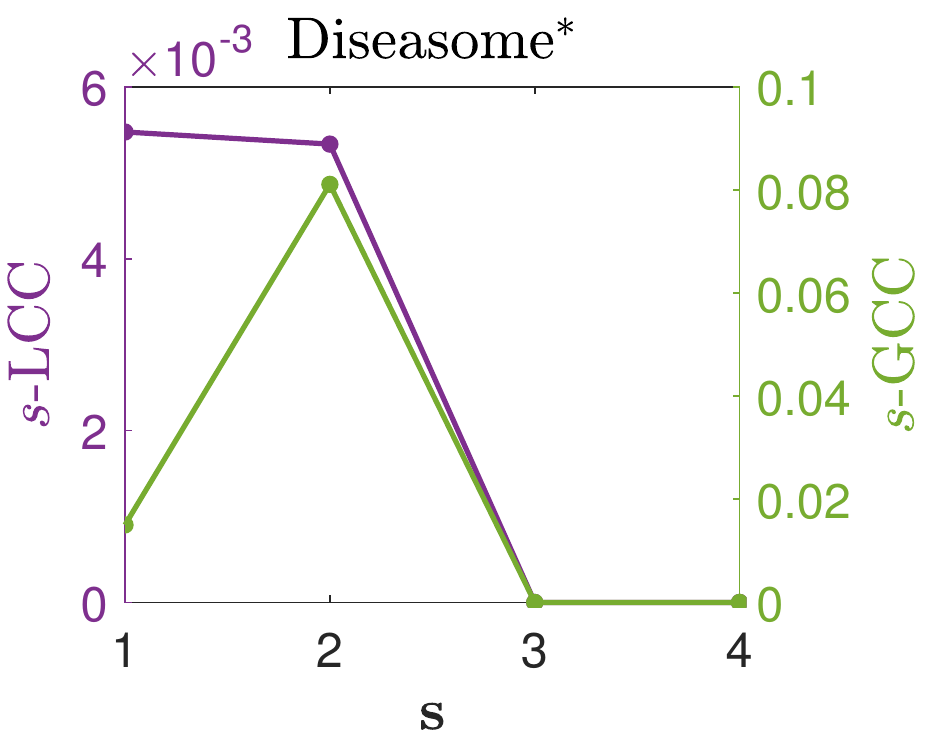} \\
\includegraphics[width = 0.5\linewidth]{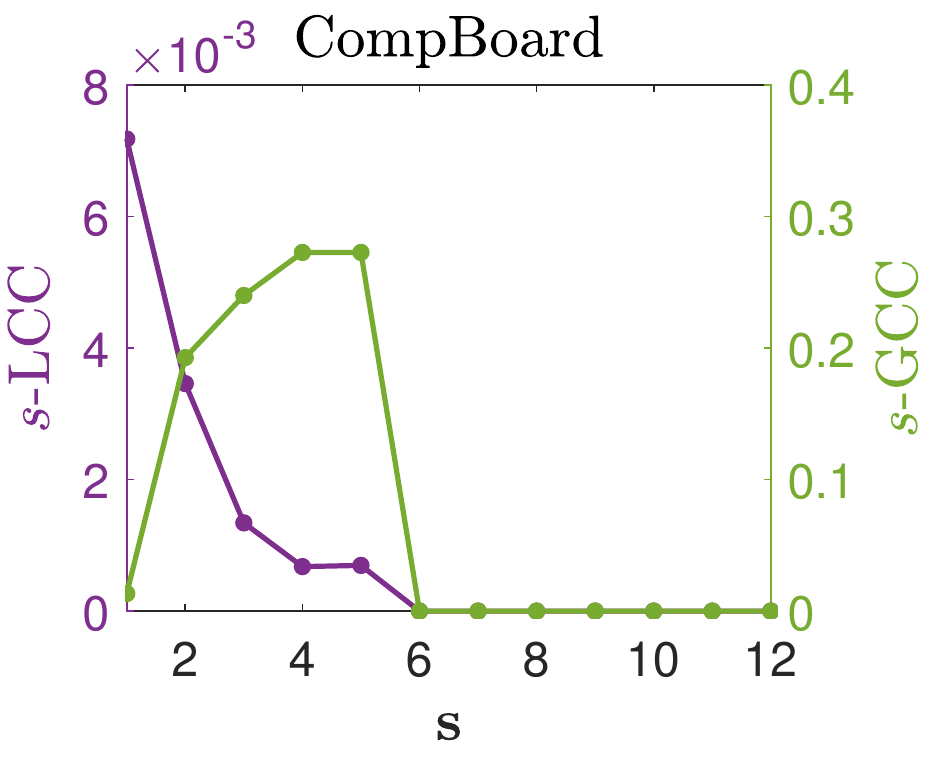}\includegraphics[width = 0.5\linewidth]{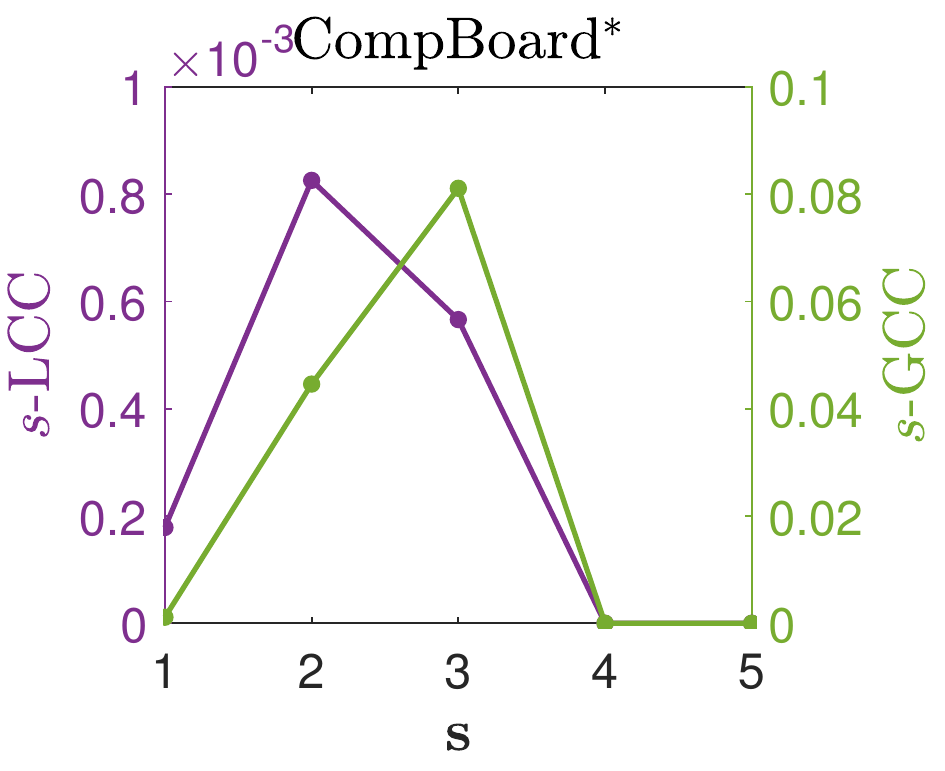} \\
\end{subfigure} 
\begin{subfigure}[b]{0.49\textwidth}
\includegraphics[width = 0.5\linewidth]{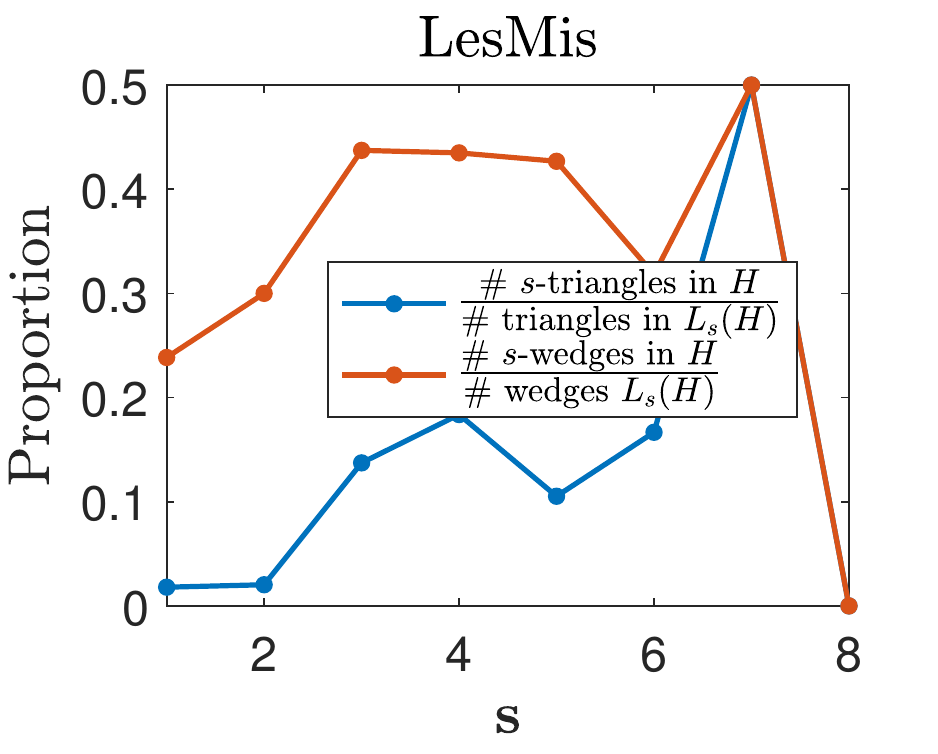}\includegraphics[width = 0.5\linewidth]{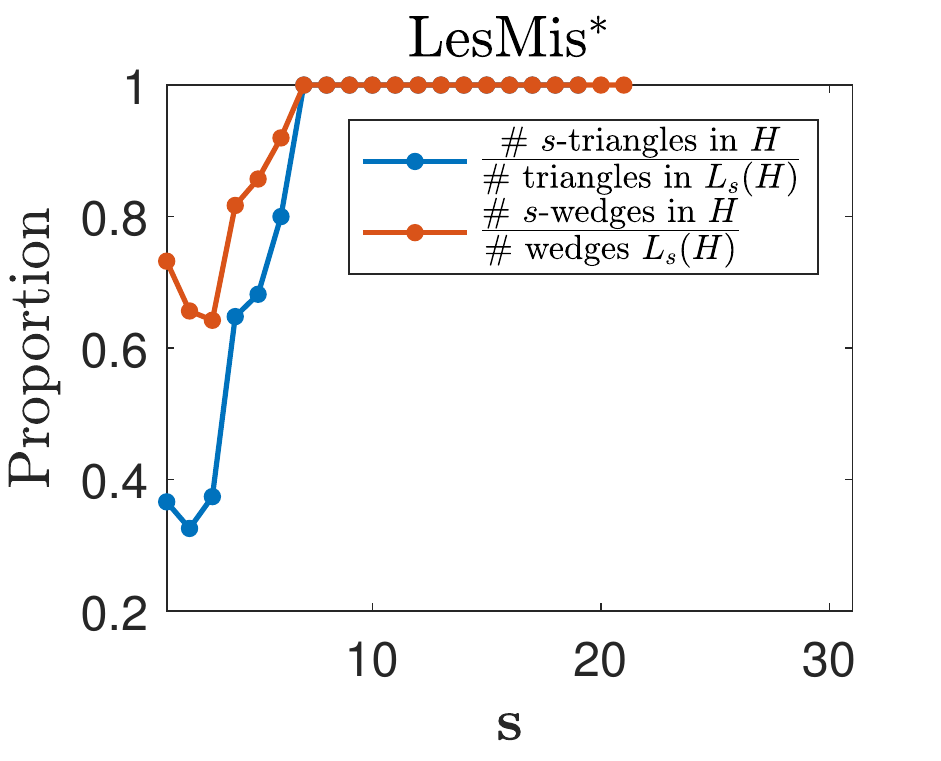} \\
\includegraphics[width = 0.5\linewidth]{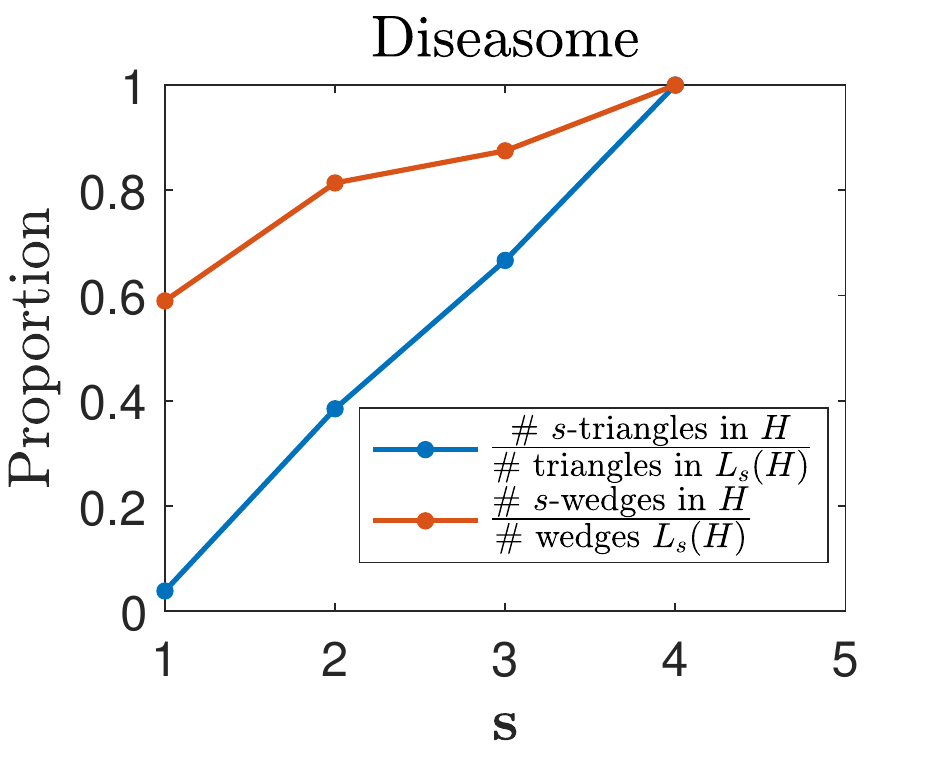}\includegraphics[width = 0.5\linewidth]{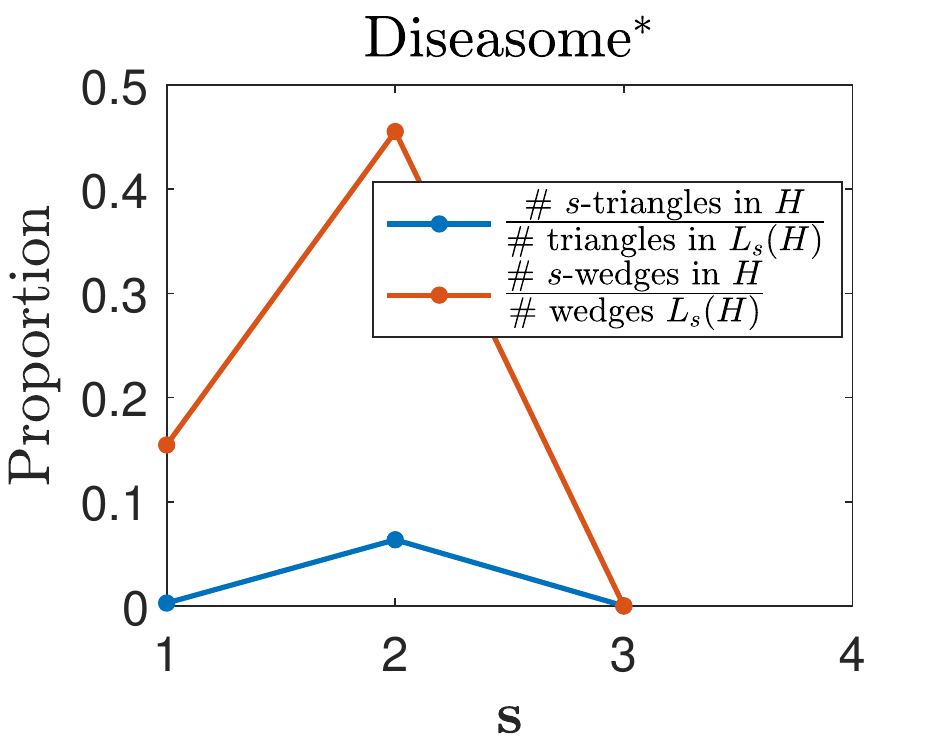} \\
\includegraphics[width = 0.5\linewidth]{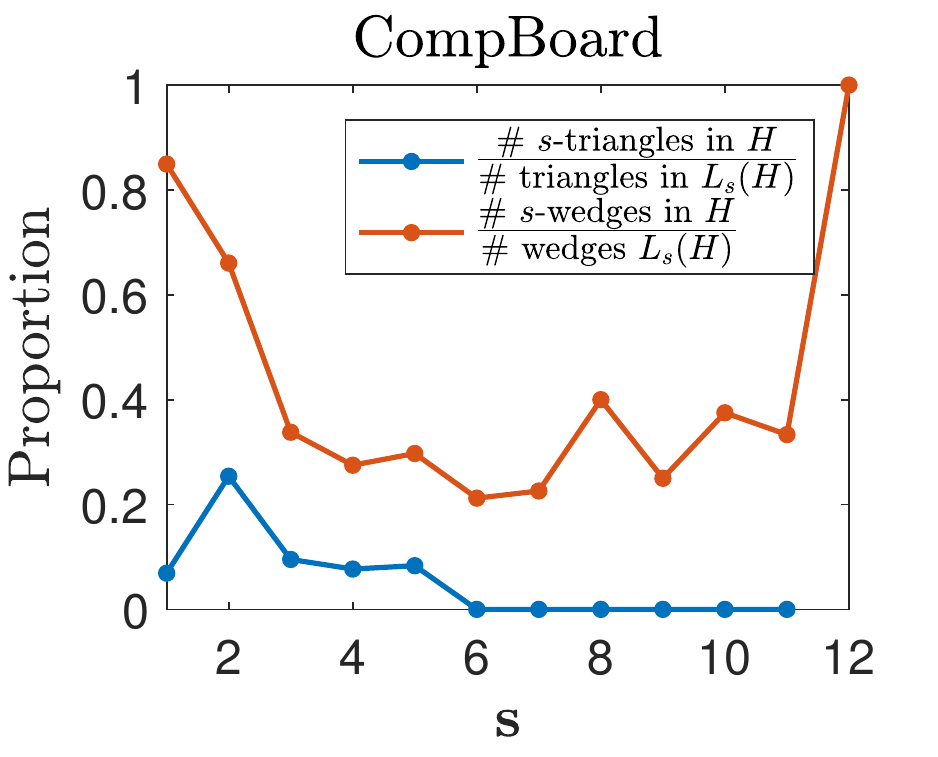}\includegraphics[width = 0.5\linewidth]{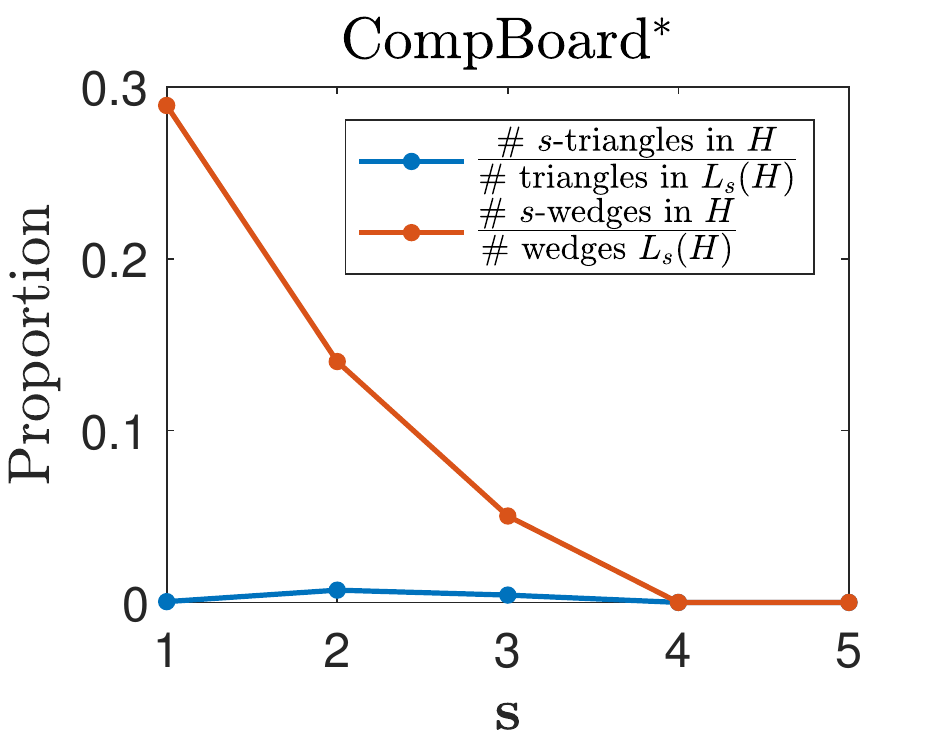} \\
\end{subfigure} 
\vspace{-5mm}
\caption{Mean local and global $s$-clustering coefficients of each hypergraph (left block), and proportion of triangles and wedges in the $s$-line graph that correspond to $s$-triangles and $s$-wedges for LesMis, Diseasome, and CompBoard, and their dual hypergraphs. } \label{fig:clus}
\end{figure}

\paragraph*{Application to data} Fig.~\ref{fig:clus} plots the mean
$s$-LCC and $s$-GCC (left block) as well as the proportion of triangles
and wedges in $s$-line graph that correspond to $s$-triangles and
$s$-wedges in the hypergraph (right block) for each of our datasets. Recall every triangle and wedge in the $s$-line graph represents a
closed $s$-walk of length 3 and $s$-trace of length 2 which, in turn,
may or may not be an $s$-triangle or $s$-wedge, respectively. For all three
datasets, a higher proportion of wedges in the $s$-line
graph correspond to $s$-wedges compared with the proportion of triangles
in $L_{s}(H)$ that correspond to $s$-triangles.

On the other hand, the datasets exhibit different behavior regarding the
absolute size of these proportions, as well as how these proportions vary
as $s$ varies. For LesMis$^*$, a relatively larger proportion of triangles
in $L_{s}$ correspond to $s$-triangles than for CompBoard$^*$. Furthermore,
the proportions of $s$-triangles to $s$-wedges are much greater, both on
average locally (given by the mean $s$-LCC) as well as globally (given
by the $s$-GCC) than for CompBoard$^*$. In contrast, CompBoard$^*$
exhibits an extremely small proportion of triangles in $L_{s}(H)$ corresponding
to $s$-triangles. This means whenever there is a triad of board members
where each pair belong to common company boards, it is almost
always the case that for at least one pair of board members, the set of
companies in common are either identical (i.e. forming $s$-trace that is
not an $s$-meander) or subsets of each other (i.e. an $s$-meander that
is not an $s$-path). In general, $s$-triangles in both CompBoard and its dual
are scarce, reflected by extremely low $s$-LCC and $s$-GCC
coefficients. 

In the context of a company-board
network an $s$-wedge is a generalization of a \textit{different represenatives'
interlock},\footnote{Defined in \cite{Axinn1984} as ``the linking of two
companies by a third company having different representatives on the board
of the two companies.''} a topic prominent in the corporate
governance literature \cite{Axinn1984,levine1979study,Robins2004}. Furthermore,
pairs of hyperedges having an $s$-distance of 1 and 2 represent so-called
``direct interlocks''\footnote{A~direct interlock occurs when two company
boards have 1 or more members in common. \cite{levine1979study}} and ``third
company interlocks'',\footnote{Defined in \cite{Axinn1984} as ``the linking
of two companies by one company having a director on the board of a second, \ldots which
has directors in common with a third company, \ldots which
in turn has directors on the board of a competitor of the first company.''}
respectively, between competing companies. This parallel illustrates how
$s$-path and $s$-distance based notions may provide a generalized framework
for describing and measuring phenomenon already important to particular
domains. For CompBoard, since the aforementioned interlocks
between competing companies are regulated by Section~8 of the Clayton Act
\cite{Axinn1984}, it is unsurprising our results show $s$-wedges (and hence $s$-triangles)
are relatively rare.

\section{Comparison with Generative Hypergraph Null Models}\label{sec:genModels}

Graph generation serves far-ranging purposes across scientific disciplines.
Generative graph models are used for benchmarking, algorithm
testing, and creating surrogate graphs to protect the anonymity of restricted
data. Here, we apply hypergraph generative models as \textit{null models}
to experimentally test the significance of the high-order properties explored
in Sect.~\ref{sec:sMetrics}. By ``null model'', we mean a generative model
that controls for certain basic features of the data. Such models may be
utilized to test whether observed measurements in the data are necessarily
consistent with controlled features. For example, in the Erd\H{o}s--R\'{e}nyi
graph model, the user specifies the desired number of vertices $n$ and
edge-probability $p$; hence, by controlling $n$ and $p$ one can generate
ensembles of random graphs with the same expected edge density. A~subsequent
comparison of measurements on given graph data against those on random
graphs with the same edge density tests whether the measured features can
be explained as sole consequences of edge density. To the extent to which
the properties of the real and synthetic graphs diverge, this provides
evidence the properties under question cannot be explained as sole
consequences of the structural properties preserved by the model.

In comparison to their graph counterparts, generative hypergraph models
are relatively few. 
While work on random uniform hypergraphs dates back to at least the 1970s, researchers have recently begun developing
a wider variety of hypergraph models, both for uniform hypergraphs
\cite{cooley2018subcritical,Cooley2015,Cooley2016ThresholdAH,Parczyk2015}
and non-uniform hypergraphs
\cite{chodrow2019configuration,Darling2005,Dewar2018,Ghoshal2009,kaminski2018clustering}.
We consider three generative hypergraph models from
\cite{Aksoy2017}, which can be thought of as hypergraph interpretations
of the graph models Erd\H{o}s--R\'{e}nyi (ER)
\cite{erdHos1960evolution}, Chung--Lu (CL) \cite{chung2006complex}, and
Block Two-Level Erd\H{o}s--R\'{e}nyi (BTER)
\cite{Kolda2014,Seshadhri2012}. These models were originally presented
as ``bipartite models'' in \cite{Aksoy2017}, with similar acknowledgment
of the bicolored graph-hypergraph correspondence discussed in Sect.~\ref{sec:Prelim}. While these models were inspired from their graph counterparts
and named as such, there may be multiple ways of conceiving these models
in the hypergraph/bicolored graph setting, as is often the case with graph-to-hypergraph
extensions. In fact, others have proposed non-uniform hypergraph analogs
of Erd\H{o}s--R\'{e}nyi and Chung--Lu (see \cite{Dewar2018} and
\cite{kaminski2018clustering}, respectively) differing to those considered
here with regard to the inputs required, the model itself, and the
definition of hypergraph assumed.

We've chosen these particular models for several reasons. First, they can generate non-uniform hypergraphs in accordance with the full
generality of Definition~\ref{def:hyp}. Notably, all three of these models
permit duplicated edges, which occur frequently
in hypergraph-structured data and are highly prevalent on our particular
data (see Fig.~\ref{fig:dd}).
One might expect duplicate edges to also
be common in author-paper networks, occurring whenever the same set of
authors write multiple papers together (e.g. papers $1$, $4$ for authors
$A$, $B$ in the leftmost network of Fig.~\ref{fig:authPap}). These joint
papers suggest stronger relationships amongst the authors in question;
disregarding duplicate edges ignores this and skews measurements, such
as $s$-centrality, meant to capture such properties. In contrast to the
models we consider, the aforementioned non-uniform hypergraph ER and CL
models proposed by \cite{Dewar2018} and \cite{kaminski2018clustering} do
not permit duplicated hyperedges, but treat hyperedges themselves as multisets.

Secondly, taken as a suite, these models provide
\textit{tiered} control over three fundamental properties: (1)~vertex-hyperedge
density, (2)~vertex degree and edge cardinality distributions, and
(3)~metamorphosis coefficients, a measure of community structure from
\cite{Aksoy2017} which we return to shortly. Specifically,
ER controls for vertex-hyperedge density, CL controls for density as well
as degree distributions, and BTER controls for all three of the aforementioned
properties. Taken in sequence, ER, CL and BTER can each be conceived
formally as a generalization of the previous model. All three
models afford scalable implementations and
\cite{Aksoy2017,jenkins2018chapel} report results on hypergraphs generated
using these models with hundreds of millions vertex-hyperedge memberships; open source implementations are available\footnote{\url{https://github.com/pnnl/chgl}}
as part of The Chapel HyperGraph Library (CHGL,
\cite{jenkins2018chapel}), a prototype HPC library
\cite{doecode_18401} for large-scale hypergraph generation and analysis
written in the emerging programming language of Chapel.

\subsection{Three Generative Hypergraph Models} \label{sec:models}

We define the generative models we consider below, and then briefly compare their properties. 

\begin{enumerate}
\item {\bf Erd\H{o}s-R\'{e}nyi}, $\mbox{ER}(n,m,p)$.
The user specifies three scalar parameters: the desired number of vertices $n$, desired number of hyperedges $m$, and vertex-hyperedge membership probability, $p \in [0,1]$. For each of the $nm$ vertex-hyperedge pairs, the probability of membership is the same, 
\[
\Pr(v \in e)=p.
\]
\item {\bf Chung-Lu}, $\mbox{CL}({{\vec{d_v}}}, {{\vec{d_e}}})$. The
user specifies a desired vertex degree sequence
$\vec{d_{v}}=(d_{v_{1}},\ldots,d_{v_{n}})$ and desired hyperedge size
sequence $\vec{d_{e}}=(d_{e_{1}},\ldots,d_{e_{m}})$, which (in order to
be realizable by a hypergraph) satisfy
$c=\sum_{i=1}^{n} d_{v_{i}}=\sum_{i=1}^{m} d_{e_{i}}$. The probability a vertex belongs to a hyperedge is proportional to the product of
the desired vertex degree and edge size, i.e.
\begin{equation*}
\Pr (v_{i} \in e_{j}) =\frac{d_{v_{i}}\cdot d_{e_{j}}}{c}.
\end{equation*}
To ensure this probability is always less than 1, one may further
require the input sequences satisfy
$\max_{i,j} d_{v_{i}}d_{e_{j}} < c$.
\item {\bf Block Two-Level Erd\H{o}s-R\'{e}nyi}, $\mbox{BTER}(\vec{d_v}, \vec{d_e}, \vec{m_v}, \vec{m_e})$ . 
In addition
to the desired vertex degree and edge size sequences mentioned in Chung--Lu,
the user also specifies desired vertex and edge \textit{metamorphosis coefficients},
$\vec{m_{v}}$ and $\vec{m_{e}}$, which, as clarified further
below, are measures of community structure based on the prevalence of small,
dense substructures in the hypergraph. The BTER model is designed to output
a hypergraph that matches the input degree distribution and metamorphosis
coefficients. The BTER model proceeds in two phases: in the first, metamorphosis
coefficients are approximately matched by grouping vertices and hyperedges
into small, disjoint sets called \textit{affinity blocks} and applying the
Erd\H{o}s--R\'{e}nyi model on each block. In the second, the degree distributions
are matched by running the Chung--Lu model on the excess desired degrees,
thereby linking the blocks. As formal details of the BTER model are complicated,
the reader is referred to \cite{Aksoy2017} for a full specification.
\end{enumerate}

For ER the expected number of vertex-hyperedge memberships is $pnm$, and hence
this simple model can be used to generate random hypergraphs with a specified
vertex-hyperedge membership density. We reported this density for our datasets
in Fig.~\ref{fig:dd}. For the CL model, each vertex $v$ achieves its
user-specified desired degree $d_{v}$ in expectation since
\begin{equation*}
\mathbb{E}\bigl(\operatorname{deg}(v)\bigr)=\sum_{e}
\Pr (v \in e) = \frac{d_{v}}{c} \sum_{e}
d_{e} = d_{v}.
\end{equation*}
An identical argument also shows each hyperedge $e$ achieves its desired
size $d_{e}$ in expectation. In this way, CL not only matches the desired
vertex-hyperedge membership density in expectation like ER, but additionally
matches the vertex degree and edge size distributions in expectation. We
reported these degree distributions for our datasets in Fig.~\ref{fig:dd}.

The CL model is a generalization of the ER model in the sense that the
ER can be obtained from CL by taking the degree and edge size sequences
to be constant, i.e.
\begin{equation*}
\operatorname{CL} \bigl((\underbrace{mp, mp ,\ldots, mp}_{n \text{ times}}), (
\underbrace{np, np,\ldots, np}_{m \text{ times}}) \bigr)= \operatorname{ER}(n,m,p).
\end{equation*}
Lastly, the BTER model (which, as explained in \cite{Aksoy2017}, utilizes
the CL model as a subroutine) can be understood as a generalization of
CL. The BTER model is designed to match not only vertex and edge size distributions,
but also per-degree\vadjust{\goodbreak}  metamorphosis coefficients. A~complete definition of
metamorphosis coefficients is involved; interested readers are referred
to \cite{Aksoy2017} for full details. Nonetheless, to elucidate how metamorphosis
coefficients are interpreted in the hypergraph setting, we provide a high-level
description.

Metamorphosis coefficients are measures of network community structure
based on counts of bipartite 4-cycles, also called \textit{butterflies}, and
bipartite 3-paths, also called \textit{caterpillars}. In the language of hypergraphs,
a butterfly is a subhypergraph consisting of two vertices and two edges
intersecting in those two vertices; a caterpillar is an edge with two vertices
intersecting with another edge in one of those vertices. The authors in
\cite{Aksoy2017} define metamorphosis coefficients for vertices within
each of the two partitions of a bipartite graph, based on the ratios of
butterfly to caterpillar counts those vertices participate in. Stated equivalently,
this defines metamorphosis for the vertices and hyperedges of a hypergraph.
If a hyperedge $e$ has a large metamorphosis coefficient, this means a
large proportion of the edges that $e$ intersects with intersect in (at
least) 2 vertices; dually, if vertex $v$ has large metamorphosis, then
a large proportion of vertices $v$ shares an edge with share (at least)
2 edges.
For example, in Fig.~\ref{fig:authPap} each author in
the leftmost network repeats a coauthorship with someone on 1 out of 3
of their other papers, and thus has metamorphosis $\frac{1}{3}$; in
the rightmost, each author repeats a coauthorship on all their other papers,
and thus has metamorphosis~1. BTER matches degree
distributions, as well as the average metamorphosis coefficients for vertices
and hyperedges of a given degree and cardinality, respectively.

Taken as a suite, these three models serve well as null-models
since each provides successively more control over hypergraph structure
than the previous, providing the flexibility to choose different tiers
of structural nuance for the generated hypergraphs. In the next section,
we run each model multiple times on each dataset, and study how well each model replicated $s$-walk properties. By (for example) ``running CL on LesMis'', we mean
extracting the model inputs (in this case, the vertex degrees and hyperedge
sizes) from the data, and using the Chung--Lu model to generate a hypergraph
under these inputs.

\subsection{Comparison}

\begin{figure}[t!]
\centering
\begin{subfigure}[b]{0.48\textwidth}
\includegraphics[width = 0.5\linewidth]{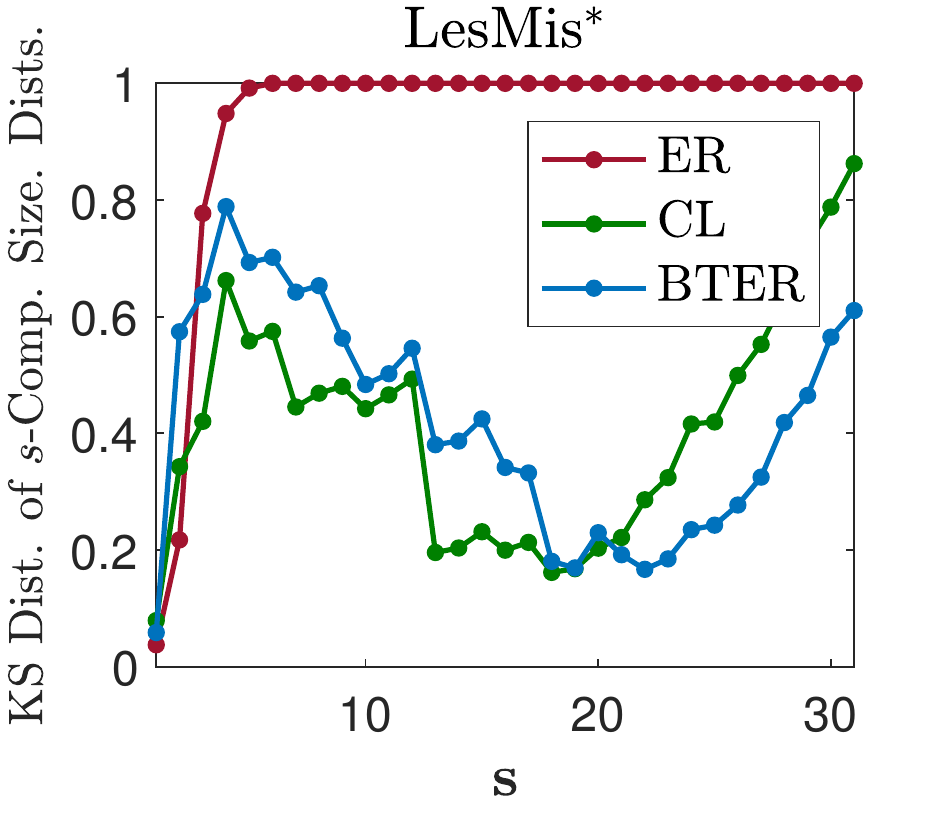}\includegraphics[width = 0.5\linewidth]{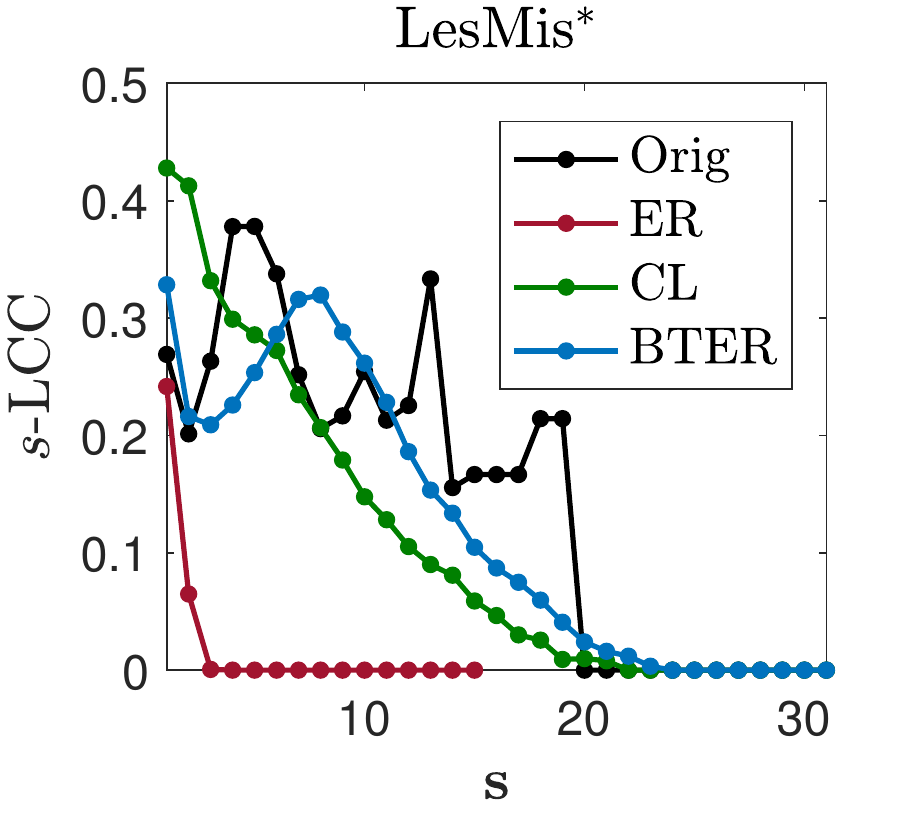} \\
\includegraphics[width = 0.5\linewidth]{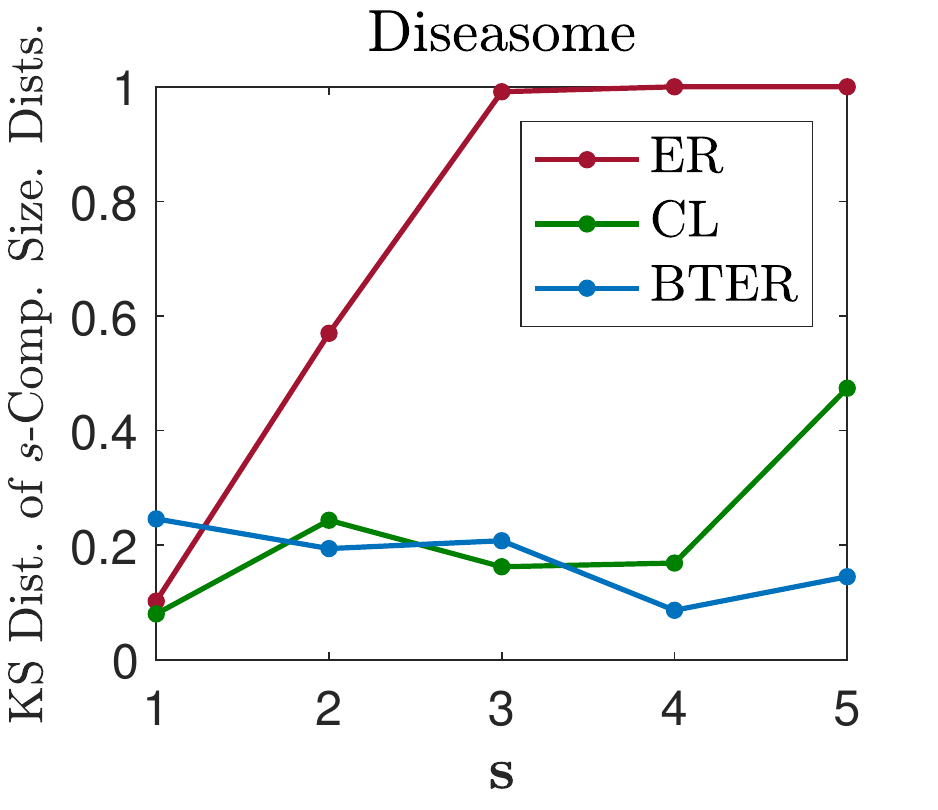}\includegraphics[width = 0.5\linewidth]{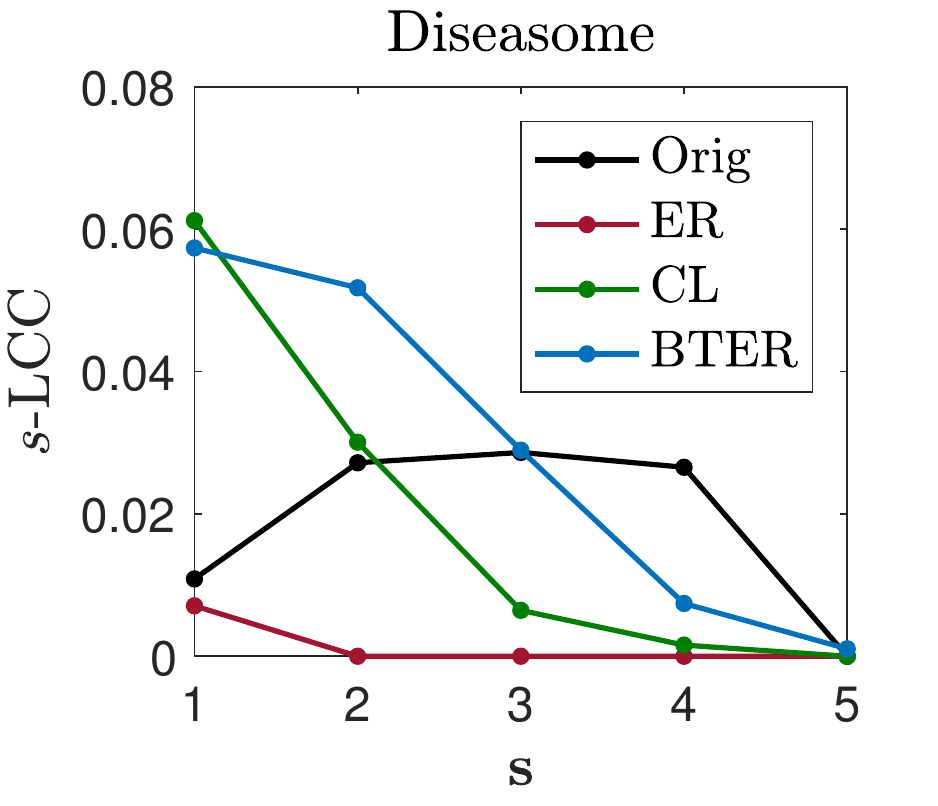} \\
\includegraphics[width = 0.5\linewidth]{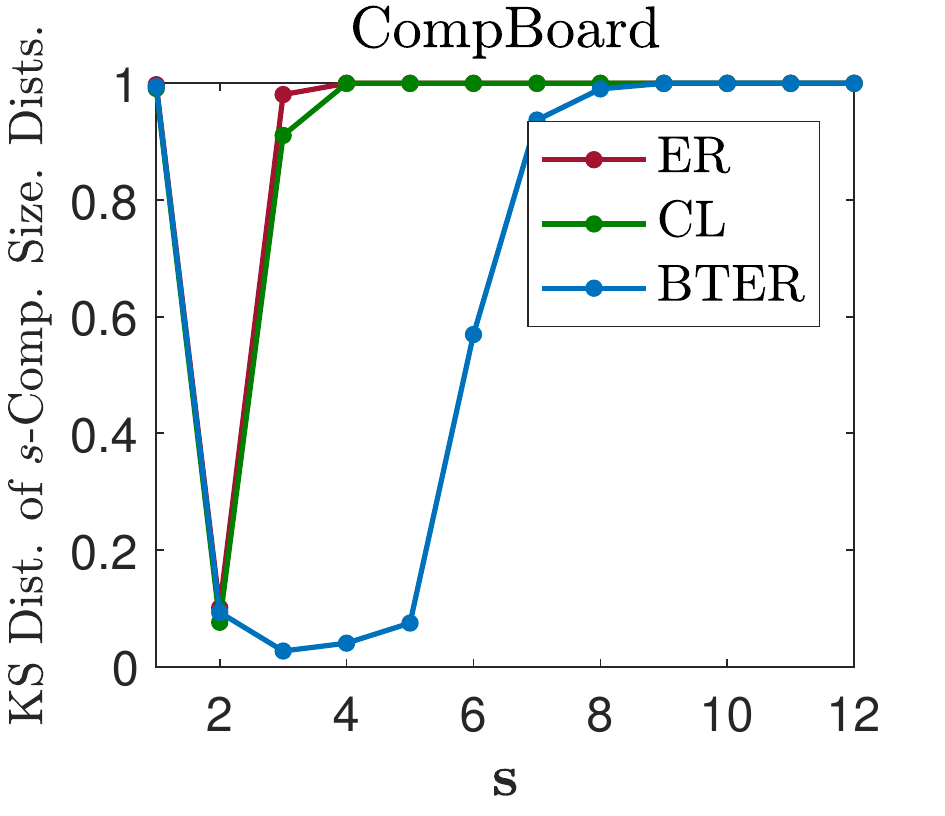}\includegraphics[width = 0.5\linewidth]{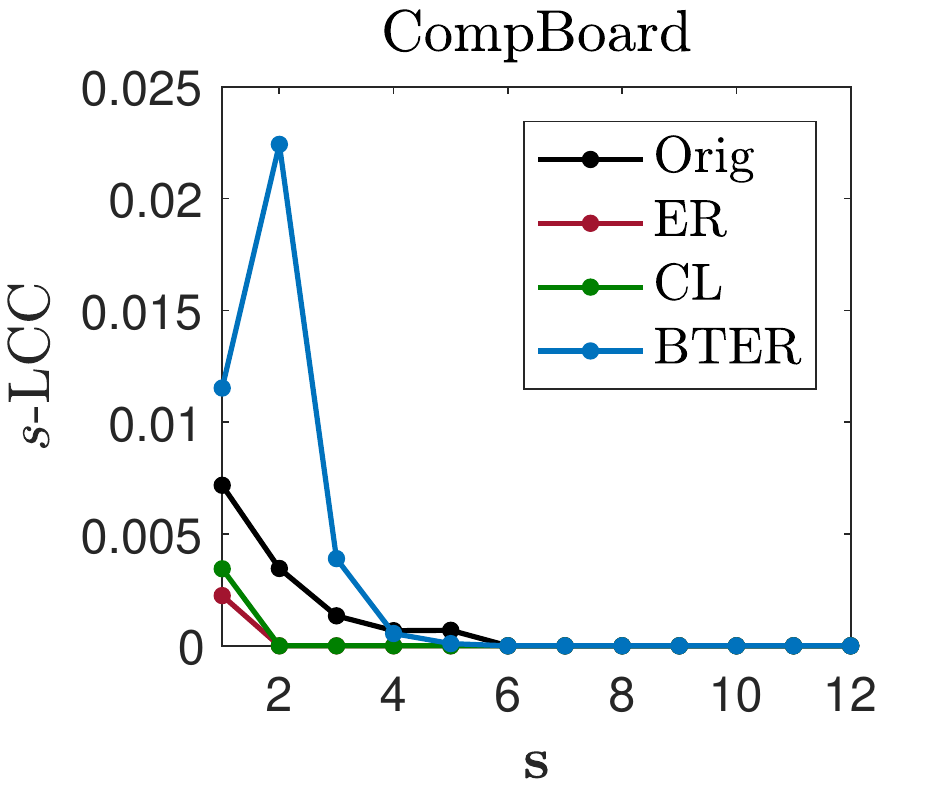} \\
\end{subfigure}
\quad 
\begin{subfigure}[b]{0.48\textwidth}
\includegraphics[width = 0.5\linewidth]{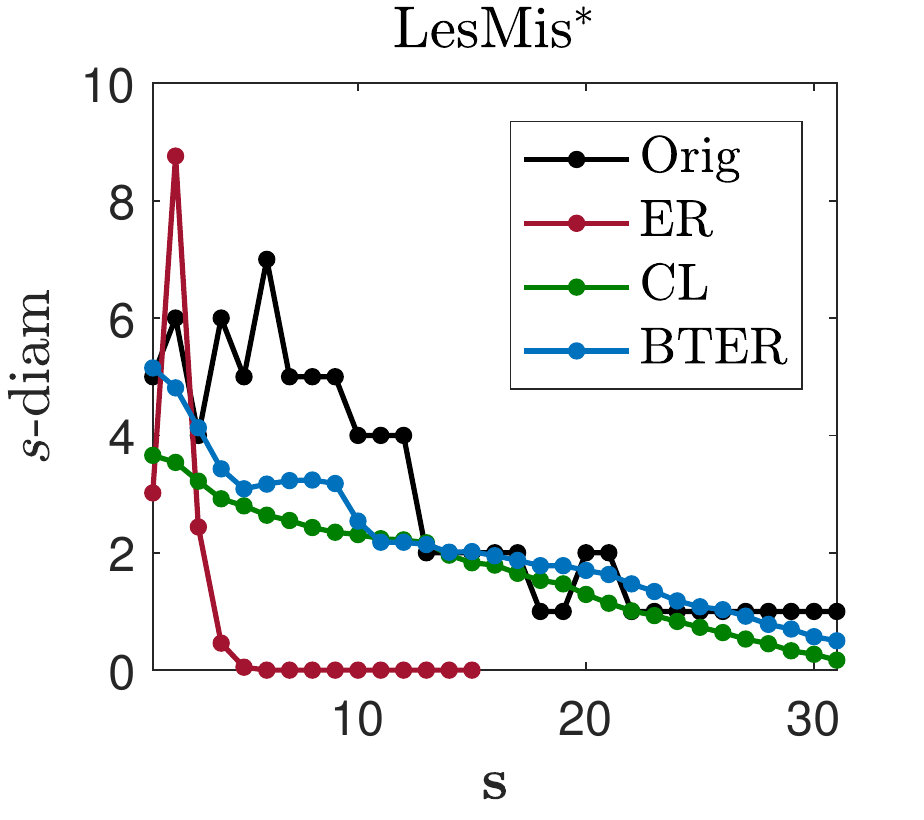}\includegraphics[width = 0.5\linewidth]{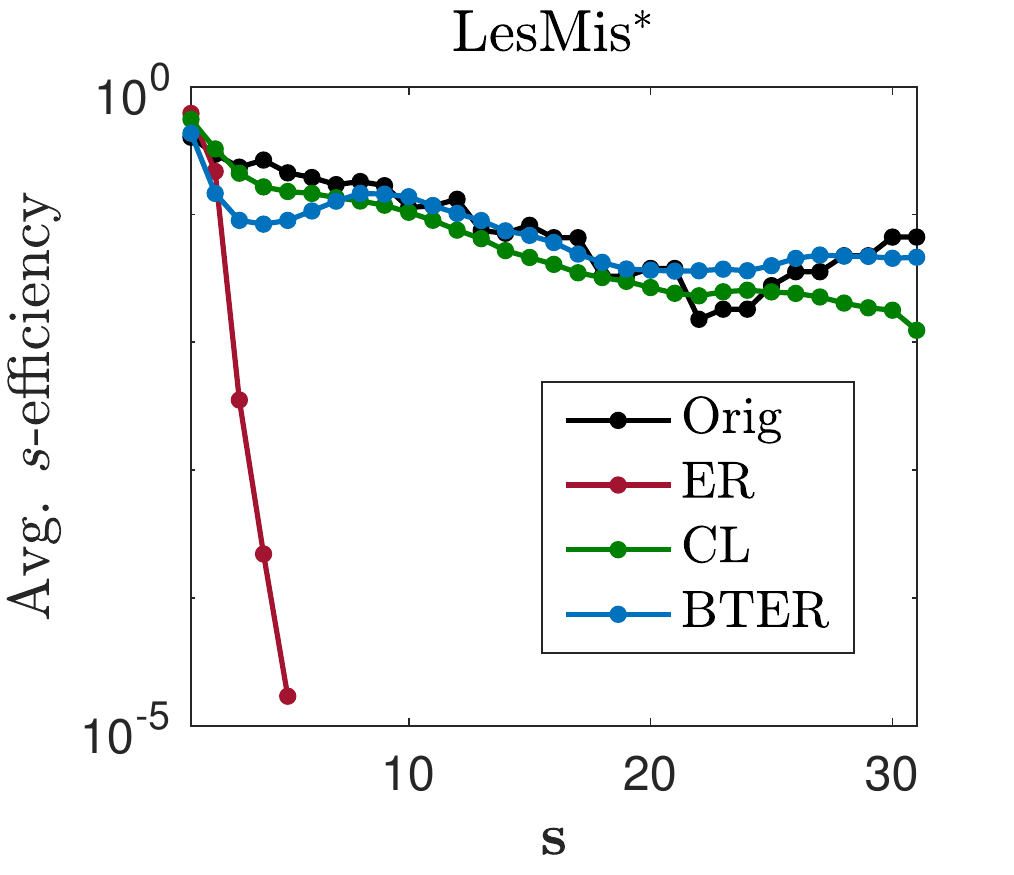} \\
\includegraphics[width = 0.5\linewidth]{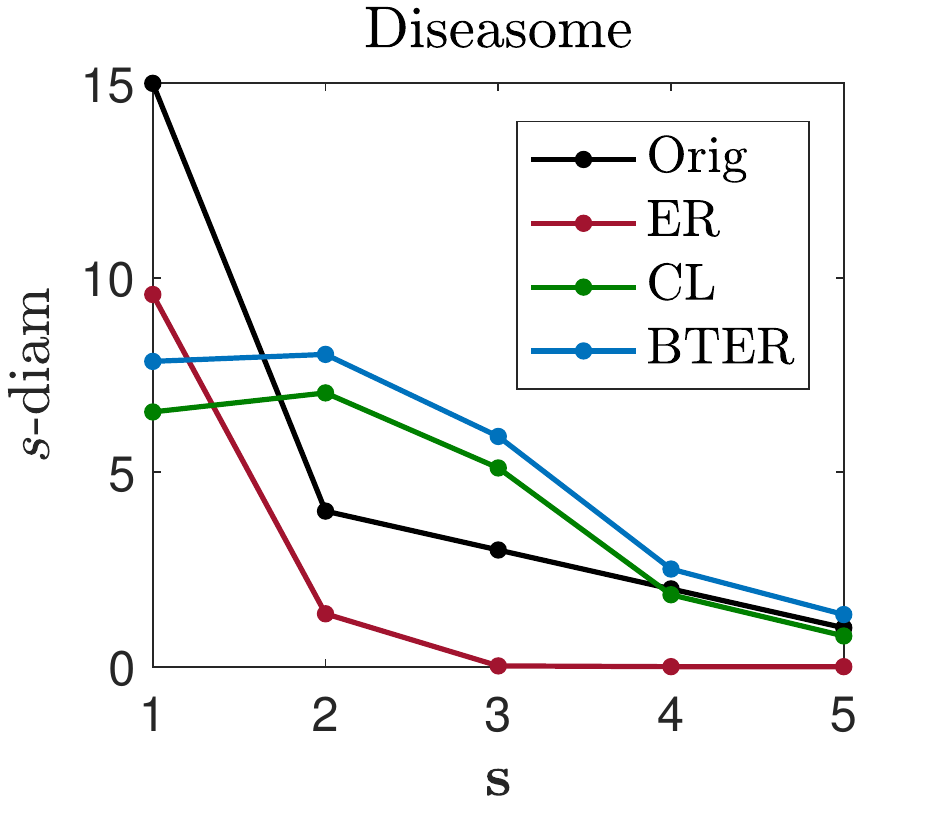}\includegraphics[width = 0.5\linewidth]{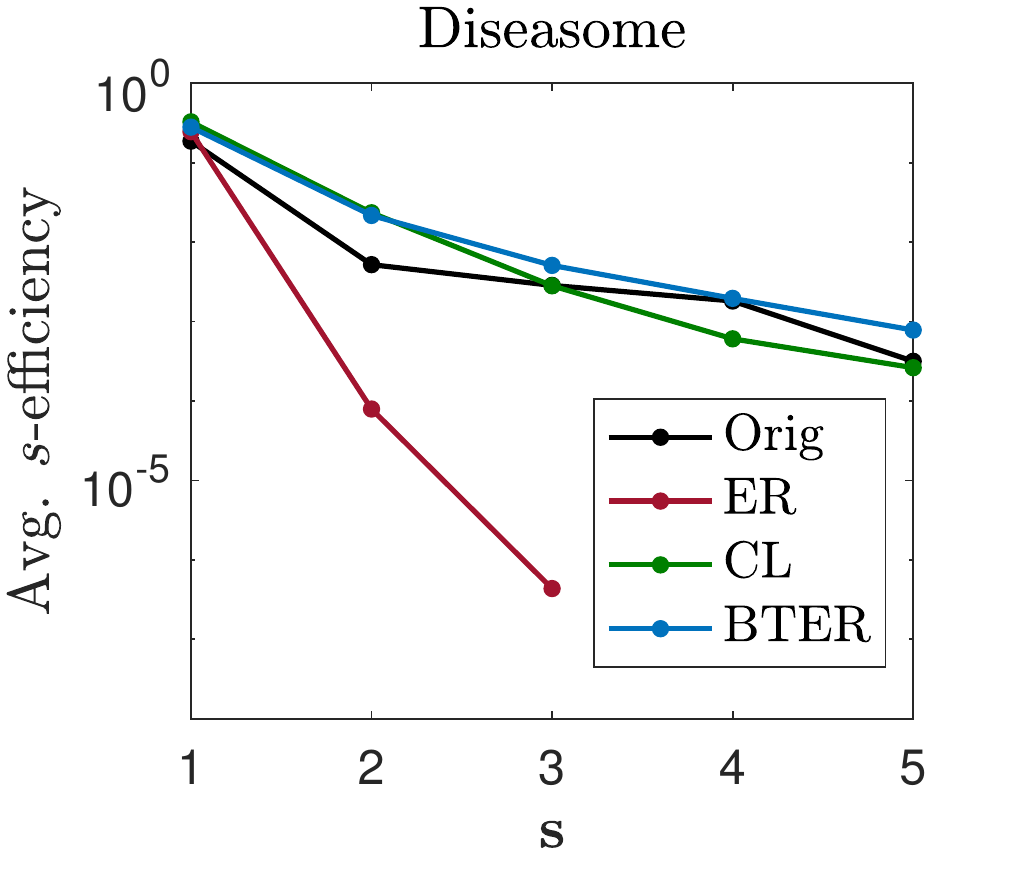} \\
\includegraphics[width = 0.5\linewidth]{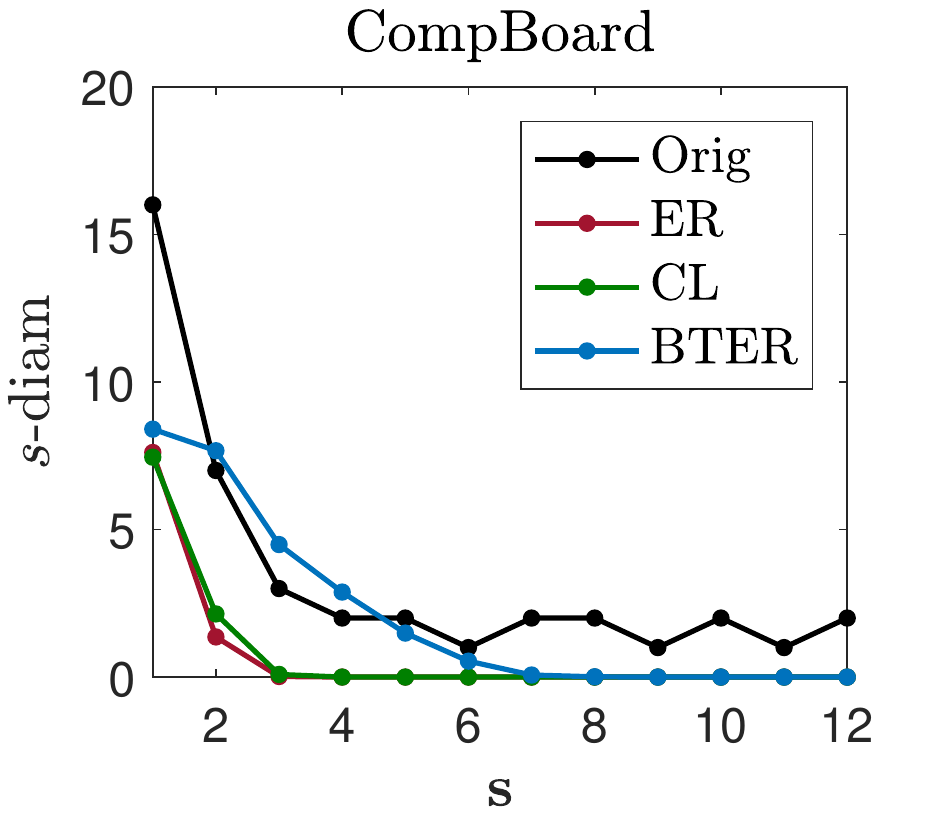}\includegraphics[width = 0.5\linewidth]{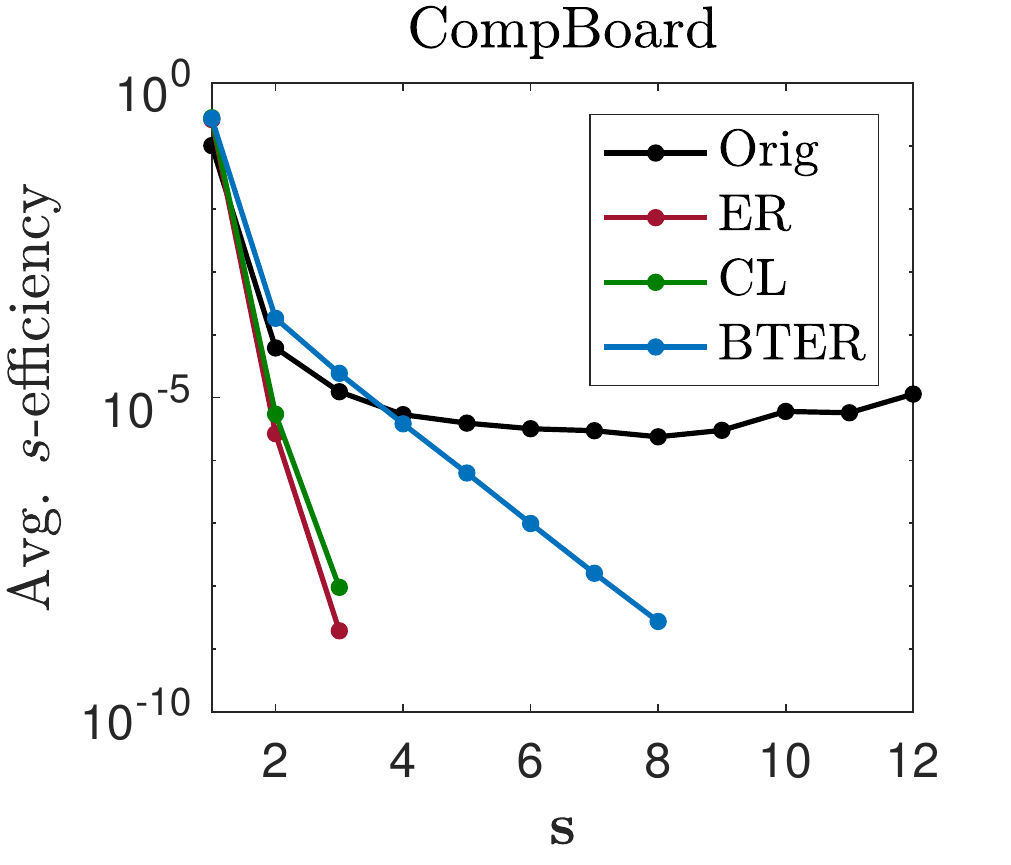} \\

\end{subfigure}
\vspace{-5mm}
\caption{Comparison of $s$-component size distributions, mean $s$-local clustering coefficients, $s$-diameter, and average $s$-efficiency of the LesMis$^*$, Diseasome, and CompBoard datasets against those of hypergraphs synthetically generated by ER, CL, and BTER.} \label{fig:nullModel}
\end{figure}

Fig.~\ref{fig:nullModel} compares $s$-walk based properties of
LesMis$^*$, Diseasome, and CompBoard against those of synthetic hypergraphs
generated by ER, CL, and BTER. For each dataset, we generate $100$ instances
of each synthetic model and compute the properties in question for each
instance. The plot reports the average values observed over the 100 trials,
for each~$s$.

In the leftmost column in Fig.~\ref{fig:nullModel}, we use \textit{Kolmogorov--Smirnov}
(KS) distance to compare the $s$-component size probability distributions
of the original and synthetic hypergraphs. KS
distance is normalized between 0 and 1, with smaller KS-values indicating
greater similarity.\footnote{For example, the green point at
$(13,0.2)$ in the LesMis$^*$ plot means that, over 100 trials, the average
KS-distance between the $13$-component size distribution of original LesMis$^*$
dataset, and that of a CL hypergraph, is $0.2$. In cases where the synthetic
graph had no hyperedges containing at least $s$ vertices (and hence an
empty $s$-component size distribution) we define KS distance between the
original $s$-comp distribution as 1 (the maximum).} Comparing the models as $s$ increases, the ER model exhibits higher
KS distance than for CL and BTER, indicating $s$-component size distributions
that are more dissimilar to the original. All three models seem to exhibit
larger KS distances for larger $s$ values, although in some cases (e.g. for
CL and BTER on LesMis$^*$) this increase is not monotonic in~$s$. One
notable exception is CompBoard, in which all models exhibit much larger
KS distance for $s=1$ than $s=2$. This can likely be attributed to the
large number of isolated hyperedges observed in $1$-components of CompBoard:
in contrast, all three models tend to output hypergraphs in which the majority
of hyperedges are contained within a single giant $1$-component.

Turning to $s$-distances, we compare the
original $s$-diameter (center right) and average $s$-efficiency (far right)
to those of the model's synthetic hypergraphs.
As the average $s$-efficiency plots are in log-scale, average values of
0 (which occur whenever no two hyperedges intersect in $s$ vertices) are
not plotted. ER tends to have lower $s$-diameter
and average $s$-efficiency as $s$ increases, when compared to CL and BTER.
For LesMis$^*$ and Diseasome, CL and BTER seem to perform comparably;
for CompBoard, however, BTER does noticeably better than both in matching
average $s$-efficiency for $s \geq 2$, although still diverging from the
original values considerably for $s \geq 5$. Lastly, we consider the model's
performance with regard to mean $s$-local clustering coefficients (center
left). For all three datasets, ER produces smaller clustering coefficients
than observed in the original data, for all~$s$. For CL and BTER, the mean $s$-local clustering coefficients
sometimes exceed those of the original data for small values of $s$ (e.g. for $s \leq 2$ on Diseasome),
while for some larger values of $s$ (e.g. for $13 \leq s \leq 19$ on LesMis$^*$),
BTER and CL produce smaller local clustering coefficients than those of
the original data.

Taking a broader view of these results, none of these three models
are able to provide a consistent, close match across values of~$s$. This
suggests the $s$-walk-based measures in question cannot be explained as
sole consequences of the model inputs (e.g. degree distributions for the
Chung--Lu model), that are preserved in expectation in the output hypergraphs.
Nonetheless, this experiment should not be extrapolated to provide generalized
guidance on which model best preserves certain $s$-walk properties. Depending on the properties of the data in question, it may be the case
that ER (the least accurate model on our data) provides a closer match
than CL or BTER. In order to provide more a comprehensive approach to such
questions, it would be of interest to determine conditions on model inputs
under which certain $s$-walk properties of the output hypergraphs can be
tightly bounded or controlled. While such work is outside the scope of
the present paper, the aforementioned research by Kang, Cooley, and Koch
\cite{cooley2018subcritical,Cooley2015,Cooley2016ThresholdAH} illustrates establishing guarantees on even basic high-order walk based
properties in random hypergraphs (such as the size of the largest
$s$-component) requires sophisticated probabilistic analysis.

\section{Conclusion}\label{sec:conc}

The prevalence and complexity of hypernetwork data necessitates analytic
methods that are both applicable and able to capture hypergraph-native
phenomena. We have proposed hypergraph $s$-walks provide a framework
under which graph analytic tools popular in network science
extend more meaningfully to hypergraphs. In applying these measures to
real data, we've explored how they may reveal varied, interpretable, and
significant structural properties of the data otherwise lost when analyzing
hypergraphs under the lens of the usual graph walk. The methods we've
focused on---connected component analyses, distance-based measures, high-order
motifs and clustering coefficients---are meant to illustrate the breadth
of tools to which this approach is relevant. However, ours is clearly far
from a comprehensive exploration. We conclude by outlining future work.

One immediate open question concerns how the methods we've developed
may be generalized further. For instance, it would be of both theoretical
and practical interest to develop tractable $s$-walk based measures for
weighted hypergraphs (with real-valued vertex and/or edge weights), directed
hypergraphs (in which each edge's vertices are either in its ``head'' or
``tail''), ordered hypergraphs (in which each edge's vertices are totally
ordered), or temporal hypergraphs (consisting of sequences of hypergraphs).
With regard to the latter topic, our work does
not address how hypergraphs, and the structural properties we observed,
evolve throughout time. The suite of generative models we considered
are effective as structural null models, but do not explicitly posit a
process or mechanism through which hypernetworks grow. In contrast, other
researchers have put forth and studied hypergraph evolution mechanisms,
such as a preferential attachment inspired model for non-uniform hypergraphs
\cite{Guo2016,Guo2016a}. An analysis of these, or the development of
new, temporal hypergraph models may shed insight into how high-order
structural properties put forth here emerge in network topology.

Lastly, another open direction lies in devising
efficient computational methods for the $s$-walk measures put forth here.
We did not explore the algorithmic aspects underlying these methods.
In some cases, the methods we utilized---while sufficient on our data---were not scalable to massive hypergraph data (e.g. computing $s$-centrality
via the $s$-line graph quickly becomes infeasible for large hypergraphs
with skewed degree distributions, as the density of $s$-line graphs increase
quadratically in the maximum vertex degree). Developing algorithms that
leverage the sparsity of the hypergraph (rather than resorting to computation
on dense $s$-line graphs) would help facilitate the application of these
methods to larger-scale data. Furthermore, just as researchers have begun
developing efficient schemes for computing atomic bipartite graph motifs
such as cycles of length 4 \cite{Sanei-Mehri2018,wang2018vertex}, work
in a similar vein would prove useful for enabling large-scale $s$-triangle
counting in hypergraphs. \\

{\bf \noindent Acknowledgements}: We would like to thank numerous colleagues for helpful discussions, including
Marcin Zalewski, Francesca Grogan, Katy Nowak, Dustin Arendt, Stephen Young,
Brett Jefferson, and Louis Jenkins. We also thank referees for thoughtful comments which improved the manuscript.   \\

{\bf \noindent Funding}: This work was partially funded under the High Performance Data
Analytics (HPDA) program at the Department of Energy's Pacific Northwest National
Laboratory. PNNL Information Release: PNNL-SA-144766. Pacific Northwest National Laboratory is operated by Battelle
Memorial Institute under Contract DE-ACO6-76RL01830.



\printbibliography

@inproceedings{Sariyuece2018,
	Address = {London, UK},
	Author = {Ahmet Erdem Sar{\i}y{\"u}ce and Ali Pinar},
	Booktitle = {Proceedings of the Eleventh {ACM} International Conference on Web Search and Data Mining - {WSDM} '18},
	Doi = {10.1145/3159652.3159678},
	Publisher = {{ACM} Press},
	Title = {Peeling Bipartite Networks for Dense Subgraph Discovery},
	Year = {2018}}

@article{Schmidt2019,
	Author = {Martin Schmidt},
	Date = {2019-07-04},
	Eprint = {1907.02574v1},
	Eprintclass = {math.CO},
	Eprinttype = {arXiv},
	Title = {Functorial Approach to Graph and Hypergraph Theory},
	Year = {2019}}

@article{Seshadhri2012,
	Author = {C. Seshadhri and Tamara G. Kolda and Ali Pinar},
	Doi = {10.1103/physreve.85.056109},
	Journal = {Physical Review E},
	Month = {may},
	Number = {5},
	Publisher = {American Physical Society ({APS})},
	Title = {Community structure and scale-free collections of Erd{\H{o}}s-R{\'{e}}nyi graphs},
	Volume = {85},
	Year = {2012}}

@article{wang1999paths,
	Author = {Wang, Jianfang and Lee, Tony T},
	Journal = {Science in China Series A: Mathematics},
	Number = {1},
	Pages = {1--12},
	Publisher = {Springer},
	Title = {Paths and cycles of hypergraphs},
	Volume = {42},
	Year = {1999}}

@article{wang2018vertex,
	Author = {Wang, Kai and Lin, Xuemin and Qin, Lu and Zhang, Wenjie and Zhang, Ying},
	Journal = {arXiv preprint arXiv:1812.00283},
	Title = {Vertex Priority Based Butterfly Counting for Large-scale Bipartite Networks},
	Year = {2018}}

@article{Watts1998,
	Author = {Duncan J. Watts and Steven H. Strogatz},
	Doi = {10.1038/30918},
	Journal = {Nature},
	Month = {jun},
	Number = {6684},
	Pages = {440--442},
	Publisher = {Springer Nature},
	Title = {Collective dynamics of `small-world' networks},
	Volume = {393},
	Year = {1998}}

@article{Whitney1932,
	Author = {Hassler Whitney},
	Doi = {10.2307/2371086},
	Journal = {American Journal of Mathematics},
	Month = {jan},
	Number = {1},
	Pages = {150},
	Publisher = {{JSTOR}},
	Title = {Congruent Graphs and the Connectivity of Graphs},
	Volume = {54},
	Year = {1932}}

@inproceedings{zhou2007learning,
	Author = {Zhou, Dengyong and Huang, Jiayuan and Sch{\"o}lkopf, Bernhard},
	Booktitle = {Advances in neural information processing systems},
	Pages = {1601--1608},
	Title = {Learning with hypergraphs: Clustering, classification, and embedding},
	Year = {2007}}

@article{Zhou2011,
	Author = {Wanding Zhou and Luay Nakhleh},
	Doi = {10.1186/1471-2105-12-132},
	Journal = {{BMC} Bioinformatics},
	Month = {may},
	Number = {1},
	Publisher = {Springer Nature},
	Title = {Properties of metabolic graphs: biological organization or representation artifacts?},
	Volume = {12},
	Year = {2011}}

@article{Newman2001,
	Author = {M. E. J. Newman and S. H. Strogatz and D. J. Watts},
	Doi = {10.1103/physreve.64.026118},
	Journal = {Physical Review E},
	Month = {jul},
	Number = {2},
	Publisher = {American Physical Society ({APS})},
	Title = {Random graphs with arbitrary degree distributions and their applications},
	Volume = {64},
	Year = {2001}}

@article{Opsahl2013,
	Author = {Tore Opsahl},
	Doi = {10.1016/j.socnet.2011.07.001},
	Journal = {Social Networks},
	Month = {may},
	Number = {2},
	Pages = {159--167},
	Publisher = {Elsevier {BV}},
	Title = {Triadic closure in two-mode networks: Redefining the global and local clustering coefficients},
	Volume = {35},
	Year = {2013}}

@article{Parczyk2015,
	Author = {O. Parczyk and Y. Person},
	Doi = {10.1016/j.endm.2015.06.083},
	Journal = {Electronic Notes in Discrete Mathematics},
	Month = {nov},
	Pages = {611--619},
	Publisher = {Elsevier {BV}},
	Title = {On Spanning Structures in Random Hypergraphs},
	Volume = {49},
	Year = {2015}}

@misc{hnxRef,
	Author = {Praggastis, Brenda and Arendt, Dustin and Joslyn, Cliff and Purvine, Emilie and Aksoy, Sinan and Monson, Kyle},
	Howpublished = {https://github.com/pnnl/HyperNetX},
	Title = {HyperNetX},
	Url = {https://github.com/pnnl/HyperNetX},
	Year = {2019}}

@inproceedings{purvine2018topological,
	Author = {Purvine, Emilie and Aksoy, Sinan and Joslyn, Cliff and Nowak, Kathleen and Praggastis, Brenda and Robinson, Michael},
	Booktitle = {International Conference on Human Interface and the Management of Information},
	Organization = {Springer},
	Pages = {90--109},
	Title = {A Topological Approach to Representational Data Models},
	Year = {2018}}

@article{Robins2004,
	Author = {Garry Robins and Malcolm Alexander},
	Doi = {10.1023/b:cmot.0000032580.12184.c0},
	Journal = {Computational {\&} Mathematical Organization Theory},
	Month = {may},
	Number = {1},
	Pages = {69--94},
	Publisher = {Springer Nature},
	Title = {Small Worlds Among Interlocking Directors: Network Structure and Distance in Bipartite Graphs},
	Volume = {10},
	Year = {2004}}

@techreport{rochat2009closeness,
	Author = {Rochat, Yannick},
	Title = {Closeness centrality extended to unconnected graphs: The harmonic centrality index},
	Year = {2009}}

@article{rodl2004regularity,
	Author = {R{\"o}dl, Vojt{\v{e}}ch and Skokan, Jozef},
	Journal = {Random Structures \& Algorithms},
	Number = {1},
	Pages = {1--42},
	Publisher = {Wiley Online Library},
	Title = {Regularity Lemma for k-uniform hypergraphs},
	Volume = {25},
	Year = {2004}}

@article{Rodriextasciiacuteguez2002,
	Author = {J.A. Rodriguez},
	Doi = {10.1080/03081080290011692},
	Journal = {Linear and Multilinear Algebra},
	Month = {jan},
	Number = {1},
	Pages = {1--14},
	Publisher = {Informa {UK} Limited},
	Title = {On the Laplacian Eigenvalues and Metric Parameters of Hypergraphs},
	Volume = {50},
	Year = {2002}}

@inproceedings{Sanei-Mehri2018,
	Address = {London, United Kingdom},
	Author = {Seyed-Vahid Sanei-Mehri and Ahmet Erdem Sariyuce and Srikanta Tirthapura},
	Booktitle = {Proceedings of the 24th {ACM} {SIGKDD} International Conference on Knowledge Discovery {\&} Data Mining - {KDD} '18},
	Doi = {10.1145/3219819.3220097},
	Publisher = {{ACM} Press},
	Title = {Butterfly Counting in Bipartite Networks},
	Year = {2018}}

@article{Latora2001,
	Author = {Vito Latora and Massimo Marchiori},
	Doi = {10.1103/physrevlett.87.198701},
	Journal = {Physical Review Letters},
	Month = {oct},
	Number = {19},
	Publisher = {American Physical Society ({APS})},
	Title = {Efficient Behavior of Small-World Networks},
	Volume = {87},
	Year = {2001}}

@incollection{levine1979study,
	Address = {Bedford, Massachusetts},
	Author = {Levine, Joel H and Roy, William S},
	Booktitle = {Perspectives on Social Network Research},
	Pages = {349--378},
	Publisher = {Elsevier},
	Title = {A study of interlocking directorates: Vital concepts of organization},
	Year = {1979}}

@inproceedings{lu2011high,
	Author = {Lu, Linyuan and Peng, Xing},
	Booktitle = {WAW},
	Organization = {Springer},
	Pages = {14--25},
	Title = {High-Ordered Random Walks and Generalized Laplacians on Hypergraphs.},
	Year = {2011}}

@article{Luxburg2007,
	Author = {Ulrike von Luxburg},
	Doi = {10.1007/s11222-007-9033-z},
	Journal = {Statistics and Computing},
	Month = {aug},
	Number = {4},
	Pages = {395--416},
	Publisher = {Springer Science and Business Media {LLC}},
	Title = {A tutorial on spectral clustering},
	Volume = {17},
	Year = {2007}}

@article{Nacher2011,
	Author = {J.C. Nacher and T. Akutsu},
	Doi = {10.1016/j.physa.2011.06.073},
	Journal = {Physica A: Statistical Mechanics and its Applications},
	Month = {nov},
	Number = {23-24},
	Pages = {4636--4651},
	Publisher = {Elsevier {BV}},
	Title = {On the degree distribution of projected networks mapped from bipartite networks},
	Volume = {390},
	Year = {2011}}

@article{naik2018recent,
	Author = {Naik, Ranjan N},
	Journal = {arXiv preprint arXiv:1809.08472},
	Title = {Recent Advances on Intersection Graphs of Hypergraphs: A Survey},
	Year = {2018}}

@article{Naik1982,
	Author = {Ranjan N. Naik and S.B. Rao and S.S. Shrikhande and N.M. Singhi},
	Doi = {10.1016/s0195-6698(82)80029-2},
	Journal = {European Journal of Combinatorics},
	Month = {jun},
	Number = {2},
	Pages = {159--172},
	Publisher = {Elsevier {BV}},
	Title = {Intersection Graphs of k-uniform Linear Hypergraphs},
	Volume = {3},
	Year = {1982}}

@article{Newman2003,
	Author = {M. E. J. Newman},
	Doi = {10.1137/s003614450342480},
	Journal = {{SIAM} Review},
	Month = {jan},
	Number = {2},
	Pages = {167--256},
	Publisher = {Society for Industrial {\&} Applied Mathematics ({SIAM})},
	Title = {The Structure and Function of Complex Networks},
	Volume = {45},
	Year = {2003}}

@article{Newman2004,
	Author = {M. E. J. Newman and M. Girvan},
	Doi = {10.1103/physreve.69.026113},
	Journal = {Physical Review E},
	Month = {feb},
	Number = {2},
	Publisher = {American Physical Society ({APS})},
	Title = {Finding and evaluating community structure in networks},
	Volume = {69},
	Year = {2004}}

@incollection{Katona1975,
	Address = {Mathematical Centre, Amsterdam},
	Author = {G. O. H. Katona},
	Booktitle = {Combinatorics},
	Doi = {10.1007/978-94-010-1826-5\textunderscore11},
	Pages = {215--244},
	Publisher = {Springer Netherlands},
	Title = {Extremal Problems for Hypergraphs},
	Year = {1975}}

@article{katona1999hamiltonian,
	Author = {Katona, Gyula Y and Kierstead, Hal A},
	Journal = {Journal of Graph Theory},
	Number = {3},
	Pages = {205--212},
	Publisher = {Wiley Online Library},
	Title = {Hamiltonian chains in hypergraphs},
	Volume = {30},
	Year = {1999}}

@article{Kirkland2017,
	Author = {Steve Kirkland},
	Doi = {10.1093/comnet/cnx039},
	Journal = {Journal of Complex Networks},
	Month = {aug},
	Number = {2},
	Pages = {297--316},
	Publisher = {Oxford University Press ({OUP})},
	Title = {Two-mode networks exhibiting data loss},
	Volume = {6},
	Year = {2017}}

@article{Klamt2009,
	Author = {Steffen Klamt and Utz-Uwe Haus and Fabian Theis},
	Doi = {10.1371/journal.pcbi.1000385},
	Editor = {J{\"o}rg Stelling},
	Journal = {{PLoS} Computational Biology},
	Month = {may},
	Number = {5},
	Pages = {e1000385},
	Publisher = {Public Library of Science ({PLoS})},
	Title = {Hypergraphs and Cellular Networks},
	Volume = {5},
	Year = {2009}}

@book{knuth1993stanford,
	Address = {New York},
	Author = {Knuth, Donald Ervin},
	Publisher = {AcM Press New York},
	Title = {The Stanford GraphBase: a platform for combinatorial computing},
	Year = {1993}}

@article{Kolda2014,
	Author = {Tamara G. Kolda and Ali Pinar and Todd Plantenga and C. Seshadhri},
	Doi = {10.1137/130914218},
	Journal = {{SIAM} J. Sci. Comput.},
	Month = {jan},
	Number = {5},
	Pages = {C424--C452},
	Publisher = {Society for Industrial {\&} Applied Mathematics ({SIAM})},
	Title = {A Scalable Generative Graph Model with Community Structure},
	Url = {http://dx.doi.org/10.1137/130914218},
	Volume = {36},
	Year = {2014}}

@article{krivelevich2003approximate,
	Author = {Krivelevich, Michael and Sudakov, Benny},
	Journal = {Journal of Algorithms},
	Number = {1},
	Pages = {2--12},
	Publisher = {Elsevier},
	Title = {Approximate coloring of uniform hypergraphs},
	Volume = {49},
	Year = {2003}}

@inproceedings{Kuang2012,
	Author = {Da Kuang and Chris Ding and Haesun Park},
	Booktitle = {Proceedings of the 2012 {SIAM} International Conference on Data Mining},
	Doi = {10.1137/1.9781611972825.10},
	Month = {apr},
	Publisher = {Society for Industrial and Applied Mathematics},
	Title = {Symmetric Nonnegative Matrix Factorization for Graph Clustering},
	Year = {2012}}

@article{Larremore2014,
	Author = {Daniel B. Larremore and Aaron Clauset and Abigail Z. Jacobs},
	Doi = {10.1103/physreve.90.012805},
	Journal = {Physical Review E},
	Month = {jul},
	Number = {1},
	Publisher = {American Physical Society ({APS})},
	Title = {Efficiently inferring community structure in bipartite networks},
	Volume = {90},
	Year = {2014}}

@article{Latapy2008,
	Author = {Matthieu Latapy and Cl{\'{e}}mence Magnien and Nathalie Del Vecchio},
	Doi = {10.1016/j.socnet.2007.04.006},
	Journal = {Social Networks},
	Month = {jan},
	Number = {1},
	Pages = {31--48},
	Publisher = {Elsevier {BV}},
	Title = {Basic notions for the analysis of large two-mode networks},
	Volume = {30},
	Year = {2008}}

@article{Guo2016,
	Author = {Jin-Li Guo and Xin-Yun Zhu and Qi Suo and Jeffrey Forrest},
	Doi = {10.1038/srep36648},
	Journal = {Scientific Reports},
	Month = {nov},
	Number = {1},
	Publisher = {Springer Science and Business Media {LLC}},
	Title = {Non-uniform Evolving Hypergraphs and Weighted Evolving Hypergraphs},
	Volume = {6},
	Year = {2016}}

@techreport{hagberg2008exploring,
	Author = {Hagberg, Aric and Swart, Pieter and S Chult, Daniel},
	Institution = {Los Alamos National Lab.(LANL), Los Alamos, NM (United States)},
	Title = {Exploring network structure, dynamics, and function using NetworkX},
	Year = {2008}}

@article{hamosh2005online,
	Author = {Hamosh, Ada and Scott, Alan F and Amberger, Joanna S and Bocchini, Carol A and McKusick, Victor A},
	Journal = {Nucleic acids research},
	Number = {suppl\_1},
	Pages = {D514--D517},
	Publisher = {Oxford University Press},
	Title = {Online Mendelian Inheritance in Man (OMIM), a knowledgebase of human genes and genetic disorders},
	Volume = {33},
	Year = {2005}}

@article{han2010dirac,
	Author = {H{\`a}n, Hi{\^e}p and Schacht, Mathias},
	Journal = {Journal of Combinatorial Theory, Series B},
	Number = {3},
	Pages = {332--346},
	Publisher = {Elsevier},
	Title = {Dirac-type results for loose Hamilton cycles in uniform hypergraphs},
	Volume = {100},
	Year = {2010}}

@inproceedings{jenkins2018chapel,
	Author = {Jenkins, Louis and Bhuiyan, Tanveer and Harun, Sarah and Lightsey, Christopher and Mentgen, David and Aksoy, Sinan and Stavcnger, Timothy and Zalewski, Marcin and Medal, Hugh and Joslyn, Cliff},
	Booktitle = {2018 IEEE High Performance extreme Computing Conference (HPEC)},
	Organization = {IEEE},
	Pages = {1--6},
	Title = {Chapel HyperGraph Library (CHGL)},
	Year = {2018}}

@misc{doecode_18401,
	Abstractnote = {CHGL is a hypergraph computation library for Chapel. It's primary contributions are data structures for hypergraphs and the algorithms to operate on these data structures.},
	Author = {Jenkins, Louis and Stavenger, Tim and Zalewski, Marcin and Joslyn, Cliff and Aksoy, Sinan and Medal, Hugh},
	Howpublished = {https://github.com/pnnl/chgl},
	Title = {pnnl/chgl},
	Url = {https://github.com/pnnl/chgl}}

@inproceedings{joslynHICSS,
	Author = {Joslyn, Cliff and Aksoy, Sinan and Arendt, Dustin and Jenkins, Louis and Praggastis, Brenda and Purvine, Emilie and Zalewski, Marcin},
	Booktitle = {HICSS 2019 Symposium on Cybersecurity Big Data Analytics},
	Title = {High Performance Hypergraph Analytics of Domain Name System Relationships},
	Year = {2019}}

@incollection{Joslyn2016,
	Address = {Santa Clara University},
	Author = {Cliff Joslyn and Emilie Purvine},
	Booktitle = {Association for Women in Mathematics Series},
	Doi = {10.1007/978-3-319-34139-2\textunderscore19},
	Pages = {379--400},
	Publisher = {Springer International Publishing},
	Title = {Information Measures of Frequency Distributions with an Application to Labeled Graphs},
	Year = {2016}}

@article{kaminski2018clustering,
	Author = {Kaminski, Bogumil and Poulin, Valerie and Pralat, Pawel and Szufel, Przemyslaw and Theberge, Francois},
	Journal = {arXiv preprint arXiv:1810.04816},
	Title = {Clustering via Hypergraph Modularity},
	Year = {2018}}

@article{erdHos1960evolution,
	Author = {Erd{\H{o}}s, Paul and R{\'e}nyi, Alfr{\'e}d},
	Journal = {Publ. Math. Inst. Hung. Acad. Sci},
	Number = {1},
	Pages = {17--60},
	Title = {On the evolution of random graphs},
	Volume = {5},
	Year = {1960}}

@article{Estrada2006,
	Author = {Ernesto Estrada and Juan A. Rodr{\'{\i}}guez-Vel{\'{a}}zquez},
	Doi = {10.1016/j.physa.2005.12.002},
	Journal = {Physica A: Statistical Mechanics and its Applications},
	Month = {may},
	Pages = {581--594},
	Publisher = {Elsevier {BV}},
	Title = {Subgraph centrality and clustering in complex hyper-networks},
	Volume = {364},
	Year = {2006}}

@article{Everett2013,
	Author = {M.G. Everett and S.P. Borgatti},
	Doi = {10.1016/j.socnet.2012.05.004},
	Journal = {Social Networks},
	Month = {may},
	Number = {2},
	Pages = {204--210},
	Publisher = {Elsevier {BV}},
	Title = {The dual-projection approach for two-mode networks},
	Volume = {35},
	Year = {2013}}

@article{Fong2018,
	Author = {Brendan Fong and David I Spivak},
	Date = {2018-06-21},
	Eprint = {1806.08304v3},
	Eprintclass = {math.CT},
	Eprinttype = {arXiv},
	Title = {Hypergraph Categories},
	Year = {2019}}

@article{Freeman1978,
	Author = {Linton C. Freeman},
	Doi = {10.1016/0378-8733(78)90021-7},
	Journal = {Social Networks},
	Month = {jan},
	Number = {3},
	Pages = {215--239},
	Publisher = {Elsevier {BV}},
	Title = {Centrality in social networks conceptual clarification},
	Volume = {1},
	Year = {1978}}

@article{Garriga2010,
	Author = {Gemma C. Garriga and Esa Junttila and Heikki Mannila},
	Doi = {10.1007/s10115-010-0319-7},
	Journal = {Knowledge and Information Systems},
	Month = {jul},
	Number = {1},
	Pages = {197--226},
	Publisher = {Springer Nature},
	Title = {Banded structure in binary matrices},
	Volume = {28},
	Year = {2010}}

@article{Ghoshal2009,
	Author = {Gourab Ghoshal and Vinko Zlati{\'{c}} and Guido Caldarelli and M. E. J. Newman},
	Doi = {10.1103/physreve.79.066118},
	Journal = {Physical Review E},
	Month = {jun},
	Number = {6},
	Publisher = {American Physical Society ({APS})},
	Title = {Random hypergraphs and their applications},
	Volume = {79},
	Year = {2009}}

@article{Goh2007,
	Author = {K.-I. Goh and M. E. Cusick and D. Valle and B. Childs and M. Vidal and A.-L. Barabasi},
	Doi = {10.1073/pnas.0701361104},
	Journal = {Proceedings of the National Academy of Sciences},
	Month = {may},
	Number = {21},
	Pages = {8685--8690},
	Publisher = {Proceedings of the National Academy of Sciences},
	Title = {The human disease network},
	Volume = {104},
	Year = {2007}}

@article{Guo2016a,
	Author = {Jin-Li Guo and Qi Suo and Ai-Zhong Shen and Jeffrey Forrest},
	Doi = {10.1038/srep33651},
	Journal = {Scientific Reports},
	Month = {sep},
	Number = {1},
	Publisher = {Springer Science and Business Media {LLC}},
	Title = {The Evolution of Hyperedge Cardinalities and Bose-Einstein Condensation in Hypernetworks},
	Volume = {6},
	Year = {2016}}

@article{Conyon2004,
	Author = {Martin J. Conyon and Mark R. Muldoon},
	Doi = {10.2139/ssrn.546963},
	Journal = {{SSRN} Electronic Journal},
	Publisher = {Elsevier {BV}},
	Title = {The Small World Network Structure of Boards of Directors},
	Year = {2004}}

@article{cooley2018subcritical,
	Author = {Cooley, Oliver and Fang, Wenjie and Del Giudice, Nicola and Kang, Mihyun},
	Journal = {arXiv preprint arXiv:1810.08107},
	Title = {Subcritical random hypergraphs, high-order components, and hypertrees},
	Year = {2018}}

@article{Cooley2016ThresholdAH,
	Author = {Oliver Cooley and Mihyun Kang and Christoph Koch},
	Journal = {Electr. J. Comb.},
	Pages = {P2.48},
	Title = {Threshold and Hitting Time for High-Order Connectedness in Random Hypergraphs},
	Volume = {23},
	Year = {2016}}

@article{Cooley2015,
	Author = {Oliver Cooley and Mihyun Kang and Christoph Koch},
	Doi = {10.1016/j.endm.2015.06.077},
	Journal = {Electronic Notes in Discrete Mathematics},
	Month = {nov},
	Pages = {569--575},
	Publisher = {Elsevier {BV}},
	Title = {Evolution of high-order connected components in random hypergraphs},
	Volume = {49},
	Year = {2015}}

@article{cooper2012spectra,
	Author = {Cooper, Joshua and Dutle, Aaron},
	Journal = {Linear Algebra and its Applications},
	Number = {9},
	Pages = {3268--3292},
	Publisher = {Elsevier},
	Title = {Spectra of uniform hypergraphs},
	Volume = {436},
	Year = {2012}}

@article{Darling2005,
	Author = {R. W. R. Darling and J. R. Norris},
	Doi = {10.1214/105051604000000567},
	Journal = {The Annals of Applied Probability},
	Month = {feb},
	Number = {1A},
	Pages = {125--152},
	Publisher = {Institute of Mathematical Statistics},
	Title = {Structure of large random hypergraphs},
	Volume = {15},
	Year = {2005}}

@article{Dewar2018,
	Author = {Megan Dewar and John Healy and Xavier P{\'{e}}rez-Gim{\'{e}}nez and Pawe{\l} Pra{\l}at and John Proos and Benjamin Reiniger and Kirill Ternovsky},
	Doi = {10.24166/im.03.2018},
	Journal = {Internet Mathematics},
	Month = {mar},
	Publisher = {Internet Mathematics},
	Title = {Subhypergraphs in non-uniform random hypergraphs},
	Year = {2018}}

@article{dinur2005hardness,
	Author = {Dinur, Irit and Regev, Oded and Smyth, Clifford},
	Journal = {Combinatorica},
	Number = {5},
	Pages = {519--535},
	Publisher = {Springer},
	Title = {The hardness of 3-uniform hypergraph coloring},
	Volume = {25},
	Year = {2005}}

@article{Dorfler1980,
	Author = {W. D\"{o}rfler and D. A. Waller},
	Doi = {10.1007/bf01224952},
	Journal = {Archiv der Mathematik},
	Month = {dec},
	Number = {1},
	Pages = {185--192},
	Publisher = {Springer Science and Business Media {LLC}},
	Title = {A category-theoretical approach to hypergraphs},
	Volume = {34},
	Year = {1980}}

@article{Barber2007,
	Author = {Michael J. Barber},
	Doi = {10.1103/physreve.76.066102},
	Journal = {Physical Review E},
	Month = {dec},
	Number = {6},
	Publisher = {American Physical Society ({APS})},
	Title = {Modularity and community detection in bipartite networks},
	Volume = {76},
	Year = {2007}}

@book{Berge1984,
	Address = {Amsterdam},
	Author = {C. Berge},
	Isbn = {9780080880235},
	Publisher = {North Holland},
	Title = {Hypergraphs: Combinatorics of Finite Sets (North-Holland Mathematical Library)},
	Year = {1984}}

@article{bermond1977line,
	Author = {Bermond, Jean-Claude and Heydemann, Marie-Claude and Sotteau, Dominique},
	Journal = {Discrete Mathematics},
	Number = {3},
	Pages = {235--241},
	Publisher = {Elsevier},
	Title = {Line graphs of hypergraphs I},
	Volume = {18},
	Year = {1977}}

@article{Bolla1993,
	Author = {Marianna Bolla},
	Doi = {10.1016/0012-365x(93)90322-k},
	Journal = {Discrete Mathematics},
	Month = {jul},
	Number = {1-3},
	Pages = {19--39},
	Publisher = {Elsevier {BV}},
	Title = {Spectra, Euclidean representations and clusterings of hypergraphs},
	Volume = {117},
	Year = {1993}}

@book{Bretto2013,
	Address = {Berlin/Heidelberg, Germany},
	Author = {Alain Bretto},
	Doi = {10.1007/978-3-319-00080-0},
	Publisher = {Springer International Publishing},
	Title = {Hypergraph Theory},
	Year = {2013}}

@article{chitra2019random,
	Author = {Chitra, Uthsav and Raphael, Benjamin J},
	Journal = {arXiv preprint arXiv:1905.08287},
	Title = {Random Walks on Hypergraphs with Edge-Dependent Vertex Weights},
	Year = {2019}}

@article{chodrow2019configuration,
	Author = {Chodrow, Philip S},
	Journal = {arXiv preprint arXiv:1902.09302},
	Title = {Configuration Models of Random Hypergraphs and their Applications},
	Year = {2019}}

@article{chung1993laplacian,
	Author = {Chung, Fan},
	Journal = {Expanding graphs (DIMACS series)},
	Pages = {21--36},
	Title = {The Laplacian of a hypergraph},
	Year = {1993}}

@book{chung2006complex,
	Address = {Providence, Rhode Island},
	Author = {Chung, Fan},
	Number = {107},
	Publisher = {American Mathematical Soc.},
	Title = {Complex graphs and networks},
	Year = {2006}}

@inproceedings{Agarwal2006,
	Author = {Sameer Agarwal and Kristin Branson and Serge Belongie},
	Booktitle = {Proceedings of the 23rd international conference on Machine learning - {ICML} {'}06},
	Doi = {10.1145/1143844.1143847},
	Publisher = {{ACM} Press},
	Title = {Higher order learning with graphs},
	Year = {2006}}

@book{Agresti2012,
	Address = {University of Michigan},
	Author = {Alan Agresti},
	Isbn = {978-0-470-08289-8},
	Publisher = {Wiley},
	Title = {Analysis of Ordinal Categorical Data (Wiley Series in Probability and Statistics Book 656)},
	Year = {2012}}

@article{Aksoy2017,
	Author = {Sinan G. Aksoy and Tamara G. Kolda and Ali Pinar},
	Doi = {10.1093/comnet/cnx001},
	Journal = {Journal of Complex Networks},
	Month = {mar},
	Number = {4},
	Pages = {581--603},
	Publisher = {Oxford University Press ({OUP})},
	Title = {Measuring and modeling bipartite graphs with community structure},
	Volume = {5},
	Year = {2017}}

@article{alon1990transversal,
	Author = {Alon, Noga},
	Journal = {Graphs and Combinatorics},
	Number = {1},
	Pages = {1--4},
	Publisher = {Springer},
	Title = {Transversal numbers of uniform hypergraphs},
	Volume = {6},
	Year = {1990}}

@article{Alvarez-Socorro2015,
	Author = {A. J. Alvarez-Socorro and G. C. Herrera-Almarza and L. A. Gonz{\'{a}}lez-D{\'{\i}}az},
	Doi = {10.1038/srep17095},
	Journal = {Scientific Reports},
	Month = {nov},
	Number = {1},
	Publisher = {Springer Nature},
	Title = {Eigencentrality based on dissimilarity measures reveals central nodes in complex networks},
	Volume = {5},
	Year = {2015}}

@book{Axinn1984,
	Address = {Chicago, Illinois},
	Author = {Stephen M. Axinn and Phillip A. Proger and Norman Yoerg},
	Isbn = {0897071514},
	Publisher = {Amer Bar Assn},
	Title = {Interlocking Directorates Under Section 8 of the Clayton Act (Monograph, American Bar Association, Section of Antitrust Law, 10) (5030057)},
	Year = {1984}}

@book{Barabasi2016,
	Address = {Cambridge},
	Author = {Albert L{\'a}szl{\'o} Barab{\'a}si},
	Isbn = {1107076269},
	Publisher = {Cambridge University Press},
	Title = {Network Science},
	Year = {2016}}


\appendix
\section{Hypergraph random walks and $s$-walks}

While not the focus of the present
work, the study of random walks is intimately related to many branches
of graph and hypergraph theory, underlying analytic methods such as PageRank,
diffusion processes such as chip-firing and load-balancing in distributed
networks, and clustering methods. One popular way of defining a
random walk on a hypergraph $H=(V,E)$ is as a discrete time Markov Chain,
$X_{1},X_{2},\ldots $\,, with state space $V$, such that if
$X_{t}=v_{t}$, then we:

\begin{enumerate}
\item Select an edge $e \ni v_t$, either at random or according to given edge weights. 
\item Select a vertex $v \in e$, either at random or according to given vertex weights; set $X_{t+1}=v$.
\end{enumerate}

This process defines a probability transition matrix, symmetrizations of
which yield certain \textit{Laplacian matrices} frequently used as inputs
to clustering methods, like spectral clustering \cite{Luxburg2007} and
non-negative matrix factorization \cite{Kuang2012}. For instance, Zhou
\cite{zhou2007learning} proposed a hypergraph Laplacian which may be used
to cluster a hypergraph's vertices according to a normalized hypergraph
cut criterion. Other hypergraph Laplacian matrices have been proposed by
Rodriguez \cite{Rodriextasciiacuteguez2002} and Bolla
\cite{Bolla1993}. However, Agarwal \cite{Agarwal2006} proved these Laplacians,
while defined on the hypergraph, are nonetheless closely related to Laplacians
of graphs derived from the hypergraph, such as the aforementioned 2-section
and bipartite graph. Consequently, neither these Laplacians, nor the clustering
methods that utilize them, make full use of the higher-order relationships
present in hypergraphs but absent in graphs. Nonetheless, recent work by
Chitra and Raphael \cite{chitra2019random}, has identified a potential
culprit underlying this shortcoming: these Laplacians are based on random
walks featuring so-called edge-independent vertex weights. They show edge-\textit{dependent}
vertex weights (i.e. each vertex has a collection of weights, one for
each hyperedge to which it belongs) is a necessary, albeit not sufficient,
criterion for defining a random walk on a hypergraph that isn't equivalent
to some random walk on the 2-section.

We note there are at least two ways of utilizing the
$s$-walk framework to define random walks: (1)~as an $s$-weighted random
walk, and (2)~as an $s$-stratified set of random walks. In the former,
the intersection cardinalities between hyperedges serve as the weights
for transitioning, which is equivalent to a weighted random walk on the
graph with adjacency matrix $S^{T}S$, where $S$ is the hypergraph incidence
matrix. In this case, Laplacian matrices derived from this walk are still
subject to Agarwal's aforementioned criticism. In the latter case, one
considers a set of random walks, one for each $s$-line graph (see Definition~\ref{def:s-line}), which may be either weighted or unweighted. However,
whether and how this set of random walks might be utilized to define a
Laplacian (or whether there is a different way to utilize $s$-walks to
define stochastic processes of interest) is an interesting topic we leave
to future work.

\begin{samepage}
\begin{table}[h]
\begin{flushleft}
\noindent \mbox{\noindent \Large{\bf B \ The $s$-connected components of the data}}
\end{flushleft}
\label{fig:sCompViz}
\centering
\scalebox{0.47}
{
\begin{tabular}{|c|c|c|c|}
\hline
& LesMis$^*$ & Diseasome & CompBoard \\
  \hline
  \hline
  \rotatebox[origin=c]{90}{1-components}
  &
  \begin{minipage}{.6\textwidth}
  \centering
    \vspace{1mm}
\includegraphics[width=\linewidth]{lm2_1_full_crop}
    \end{minipage}
    &
      \begin{minipage}{.6\textwidth}
  \centering
\includegraphics[width=\linewidth,height=0.8\linewidth]{dis2_1_full_crop}
    \end{minipage}
    &
      \begin{minipage}{.6\textwidth}
  \centering
\includegraphics[width=\linewidth,height=0.8\linewidth]{cb2_1_full_crop}
    \end{minipage}
    \\
    \hline
      \rotatebox[origin=c]{90}{2-components}
  &
  \begin{minipage}{.6\textwidth}
    \vspace{1mm}
  \centering
\includegraphics[width=\linewidth]{lm2_2_full_crop}
    \end{minipage}
    &
      \begin{minipage}{.6\textwidth}
  \centering
\includegraphics[width=\linewidth, height=0.8\linewidth]{dis2_2_full_crop}
    \end{minipage}
    &
      \begin{minipage}{.6\textwidth}
  \centering
\includegraphics[width=\linewidth,height=0.75\linewidth]{cb2_2_full_crop}
    \end{minipage}
    \\
    \hline
      \rotatebox[origin=c]{90}{3-components}
  &
  \begin{minipage}{.6\textwidth}
    \vspace{1mm}
  \centering
\includegraphics[width=\linewidth]{lm2_3_full_crop}
    \end{minipage}
    &
      \begin{minipage}{.6\textwidth}
  \centering
\includegraphics[width=\linewidth]{dis2_3_full_crop}
    \end{minipage}
    &
      \begin{minipage}{.6\textwidth}
  \centering
\includegraphics[width=\linewidth]{cb2_3_full_crop2}
    \end{minipage}
    \\
    \hline
      \rotatebox[origin=c]{90}{4-components}
  &
  \begin{minipage}{.6\textwidth}
    \vspace{1mm}
  \centering
\includegraphics[width=\linewidth]{lm2_4_full_crop}
    \end{minipage}
    &
      \begin{minipage}{.6\textwidth}
  \centering
\includegraphics[width=\linewidth]{dis2_4_full_crop}
    \end{minipage}
    &
      \begin{minipage}{.6\textwidth}
  \centering
\includegraphics[width=\linewidth]{cb2_4_full_crop}
    \end{minipage}
    \\
    \hline
      \rotatebox[origin=c]{90}{5-components}
  &
  \begin{minipage}{.6\textwidth}
    \vspace{1mm}
  \centering
\includegraphics[width=\linewidth]{lm2_5_full_crop}
    \end{minipage}
    &
      \begin{minipage}{.6\textwidth}
  \centering
\includegraphics[width=\linewidth]{dis2_5_full_crop}
    \end{minipage}
    &
      \begin{minipage}{.6\textwidth}
  \centering
\includegraphics[width=\linewidth]{cb2_5_full_crop}
    \end{minipage}
    \\
    \hline

    \end{tabular}
}
\end{table}

\end{samepage}

\end{document}